\documentclass{aa}  

\usepackage{graphicx}
\usepackage{txfonts}
\usepackage{lipsum}
\usepackage{ulem}
\usepackage[]{hyperref}
\hypersetup{
    colorlinks=true,
    linkcolor=blue,
    filecolor=magenta,
    citecolor=blue,
    urlcolor=blue,
    pdftitle={Overleaf Example},
    pdfpagemode=FullScreen,
    }

\newcommand{\MESA}{\textsc{mesa}}
\newcommand{\JKTEBOP}{\textsc{jktebop}}
\newcommand{\ALLESFITTER}{\textsc{allesfitter}}
\newcommand{\WOTAN}{\textsc{wotan}}
\newcommand{\ISPEC}{\textsc{ispec}}
\newcommand{\teff}{$T_{\rm eff}$}

\begin{document}

   \title{UX~Men: Detached Eclipsing Binary as a Benchmark Candidate}

   \subtitle{}

\author{G. N. Pawar\inst{1}\fnmsep\thanks{E-mail: pawarganeshwork@gmail.com}
\and
K. G. He{\l}miniak\inst{1}
\and 
N. J. Miller\inst{2}
\and
F. Marcadon\inst{3}
\and
A. Moharana\inst{4}
\and
T. B. Pawar\inst{5}
\and
A. Kalsi\inst{1}
\and
M.~Konacki\inst{1}
}

\institute{Nicolaus Copernicus Astronomical Center, Polish Academy of Sciences, ul. Rabia\'{n}ska 8, 87-100 Toru\'{n}, Poland
\and
Department of Physics and Astronomy, Uppsala University, Box 516, S-75120 Uppsala, Sweden
\and
Private Researcher, Orléans, France
\and
Astrophysics Group, Keele University, Staffordshire, ST5 5BG, U.K.
\and
Villanova University, Dept.\ of Astrophysics and Planetary Sciences, 800 East Lancaster Avenue, Villanova, PA 19085, USA
        }

   \date{Received June 05, 2024; accepted }

 
  \abstract
   {Detached eclipsing binaries (DEBs) provide model‑independent measurements of stellar masses and radii, essential for testing stellar‑evolution models and calibrating space‑based photometric missions.}
   {We aim to deliver ultra-precise fundamental parameters for the DEB UX Men and to assess its suitability as a benchmark for upcoming missions such as \textit{PLATO}. We further evaluate the temporal stability of the photometric solutions, evaluate how intrinsic stellar variability and instrumental systematics affect the derived parameters.}
   {We analysed 31 {\it TESS} sectors (\textasciitilde 882 d) with the {\sc lightkurve} package, and detrended spot‑induced variability. Light‑curve modelling was performed independently with {\sc jktebop} and {\sc allesfitter}. Radial velocities from HARPS (HAM/EGGS), FEROS, CORALIE, and UCLES were extracted via {\sc todcor} and fitted with a Keplerian orbit. Spectral disentangling (5350–5800 Å) yielded individual component spectra, which were processed with {\sc ispec} to determine \teff{}, $\mathrm{[M/H]}$, micro‑ and macroturbulence. Eclipse‑timing variations were examined through O–C diagrams. We have also used multi-band photometry to determine the fluxes, flux ratios and ultimately to measure the photometric \teff{}.}
   {We obtain the masses $M_{\rm A}$ = $1.2300 \pm{0.0012}$ M$_{\odot}$, $M_{\rm B}$ = $1.1946\pm{0.0012}$ M$_{\odot}$ and radii $R_{\rm A}$ = $1.3605\pm{0.0031}$ R$_{\odot}$, $R_{\rm B}$ = $1.2801\pm{0.0069}$ R$_{\odot}$ with sub-percent precision. { Uncertainties were found to be dominated by systematic effects coming from (weak) stellar activity.} Metallicity estimates are $[M/H]_{\rm A} = 0.02\pm{0.13}$ dex and $[M/H]_{\rm B}= 0.07\pm{0.10}$ dex. Isochrone fitting yields a consistent age of $\tau$ = $2.75\pm{0.13}$ Gyr. The system exhibits a \textasciitilde 673-day periodicity in eclipse‑timing variations, likely due to stellar activity, with no evidence for a third body.
   }
   {UX~Men is a high‑precision benchmark binary, providing stringent tests of stellar models and a robust calibration source for {\it PLATO}’s determination of stellar and exoplanet properties. We found no dependence of the results on the code used. However, without a proper treatment of systematics caused by activity, the uncertainties would be underestimated. Expanding the sample of similarly well‑characterised DEBs will further strengthen the empirical foundation for stellar physics and improve the calibration of future photometric missions.}


   \maketitle
%

\section{Introduction}

Detached eclipsing binaries (DEBs) are the cornerstone of stellar astrophysics because their geometric configuration allows for direct, model-independent measurements of fundamental stellar properties, such as masses, radii, and luminosities. DEBs are generally long-period systems in which the component stars can be assumed to have evolved independently. Except for the Sun and a few nearby stars such as $\alpha$ Cen, Sirius A, $\tau$ Ceti, etc; whose masses and/or radii can be obtained model-independently through interferometry \citep{2003A&A...408..681K, 2017ApJ...840...70B, 2017A&A...597A.137K,2021AJ....162...14A}, astrometric orbits, or asteroseismology and which serve as fundamental benchmark stars \citep{GBS2015A&A...582A..49H, 2019ARA&A..57..571J}, DEBs are the \textbf{primary} source of precise, model-independent mass and radius measurements \citep{2010A&ARv..18...67T,maxted2020,2021Univ....7..369S}.  Over the past two decades, large surveys and space-based photometry (e.g., {\it TESS, Kepler}) have increased the number of well-characterised DEBs.\footnotemark[1] \footnotetext[1]{\url{https://www.astro.keele.ac.uk/jkt/debcat/}} Along with spectroscopy from high-resolution échelle spectrographs, it is possible to measure the radii and masses of DEB components with precision and accuracy better than 1\% \citep{maxted2020}. The high‑precision stellar parameters from DEBs provide stringent tests of stellar‑strcture and evolution models. Sub-percent masses and radii constrain the treatment of convection and core-boundary mixing in low- and intermediate-mass stars, as shown by \cite{2014ApJ...797...31T} for V530 Ori, where such measurements reveal deficiencies in the modelling of convection. At lower masses they expose the well-known radius discrepancy of K and M-type stars, whose observed radii exceed model predictions by 5-10\% at fixed mass \citep{2013AN....334....4T, 2018AJ....155..225K}. Eccentric DEBs additionally probe the internal density concentration through apsidal motion, comparing the observed internal-structure constant $k_{2}$ against model predictions \citep{Claret2010A&A...519A..57C, Claret2021A&A...654A..17C}, and their fundamental radii help calibrate the limb-darkening laws used in light-curve and transit modelling \citep{maxted2018A&A...616A..39M}. DEBs also calibrate surface‑brightness–colour relations {\citep{2021A&A...649A.109G}, and serve as benchmark systems for distance scales \citep{2014ApJ...780...59G}. Their model-independent masses, radii, together with the precise ages derived from isochrone fitting, anchor stellar models used to characterise exoplanet host stars, making well-characterised solar-type DEBs particularly valuable for missions such as PLATO.

UX Mensae (here by UX Men) also known as HD 37513, HIP 25760, {\it Gaia} DR3 4648210527690481152 is a DEB and a double-lined spectroscopic binary (SB2) that has been studied over the years { (see below)}. The pair consists of two near‑solar‑type components with precisely measured fundamental parameters. The astrometric and photometric properties are available in the Table~\ref{tab:properties}.

\begin{table}
\centering
  \caption{Astrometric and photometric properties of UX~Men.}
   \begin{tabular}{lccc}
        \hline\hline
        
        Parameter & Value & Reference\\
        \hline
        RA J2000 & 05 30 03.1855 &  \citet{2022yCat.1355....0G} \\
        Dec J2000 & -76 14 55.3503 & \citet{2022yCat.1355....0G} \\
        {  $\mu_{\rm RA}$ (mas/yr)} & $35.043\pm0.021$ & \citet{2022yCat.1355....0G} \\
        {   $\mu_{\rm Dec}$ (mas/yr)} &  $7.303\pm0.019$ & \citet{2022yCat.1355....0G} \\
        {  $\pi$ (mas)} & $9.6864\pm0.0146$ & \citet{2022yCat.1355....0G} \\
        {  RUWE \tablefootmark{a}} & 0.906 & \citet{2022yCat.1355....0G} \\
        $B$ (mag) & $8.80\pm0.02$ & \citet{tycHog}\\
        $V$ (mag) & $8.24\pm0.01$ &  \citet{tycHog}\\
        $G$ (mag) & $8.1000\pm0.0028$ & \citet{2022yCat.1355....0G} \\
        $I$ (mag) & $7.608\pm0.115$ & \citet{KiragASAS}\\
        $J$ (mag) & $7.195\pm0.027$ & \citet{2mass} \\
        $H$ (mag) & $6.976\pm0.027$ & \citet{2mass} \\
        $K$ (mag) & $6.915\pm0.024$ & \citet{2mass} \\
        \hline 
    \end{tabular}      
    \label{tab:properties}
    \tablefoot{
    \tablefoottext{a}{RUWE stands for renormalized unit weight error \citep{2021A&A...649A...2L}.}
    }
\end{table}

Its variability was first discovered in Bamberg patrol plates and reported by \cite{1966IBVS..140....1S}. Its orbital elements were first obtained by \cite{1974A&A....32..429I} and the first complete physical analysis with multi-band photometry and radial velocities (RVs) from \cite{1974A&A....32..429I} was done by \cite{1976A&A....48...49C}. Later, this photometric data set was reanalysed together with the new CORAVEL RVs by \cite{1989A&A...211..346A}. \cite{2006A&A...450..681T} has found a probable third body in NACO images from 2004, with an angular separation of 0.751 arcsec; however, we did not find any trends in the RV solutions for the observations spanning over \textasciitilde 1880 days. The target was also analysed using ASAS photometry and échelle spectroscopy for the latest orbital solution available by \cite{krisuxmen2009MNRAS.400..969H}, with unprecedented precision in the masses of 0.165\% and 0.168\% for the primary and secondary, respectively. \cite{2017ApJ...837....7G} has discussed that surface-brightness-colour (SBC) relations on EBs can achieve precision better than 1\% in predicting angular diameters for UX~Men-like systems. It has a relatively high apparent magnitude and precise geometric parameters. At the same time, its orbital period of { $P$ }$= 4.181093974$~d is long enough to reduce tidal distortion, making the components appear close to that of isolated single stars. Its high‑accuracy, model‑independent parameters, together with the system’s well‑characterised photometry and modest interstellar extinction, make UX~Men an excellent benchmark binary for calibrating stellar models, testing SBC relations, and validating upcoming high‑precision {\it Gaia} parallaxes. It is important to check the reliability of these precisions with high-precision photometry like {\it TESS} and high-resolution spectroscopy.

In this paper, we present a comprehensive analysis of the DEB UX~Men, exploiting the unprecedented time coverage of {\it TESS} (31 sectors, \textasciitilde 882 d) and a suite of high‑resolution spectrographs (HARPS, FEROS, UCLES). 
In Section~\ref{sec:observations} we describe observations, the {\it TESS} photometry (31 sectors,\textasciitilde882 days) and the high‑resolution spectroscopy, which were used for RV extraction and spectral disentangling. Section \ref{sec:analysis} outlines the complete analysis workflow: light curve reduction and detrending, independent light curve modelling with \JKTEBOP{} and \ALLESFITTER{} in \ref{sec:jktebop} and \ref{sec:allesfitter}, eclipse timing analysis in \ref{sec:etv}, Keplerian fitting of the RVs in \ref{sec:rvmodel} and determination of atmospheric parameters via \ISPEC{}. Section \ref{sec:results} presents the ultra‑precise masses, radii, and effective temperatures, and examines the \textasciitilde673‑day activity‑driven eclipse‑timing variations (ETVs). Section \ref{sec:conclusion} summarises our conclusions.

\section{Observations}\label{sec:observations}

\subsection{{\it TESS} Photometry}
UX~Men was observed in the Transiting Exoplanet Survey Satellite ({\it TESS}) mission \citep{2015JATIS...1a4003R} as TIC~141268467. textit{TESS} has a single bandpass filter spanning 600 nm to 1000 nm with a central wavelength at 786.5 nm. UX~Men resides in the southern continuous viewing zone (CVZ). We used observations that span a total of 31 sectors in cycles 1, 3, and 5, covering \textasciitilde 882 d (details are in Table \ref{tab:tess_obs_table}) with a cadence of 120~s\footnotemark[2]\footnotetext[2]{The target has been requested in the Guest Investigator (GI) programmes G011048, G011083, G011154, G03251, G05003, and G05155.}. { A single sector covers more than 6.5 orbital periods.}

\begin{table}
    \centering
     \caption{{\it TESS} observation log of UX~Men (TIC~141268467).}
    \begin{tabular}{lccc}
        \hline
        Sectors & Obs. Date  & No. of days \\
         \hline
         1-2   & 2018 Jul 25 -- 2018 Sep 20 & 58 \\
         4-7   & 2018 Oct 19 -- 2019 Feb 01 & 106 \\
         9-12  & 2019 Feb 28 -- 2019 Jun 18 & 111 \\
         27-32 & 2020 Jul 05 -- 2020 Dec 16 & 165 \\
         34-39 & 2021 Jan 14 -- 2021 Jun 24 & 162 \\
         61-62 & 2023 Jan 18 -- 2023 Mar 10 & 52 \\
         64-69 & 2023 Apr 06 -- 2023 Sep 20 & 168 \\
         \hline 
               &  Total & 882 \\
         \hline
    \end{tabular}
    \vspace{1em}
    \label{tab:tess_obs_table}
\end{table}

\begin{figure*}
    \centering
    \includegraphics[width=1\textwidth]{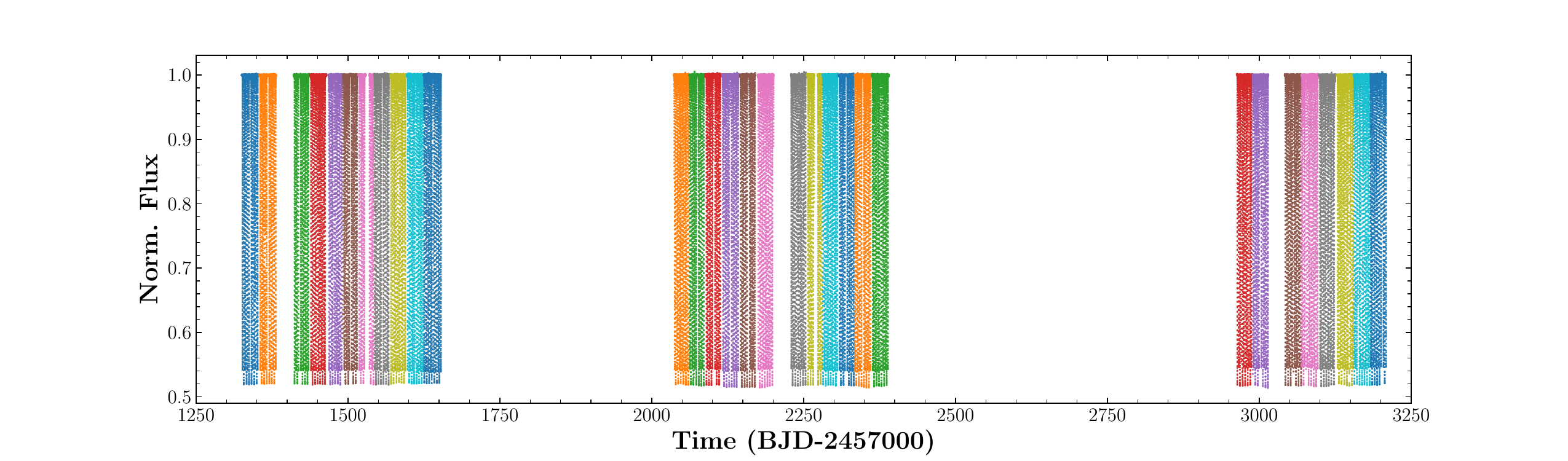}
    \caption{
     Normalised TESS light curve of UX Men obtained from all 31 observed sectors, colour‑coded by sector.
    }
    \label{fig:lc}
\end{figure*}

\subsection{High-resolution Spectroscopy}
UX~Men was observed with HARPS in two modes: High RV accuracy mode (HAM) with the resolution of R\textasciitilde 115,000  (PI: He{\l}miniak, K.)\footnotemark[3] and High-efficiency mode (EGGS) is R\textasciitilde 80,000 (PI: Gieren, W. \& Pietrzynski, G.)\footnotemark[3]. The average signal-to-noise ratio (SNR) of the spectra is \textasciitilde 73 (HAM) and \textasciitilde 40 (EGGS). The spectra range from 3781-6913 \AA. HARPS is a fibre-fed \'echelle spectrograph installed on the ESO 3.6-m telescope at the ESO's La Silla Observatory \citep{harps}. The high-resolution spectroscopic observations are acquired from the ESO data archives.

The system was also observed with FEROS on the MPG 2.2-m telescope located at ESO’s La Silla Observatory \citep{feros} with a spectral resolution of R\textasciitilde 48,000 (PI: He{\l}miniak, K. \& Graczyk, D.)\footnotemark[3] with an average SNR of \textasciitilde 90. \footnotetext[3]{Based on observations made with ESO Telescopes at the La Silla Paranal Observatory under programmes ID: 082.D-0933, 083.D-0549, 084.D-0591, 085.C-0614, 086.D-0078, 087.C-0112, 089.C-0415, 089.D-0097, 090.D-0061, 091.D-0414 and 190.D-0237.} Additional observations come from the CORALIE spectrograph at the 1.2-m~Euler telescope \citep{quel00}, also located in La~Silla, with R\textasciitilde 60,000, and a substantially lower SNR of $\sim$30.

Finally, we made use of observations made with the UCLES, a high-resolution echelle spectrograph located at the Anglo-Australian Telescope, with a spectral resolution of R\textasciitilde 60,000. The observation details are available in \cite{krisuxmen2009MNRAS.400..969H}.

All spectroscopic observations span over \textasciitilde1880 days, i.e. from 2008 Sep. 16 to 2013 Nov. 10. In total, we use 38 spectroscopic epochs (Table \ref{tab:RV_data}). Because the TESS photometry was obtained later, in 2018 - 2023, there is
no temporal overlap between the photometric and spectroscopic datasets. The RV Table \ref{tab:RV_data} shows the epoch of the observation from each spectrograph.
\begin{figure}[!ht]
    \centering
    \includegraphics[width=0.98\linewidth]{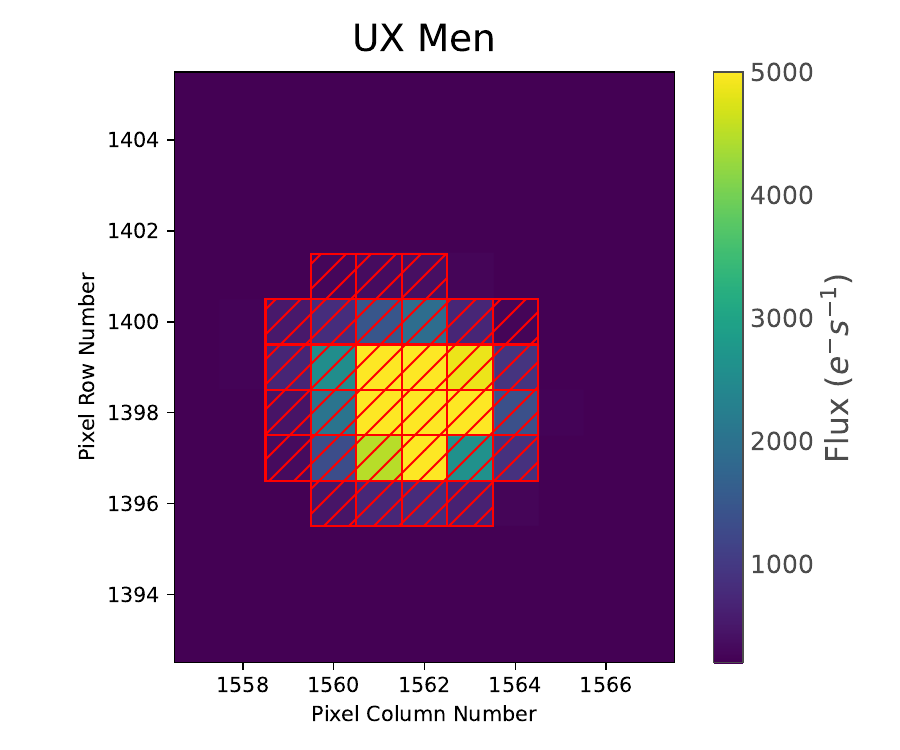}
    \caption{\textit{TESS} target‑pixel file for sector 29 with the customised photometric aperture used to extract the light curve. The aperture in the red mask isolates UX Men and excludes flux from the nearby contaminating star. The faint field stars (V-mag = 13+) are present but invisible on this scale; “contamination” here refers chiefly to scattered light from nearby bright Solar-System bodies, which the custom aperture is designed to suppress.}
    \label{fig:tpf}
\end{figure}

\section{Analysis}\label{sec:analysis}

\subsection{Light Curve Reduction}
We have extracted {\it TESS} photometric data using the {\sc lightkurve} package \citep{lightkurve2018ascl.soft12013L}, applying a customised aperture to prevent light leakage from the nearby contamination star and to prevent spillovers from the target star. In Fig. \ref{fig:tpf}, the aperture is derived from a single representative frame of Sector 29 (no frames were stacked) and is defined independently for
each sector, since the detector orientation and the level of scattered light (from Earth, the Moon, Jupiter or Venus) vary between
sectors. The extracted light curve had some spot variations, which were removed by detrending with \WOTAN{} \citep{wotan2019AJ....158..143H} and a Tukey’s bi-weight filter by selecting a sufficient window without losing the signals from the eclipses. The final output is a normalised light curve (LC). Figure \ref{fig:lc} shows normalised light curves over all 31 sectors.

\subsection{Light Curve Modelling}\label{sec:lcmodelling}
We have modelled individual sectors independently and to check the accuracy of the models, we have used two different codes with the same configuration. \JKTEBOP{} version 40 \citep{jktebop2004MNRAS.351.1277S, jktebop2013A&A...557A.119S} and \ALLESFITTER{} \citep{allesfitter-paper, allesfitter-code}. Modelling each sector independently is deliberate: the dispersion of the fitted parameters from sector-to-sector exposes the small, activity-driven systematic that a single global fit would otherwise absorb, yielding more realistic uncertainties (Section \ref{sec:resultsstellaractivity}). We obtained a total of 62 individual sets of parameters, as each of the 31 sectors was modelled independently with both codes (31 $\times$ 2 = 62 parameter sets) to assess consistency. As the adopted value we took the weighted average of each parameter (individual errors were used as weights), and as the adopted uncertainty we added in quadrature the error of the weighted average, and the $rms$ of 62 individual values.

\subsubsection{JKTEBOP}\label{sec:jktebop}
\JKTEBOP{} is suited to a well-detached system and is based on the EBOP code \citep{ebop1981ASIC...69..111E, ebop21981AJ.....86..102P}.
It assumes the stars are biaxial spheroids and computes light curves through numerical integration of concentric circles over each star. Parameters adjusted for \JKTEBOP{} models are the sum of fractional radii $(r_{A}+ r_{B})$ (where $r_{A}= R_{A}/a$ and $r_{B}= R_{B}/a$ and $a$ is the semi-major axis), radius ratio $(k=R_B/R_A=r_{B}/r_{A})$, orbital inclination $(i)$, orbital period $(P)$, surface brightness ratio (J), and reference time (associated with the mid-time of primary eclipse, $T_{0}$). The \JKTEBOP{} also allows to fit for the third light \textbf{$l_3$}, orbital eccentricity ($e$) and argument of the pericentre $(\omega)$, combined into $e\sin(\omega)$ and $e\cos(\omega)$.

\JKTEBOP{} uses a single-band light curve for the models, unlike other modelling codes like {\sc phoebe} \citep{phoebe}, thus it does not need the use of a model atmosphere. For limb darkening, we adopted the logarithmic law with coefficients \citep{claret2004A&A...428.1001C} that were specified from the estimates of $T_{\rm eff}$, $\log(g)$ and metallicity. 
Uncertainties were estimated with \JKTEBOP{}’s Monte Carlo task (10,000 simulations), in which the free parameters ($J, r_A+r_B, k, i, P, T_0, e\cos(\omega), e\sin(\omega)$ and $l_3$) are perturbed and re-fitted; each was sampled with a uniform prior around the best fit, and the reported errors are the standard deviations of the resulting fits. The figure \ref{fig:lcfit} shows an example of the LC model fit with the residuals for {\it TESS} sector 29.

\begin{figure}
    \centering
    \includegraphics[width=0.97\linewidth]{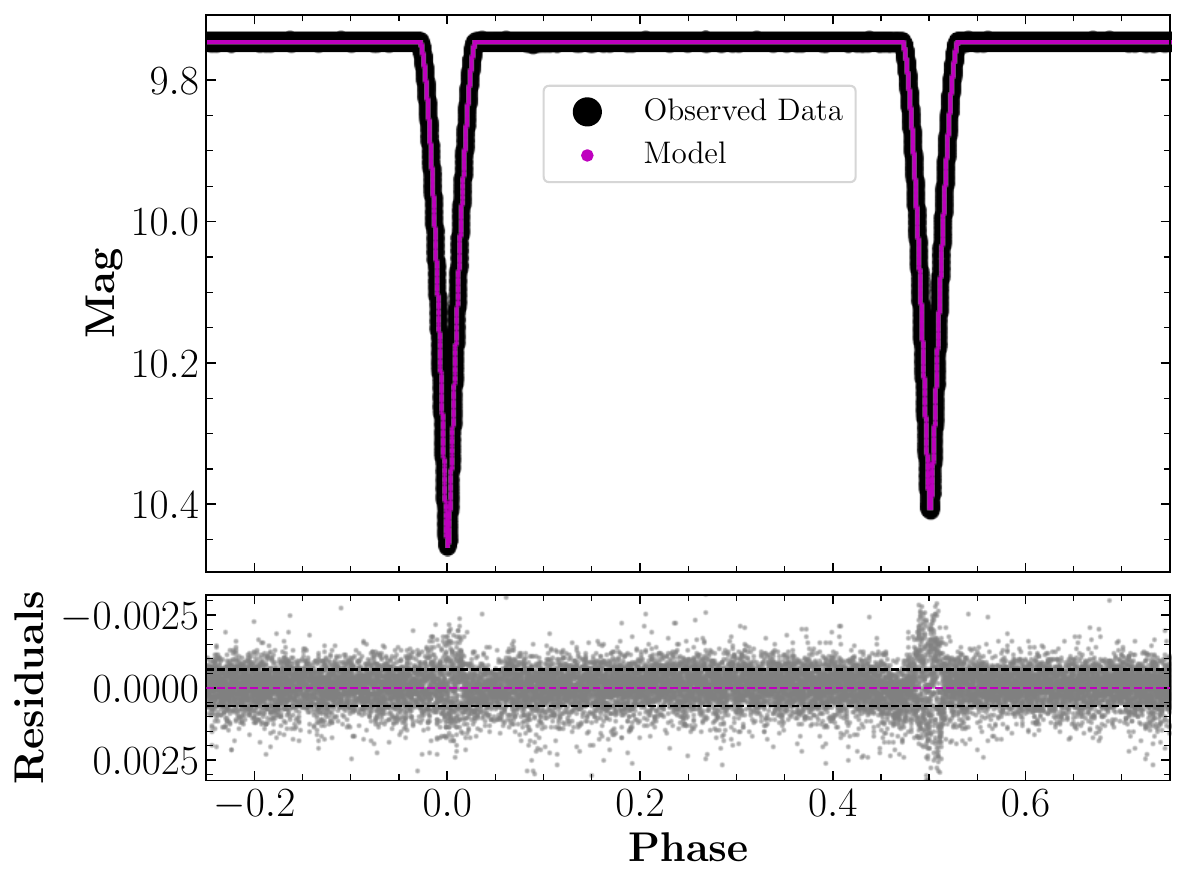}
    \caption{The phase light‑curve model fit for \textit{TESS} sector 29 (single‑sector view). The top panel displays the data points in black overlaid with the best-fit model in magenta. The bottom panel shows the residuals of the fit, with dotted black lines as the 1-sigma range. The light curve is the same detrended, normalised data as in Fig. \ref{fig:lc}, converted here in magnitudes (\JKTEBOP{}'s working unit); the magnitude zero-point is arbitrary.}
    
    \label{fig:lcfit}
\end{figure}

\subsubsection{ALLESFITTER}\label{sec:allesfitter}
\ALLESFITTER{} is an integration of many useful Python packages frequently used in modelling stellar or planetary systems. It is based on {\sc ellc} \citep{ellc2016A&A...591A.111M}, which generates lightcurves, and for Gaussian processes (GPs) it uses {\sc celerite} \citep{celerite2017AJ....154..220F}. It has two samplers that estimate the maximum likelihood of parameters: a nested sampler \citep[{\sc dynesty};][]{dynesty2020MNRAS.493.3132S} and a Markov Chain Monte Carlo (MCMC) sampler \citep[{\sc emcee};][]{emcee2013PASP..125..306F}.
\ALLESFITTER{} uses by default the differential-evolution move \textit{DEMove} \citep{demove2006S&C....16..239T}, in which trial steps are drawn from the difference vectors of randomly chosen walker pairs; this improves mixing and convergence for correlated parameters at low computational cost.

In \ALLESFITTER{} the orbital eccentricity ($e$) is re-parameterised as $\sqrt{e}\cos{\omega}$, $\sqrt{e}\sin{\omega}$, which imposes a uniform prior on $e$ and avoids the
boundary at $e$ = 0; \JKTEBOP{} instead fits ${e}\cos{\omega}$, and ${e}\sin{\omega}$, directly. For our near-circular orbit the two choices are equivalent. The third light ($l_{3}$) is expressed as dilution ratio ($D_{0}$) which relates as $D_{0}= l_{3}/(1+l_{3})$.

\ALLESFITTER{} is used to derive the full parameter set independently of \JKTEBOP{}, allowing a check for code-dependent biases and a different method of sampling with MCMC.
In our analysis to estimate the errors, the MCMC has 100 walkers with 10,000 steps (2,000 discarded as burn-in). These settings were chosen so that the autocorrelation length was many times shorter than the post-burn-in chain for every parameter and the chains satisfied the Gelman–Rubin convergence criterion; doubling the chain length left the posteriors unchanged. The agreement between these posteriors and the independent \JKTEBOP{} MC errors provides a further convergence check. We used the MCMC sampler rather than the nested sampler {\sc dynesty} because it was substantially faster here.

\subsection{Eclipse Timing Analysis}
\label{sec:etv}
We recalculate the orbital period $P$ and reference time $T_0$ with the eclipse timing analysis, based on the traditional prescription of the mid-time $T$ of a given eclipse in the cycle number $E$ to be: $T = T_0 + PE$.

In order to determine the mid-eclipse times of UX~Men, we used the procedure developed by \citet{2024ApJ...976..242M}. We summarize here the main steps of the procedure, and we refer the reader to \citet{2024MNRAS.527...53M}, \citet{2026A&A...705A.251P}, and \citet{2026A&A...710A.394M} for recent applications.
We adopted the phenomenological model of \citet{2015A&A...584A...8M}, in which the eclipse profile function is written as
\begin{multline}
    \psi(t_i,T,d,\Gamma,C) = \Bigg\{ 1 + C \bigg( \frac{t_i-T}{d} \bigg)^2 \Bigg\} \\ \times \Bigg\{ 1-\bigg\{ 1-\exp \bigg[ 1-\cosh \bigg( \frac{t_i-T}{d} \bigg) \bigg] \bigg\}^\Gamma \Bigg\}.
\end{multline}
Here, $T$, $d$, $\Gamma$, and $C$ are the time of minimum, the eclipse width, the kurtosis, and the scaling parameter, respectively.

The times of minima derived from our fitting procedure are listed in Table~\ref{tab:ecl_times} in Appendix~\ref{app:ecl_times}, along with the errors representing 1$\sigma$ percentiles from the MCMC posteriors. Our updated ephemeris are 
\begin{equation}
    T = 2459967.5382054(30) + 4.181093974(94) E.
\end{equation}

The final step consists of computing the difference between the observed
and calculated times of minima, i.e.,
\begin{equation}
    \Delta_{\rm obs} = T_{\rm o}(E) - T_{\rm c}(E) = T_{\rm o}(E) - T_0 - P E,
\end{equation}
where $T_{\rm o}(E)$ and $T_{\rm c}(E)$ refer to the observed and calculated times of minima at cycle $E$, respectively. The cycle number $E$ is counted from the reference epoch $T_{0}$, so that $E = 0$ corresponds to the eclipse at $T_{0}$; integer values denote primary eclipses and half-integer values secondary eclipses. The choice of reference epoch is arbitrary and does not affect the results.

To estimate the orbital eccentricity, we modelled the time differences between primary and secondary eclipses as
\begin{equation}
    \Delta_{\rm mod} = c_0 + c_1 E - (T_{\rm sec} - T_{\rm pri}) + 0.5\,(P+c_1),
    \label{eq:oc_model}
\end{equation}
where $c_0$ and $c_1$ are factors that correct the respective values of $T_0$ and $P$. Here, the term $(T_{\rm sec}-T_{\rm pri})$ corresponds to the time interval between the primary and secondary eclipses (\citealt{1959cbs..book.....K,2001icbs.book.....H}):
\begin{equation}
    \frac{2 \pi \, (T_{\rm sec}-T_{\rm pri})}{P+c_1} = \pi + 2 \tan^{-1} \frac{e \cos \omega}{(1-e^2)^{1/2}} + \frac{2 e \cos \omega \, (1-e^2)^{1/2}}{(1-e^2 \sin^2 \omega)},
    \label{eq:time_interval}
\end{equation}
    where $e$ and $\omega$ are the eccentricity and the argument of periastron, respectively. In Fig.~\ref{fig:ETV}, each minimum is plotted against its time of occurrence ($BJD
-2457000$), and the lower panel shows the normalised residuals $(O-C)/\sigma$ with $O-C = \Delta_{\rm obs}-\Delta_{\rm mod}$. In Table~\ref{tab:ecl_times}, the ``$O-C$'' column gives $\Delta_{\rm obs}$ for the primary eclipses ($\Delta_{\rm mod}=0$) and $\Delta_{\rm obs}-\Delta_{\rm mod}$ for the secondary eclipses. By fitting equations~(\ref{eq:oc_model}) and~(\ref{eq:time_interval}), we derived the values of the eccentricity components $e \cos \omega$ and $e \sin \omega$. We found $e \cos \omega = 0.000551 \pm 0.000012$ and $e \sin \omega = 0.00070^{+0.00044}_{-0.00011}$. These results are compared in Section~\ref{sec:corr_param} with those obtained from the LC analysis.

\begin{figure}[tbp]
    \includegraphics[width=0.98\columnwidth]{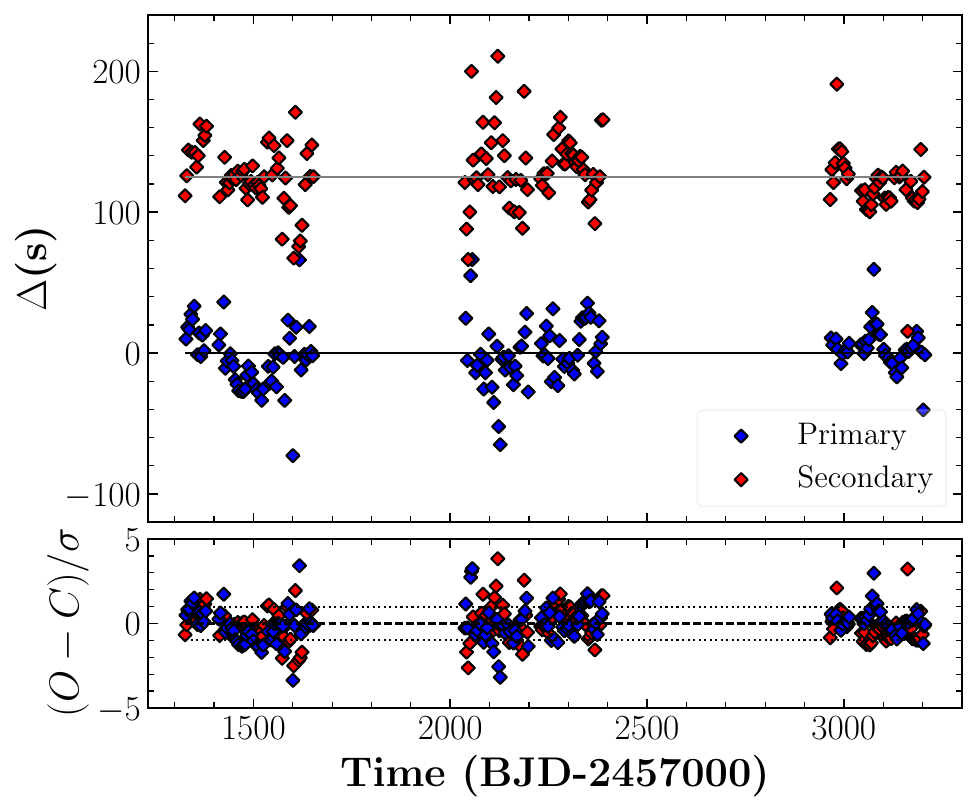}
    \caption{Eclipse timing of UX~Men. The primary and secondary eclipse times are indicated by blue and red symbols, respectively, while the linear primary and secondary eclipse ephemerides are shown as black and grey solid lines, respectively. Fitting residuals are shown in the lower panel.}
    \label{fig:ETV}
\end{figure}

\subsection{Radial Velocities -- extraction and fitting}\label{sec:rvmodel}
RVs are extracted from the spectrum of double-lined spectroscopic binaries using the {\sc todcor} algorithm \citep{zuckermazeh}, with a synthetic spectrum as a template. The details are described by \cite{todcor2005ApJ...626..431K}. In case of UX~Men we used spectra of a $T_{\rm eff}=6000$~K, $\log(g)=4.5$~dex, [$M/H$]=0.0 dex star. The extracted RVs were fitted with the \textsc{v2fit} code \cite{v2fit2010ApJ...719.1293K}, which performs a Levenberg–Marquardt $\chi^2$ minimisation. Both RV curves share a single set of orbital elements ($P$,$T_{\rm 0}$, $e$, $\omega$, $\gamma$) but have distinct semi-amplitudes $K_{A}$ and $K_{B}$, from which $q$, $M \sin^{3}{i}$ and $a_{} \sin{i}$ are estimated; the adopted values are listed in Table \ref{tab:v2fit}.
The detail fitting procedure is described e.g. in \cite{krishides2017}. All the RV measurements are given in the Table~\ref{tab:RV_data}. The orbital fit is shown in the Figure~\ref{fig:RVfit}.  

\begin{figure}[htb]
    \centering
    \includegraphics[width=1\linewidth]{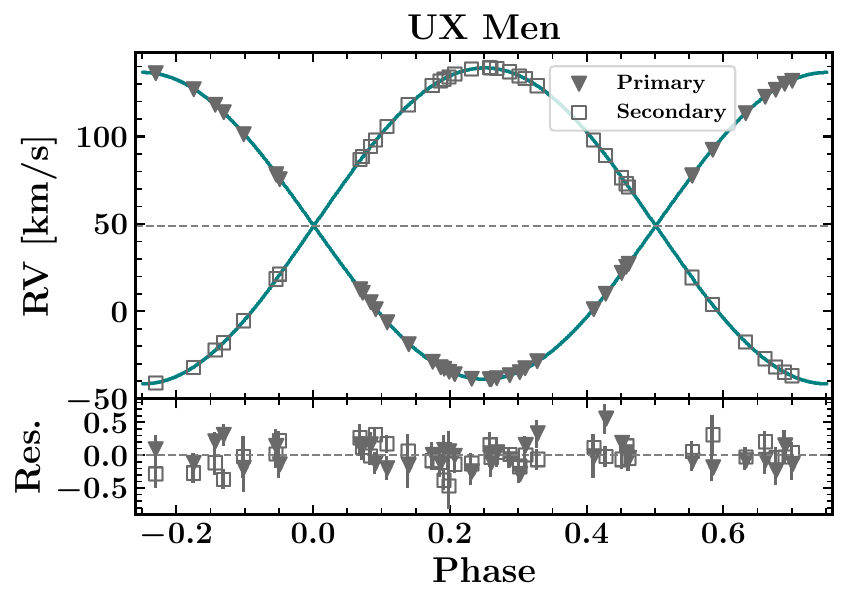}
    \caption{RV measurements of UX Men plotted over the Keplerian orbital solution for the primary (triangle) and secondary (open square). The lower panel displays the residuals of the fit for both components. The phase 0 is set on $T_0$, i.e. the primary eclipse mid-time.}
    \label{fig:RVfit}
\end{figure}

\subsection{Broadening function}\label{sec:BF}
The stellar line broadening can be used to measure RV and rotational velocities of the stars. For the case of the SB2 system, we have two sets of spectral lines present in the observed spectrum. Broadening function (BF, \citealt{bfsvd}) is a transformation from a narrow line spectrum (template) into a broad line spectrum (observed spectrum of a rotating star) to find the best fit in the velocity space.
BF works even if the individual RVs of an SB2 system are very close, e.g. in proximity of the eclipses, resulting in blended line profiles. The ratio of the areas below BF profiles represents the flux ratio (light fraction) of the two stars. Here the light fraction is the fractional contribution of each component to the total system flux in the relevant band (equivalently $l_{B}/l_{A}$ expressed as fractions). BF also contains signatures from RV shifts, stellar spots and rotational velocities. 
We estimated the projected rotational velocities of $v\sin{i}_{A} = 16.85\pm{3.61}$ km/s and $v\sin{i}_{B} = 15.19\pm{3.89}$ km/s, using BF of the HARPS (HAM and EGGS modes) and FEROS spectra, together with the flux ratio from Table \ref{tab:Stromgren}. This method has been described in detail by \citet{2023MNRAS.521.1908M}.

\subsection{Spectra Disentangling}
\label{sec:spec_disen}

To study the atmospheric parameters of the binary star, we extracted individual spectra of each star. In this stage, we did not use the CORALIE data -- a fly in the bottle of wine -- due to their relatively low individual SNRs, and insufficient number to increase the SNR of the final product to an acceptable value; they are retained for the RV analysis only.

To separate the individual spectra from the composite binary star spectra we used the spectral disentangling. Spectral disentangling assumes that the line profiles are intrinsically invariant. Thus, we have avoided spectra taken during the eclipses that cause artificial variability in the line profiles. To avoid the broad lines, we selected a wavelength range of 5350-5800~\AA, after masking the wavelength regions affected by telluric absorption.  
We used the \textsc{disentangling\_shift\_and\_add}\footnotemark[4] \footnotetext[4]{\url{https://github.com/TomerShenar/Disentangling_Shift_And_Add}} code \citep{shiftandadd2020A&A...639L...6S, shiftandadd22022A&A...665A.148S} based on the algorithm of \cite{shiftandaddalgo2006A&A...448..283G}, with 10,000 iterations. We examined cross-correlation plots on the final spectra to identify any potential contaminant lines produced during the disentangling. The result spectra had some trends in the continuum due to normalisation bias, which were cleaned by modelling this additive trend and subtracting it from the individual component spectra, as described in \cite{2008A&A...482.1031H}. \textsc{disentangling\_shift\_and\_add} recovers the shapes of the component spectra but leaves their relative continuum level (the light, or flux ratio) degenerate, since a constant flux ratio cannot be retrieved from the spectra alone. We therefore scale the disentangled spectra a posteriori using the light fraction obtained independently from the BF (see Table \ref{tab:Stromgren}). The Figure~\ref{fig:disentangled} shows the resulting component spectra after disentangling. 

\begin{figure*}[htb]
    \centering
    \includegraphics[width=0.98\textwidth]{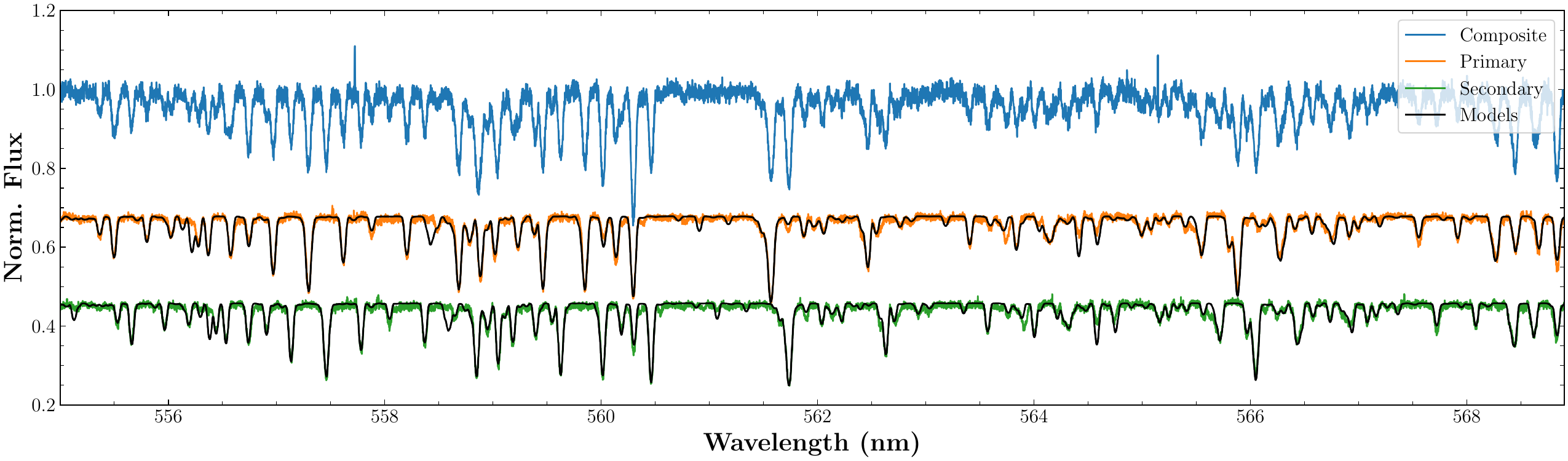}
    \caption{The composite observed HARPS-HAM spectrum from JD = 2455811.878 (blue), and the disentangled spectra of the primary (orange) and secondary (green), along with the model spectra (black). The disentangled spectra were shifted in RV to match the observed one above.}
    \label{fig:disentangled}
\end{figure*}

\subsection{Spectroscopic analysis}
\label{sec:spec_analy}

The disentangled spectra (from HARPS, FEROS and UCLES) were analysed using \ISPEC{} \citep{ispec12014A&A...569A.111B, ispec22019MNRAS.486.2075B} to measure the atmospheric parameters.
All the disentangled spectra were corrected for any RV offsets. We used the method to fit individual lines by generating synthetic spectra in \ISPEC{} with radiative transfer code {\sc spectrum} \citep{spectrum1994AJ....107..742G}. To generate the spectral model with the MARCS atmospheric model \citep{marcs2008A&A...486..951G}, a line-list from GES ({\it Gaia} ESO Survey) version 6 \citep{GaiaESOsurvery} and the solar abundances from \cite{asplund2009ARA&A..47..481A}.

The best fit is determined by minimising ${\chi}^2$. The limb-darkening was held fixed using the logarithmic-law coefficients of \citet{claret2004A&A...428.1001C}, 
spectral resolution to the instrument's resolution, and $\log g$ to reduce the degeneracy, and fitted $T_{\rm eff}$, metallicity [$M/H$], $\alpha$-elements enhancement [$\alpha/Fe$], micro- and macroturbulence velocities $v_\mathrm{mic}$, $v_\mathrm{mac}$ using the best-suited line region in \ISPEC{} (\texttt{spectrum\_synth\_good\_for\_params\_all\_extended}) within the 5350–5800 \AA{} disentangling window to estimate atmospheric parameters. The $\log g$ was adopted from LC-RV modelling and  $v \sin i$ from BF.
The Figure~\ref{fig:disentangled} shows the synthetic modelled spectra with observed spectra of both primary and secondary stars after disentangling. We repeated the same procedure for the spectra observed from all four spectrographs/modes and calculated the weighted average of the atmospheric parameters after the fit. The measured atmospheric parameters are shown in the Table \ref{tab:iSpec_wtavgparams}.

\subsection{$T_{\rm eff}$ estimates from photometry}\label{sec:teff_photo}

We measured the fundamental effective temperatures for the two components in UX~Men using \textsc{teb}\footnotemark[5]\footnotetext[5]{\url{https://github.com/nmiller95/teb}}, which implements the method described in \cite{2020MNRAS.497.2899M} and updated in \cite{2025MNRAS.544.4611M}. This technique uses catalogue photometry, observed flux ratios in multiple photometric bands, Gaia DR3 parallax \citep{2023A&A...674A...1G}, stellar radii and an estimate of the interstellar reddening to obtain a direct measurement of $T_{\rm eff}$. It uses a model spectral energy distribution (SED) to inform the small-scale features of the individual flux distribution for each star, but the overall shapes of these are distorted using low-order Legendre polynomials to match the data, so the dependence on stellar models is minimal. Loose priors are placed on the flux ratio of the binary in the near-infrared using empirical colour-$T_{\rm eff}$ relations derived in \cite{2025MNRAS.544.4611M}, in order to help constrain the fitted flux distributions to physically meaningful limits. The posterior distribution is sampled using a Markov-chain Monte Carlo (MCMC) algorithm.
In this work we fitted each of the Str\"{o}mgren $uvby$ light curves presented in \cite{1976A&A....48...49C} using \textsc{jktebop} TASK8 (Monte Carlo) with 1000 simulations. We fixed the parameters $r_{\rm sum}$, $k$, $i$, $e\cos{\omega}$, $e\sin{\omega}$ and $P$ at the best values from the TESS light curve fits, and varied $J$, \textbf{$l_{3}$}, the linear limb darkening coefficients \citep[fixing the quadratic coefficients at the theoretical values from][]{2000A&A...363.1081C}, scale factor and $T_0$. 

In addition to the Str\"{o}mgren light ratio from Table \ref{tab:Stromgren}, we used the TESS light ratio from Table~\ref{tab:params_lc}. 
We used model SEDs interpolated from a grid of BT-Settl model atmospheres \citep{2013MSAIS..24..128A}, with $T_{\rm eff}$, $\log{g}$ and [M/H] matching the best values from the combined RV-LC results and the spectroscopic fits presented in Tables~\ref{tab:iSpec_wtavgparams} and \ref{tab:v2fit}.
\begin{figure}[ht!]
    \centering
    \includegraphics[width=0.98\linewidth]{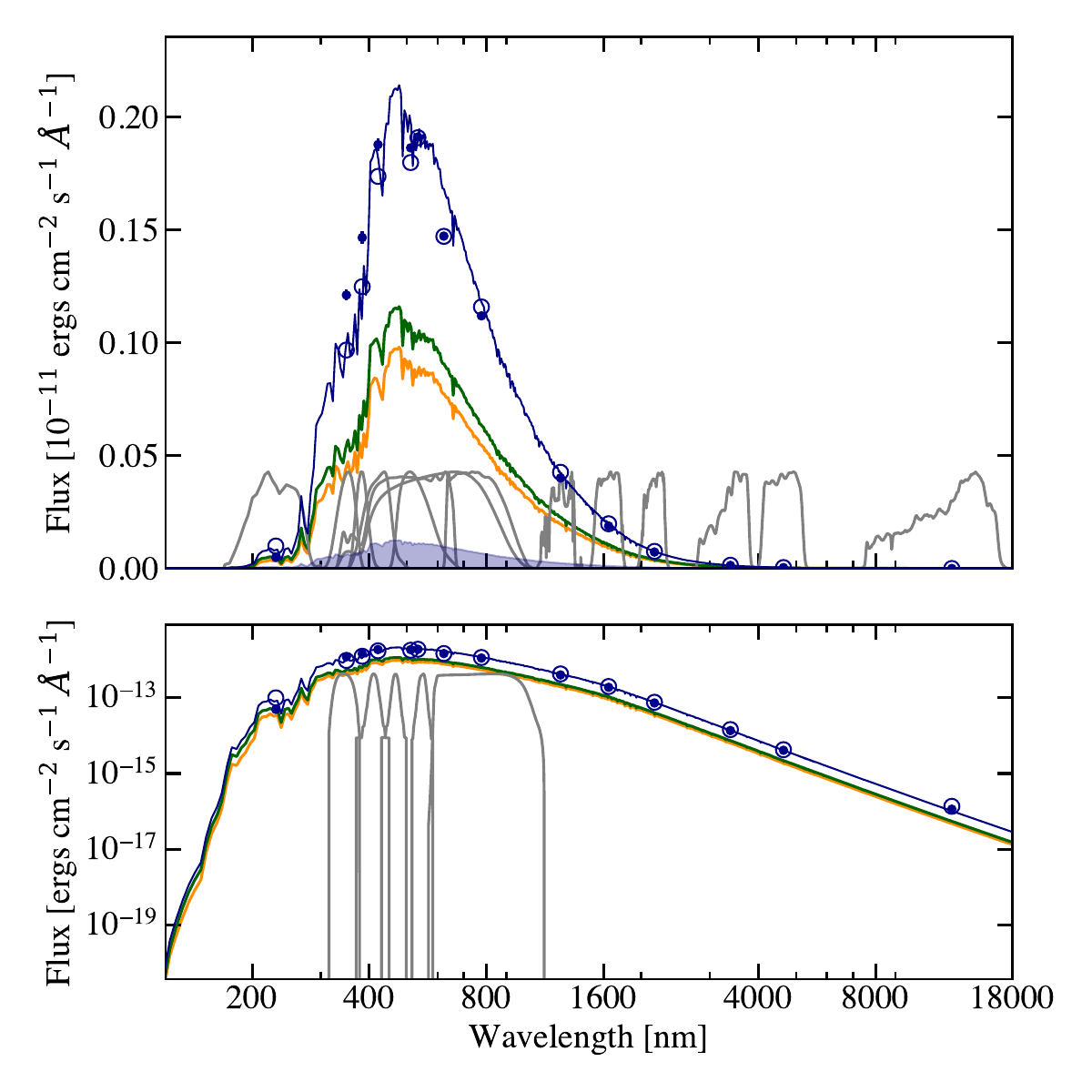}
    \caption{Upper panel: The SED of UX Men. The best-fitting combined SED is plotted as a line and the mean SED $\pm{1-\sigma}$ is plotted as a filled region. The observed fluxes are plotted as points with error bars and predicted fluxes for the best-fitting SED integrated over the response functions shown are plotted with open circles. The SEDs of the two stars are also plotted with green and orange lines. Lower panel: Same as the upper panel but with fluxes plotted on a logarithmic scale. Only filters used to measure flux ratios are plotted here.}
    \label{fig:uxmen_sed}
\end{figure}

For the estimate of $E(B-V)$, we tried both the value from the best-fitting isochrones in Section~\ref{sec:stellarage}, and by using a semi-empirical relation which connects $E(B-V)$ with the equivalent width of the Na\,D\,I lines. In the latter case, we measured the equivalent width using the available HARPS spectra of UX~Men and used the relation by \cite{2025RNAAS...9..146M}, which is a re-calibration of the original by \cite{1997A&A...318..269M}, to obtain $E(B-V)=0.0217\pm0.0099$. Both gave a satisfactory fit to the data, but the isochrone value resulted in the fit with the highest optimised log-likelihood.
For our \textsc{teb} analysis, we used 3 distortion coefficients with 256 walkers over 512 steps, with 1024 burn-in steps. 
The best fit results in effective temperatures of $T_{\rm eff,A}=6267\pm 53$\,K and $T_{\rm eff,B}=6198\pm 55$\,K. The resulting SED can be seen in  Fig. \ref{fig:uxmen_sed}, overplotted with some of the observed photometric data. These values carry an additional systematic uncertainty of 9\,K due to uncertainty in the flux scale \citep{2014AJ....147..127B}.

\begin{table}[]
    \centering
    \caption{\textbf{The predicted and observed light ratios are quoted here.}}
    \begin{tabular}{cccc}
    \hline
    Filter& $(l_{B}/l_{A})_{obs}$ & $(l_{B}/l_{A})_{obs_{err}}$& $(l_{B}/l_{A})_{syn}$\\ 
    \hline
    Str\"{o}mgren u & 0.81786 & 0.00725 & 0.82747\\
    Str\"{o}mgren v & 0.82712 & 0.00536 & 0.82489\\
    Str\"{o}mgren b & 0.84397 & 0.00491 & 0.84969 \\
    Str\"{o}mgren y & 0.84953 & 0.00505 & 0.85095\\
    Broadening function \tablefootmark{a} & 0.86098 & 0.05891 &--\\
         \hline 
    \end{tabular}
        \tablefoot{
            \tablefoottext{a}{measured average value for all the 3 spectrographs.}} 

    \label{tab:Stromgren}
\end{table}

\section{Results and Discussion}\label{sec:results}

The adopted stellar and orbital parameters are collected in Table \ref{tab:v2fit}. The light curve and RV modelling were performed independently and then combined to derive the absolute dimensions using the {\sc jktabsdim}\footnotemark[6]\footnotetext[6]{\url{http://www.astro.keele.ac.uk/jkt/codes.html}} procedure available with \JKTEBOP{}, which takes the spectroscopic and light curve solutions as input and returns the stellar masses, radii, $\log g$ and luminosities together with related quantities through the standard relations. A fully simultaneous LC-RV fit over all 31 sectors is computationally prohibitive; we therefore verified this separate treatment on a single sector (Sector 28) using \JKTEBOP{}, for which a joint LC-RV fit yields masses, radii, and uncertainties consistent with the combined approach (Fig.~\ref{fig:corel1}), confirming that fitting the two datasets separately does not underestimate the parameter uncertainties. A direct comparison of the fundamental parameters obtained here with those of \cite{krisuxmen2009MNRAS.400..969H} and \cite{1989A&A...211..346A} is given in Table \ref{tab:comparison}.

\subsection{Underestimated errors due to stellar activity}\label{sec:resultsstellaractivity}

\begin{table}[h!]
\centering
    \caption{Weighted average of all the measured parameters over all the sectors and methods. The $wt_{err}$ and $rms_{err}$ are shown separately for easier comparison of their magnitudes. The $f_{err}$ is the final error.}
    \label{tab:params_lc}
\begin{tabular}{lccccc}
\hline
Parameter &$wt_{avg}$ & $wt_{err}$ & $rms_{err}$ & $f_{err}$\\
\hline
$J$           &0.9713&0.0011&0.0132&0.0133\\
$r_{A}+ r_{B}$&0.180199&0.000079&0.000542&0.000548\\
$k$           &0.94054&0.00073&0.00484&0.00490\\
 $i\ (^\circ)$&89.61442&0.0059&0.0261&0.0268\\
$e\sin{\omega}$&0.00095&0.00004&0.00205&0.00205\\
$e\cos{\omega}$&0.000572&0.000002&0.000037&0.000037\\
$l_{3}$        &0.0125&0.00094&0.00622&0.00629\\
    \hline
   \multicolumn{5}{c}{Derived parameters} \\
    \hline

$r_A$        &0.092761&0.000073&0.000213&0.000225\\
$r_B$        &0.087276&0.000078&0.000472&0.000478\\
$l_{B}/l_{A}$ &0.8620&0.0024&0.0093&0.0096\\
$e$          &0.001107&0.000034&0.001755&0.001755\\
$\omega\ (^\circ)$ &58.858&0.019&0.958&0.959\\

\hline

\end{tabular}
\end{table}

The results of individual \JKTEBOP{} runs are listed in Table~\ref{tab:params_jktebop} \& \ref{tab:params_jktebop2}, together with their MC errors. { The \ALLESFITTER{} results are shown similarly in Tables~\ref{tab:params_lc_alles} and \ref{tab:alles_continue}.} In Table~\ref{tab:params_lc} we present the adopted lighcturve parameters of UX~Men and their error budget. 
For each quantity we computed the weighted average $wt_{\rm avg} = \sum_i w_i x_i / \sum_i w_i$, using the individual MC/MCMC errors as weights, $w_i = 1/\sigma_{{\rm MC},i}^{2}$. The uncertainty of the weighted mean and the spread across the sectors are
\begin{equation}
    wt_{err} = \left(\sum_i \frac{1}{\sigma_{{\rm MC},i}^{2}}\right)^{-1/2},
    \qquad
    rms_{err} = \left[\frac{1}{n-1}\sum_i \left(x_i - wt_{\rm avg}\right)^{2}\right]^{1/2},
    \label{eq:errs}
\end{equation}
where $n$ is the number of individual determinations. The $rms_{err}$ term dominates over $wt_{err}$ by a factor of $\sim$5-10: the former traces the between-sector, activity-driven spread, the latter the within-sector statistical precision. As these probe effectively independent sources of uncertainty, the final error is $f_{err} = \sqrt{wt_{err}^{2} + rms_{err}^{2}}$.

In Figure~\ref{fig:parmssector} we show the subtle sector-to-sector variations in the lightcurve parameters of UX Men, with their individual (MC) errors marked with errorbars, as a function of time. We run a Lomb-Scargle periodogram on them, which we show in Fig.~\ref{fig:etvperiodogram} together with the periodogram of the ETVs.
The periodicities observed in the eclipse‑timing variations and in the fitted radii, surface‑brightness ratios and inclination of the orbital plane coincide with a \textasciitilde673‑day signal that we attribute to spot‑modulated activity. Among these, the surface-brightness ratio (J) shows the largest sector-to-sector dispersion (Fig. \ref{fig:parmssector}) and the strong corresponding peak in the periodogram (Fig. \ref{fig:etvperiodogram}), as it is the parameter most sensitive to spot-induced light-curve features; this motivates the additional error term. When the data are treated as a single, homogeneous set, the formal uncertainties on the radii, inclination, and other stellar parameters ($wt_{err}$) are unrealistically small; accounting for the activity‑driven scatter inflates these errors. Even after this correction, the masses and radii remain exceptionally precise, with errors of better than 0.1\% in mass and \textasciitilde0.5\% in radius. Since a single-sector joint LC-RV fit reproduces these values and their uncertainties (Fig. \ref{fig:corel1}), the precision is not an artefact of fitting the two datasets separately, and UX Men can serve as a benchmark system, provided that the additional error budget imposed by stellar activity is explicitly taken into account.

\begin{figure*}[htb]
    \centering
    \includegraphics[width=1\textwidth]{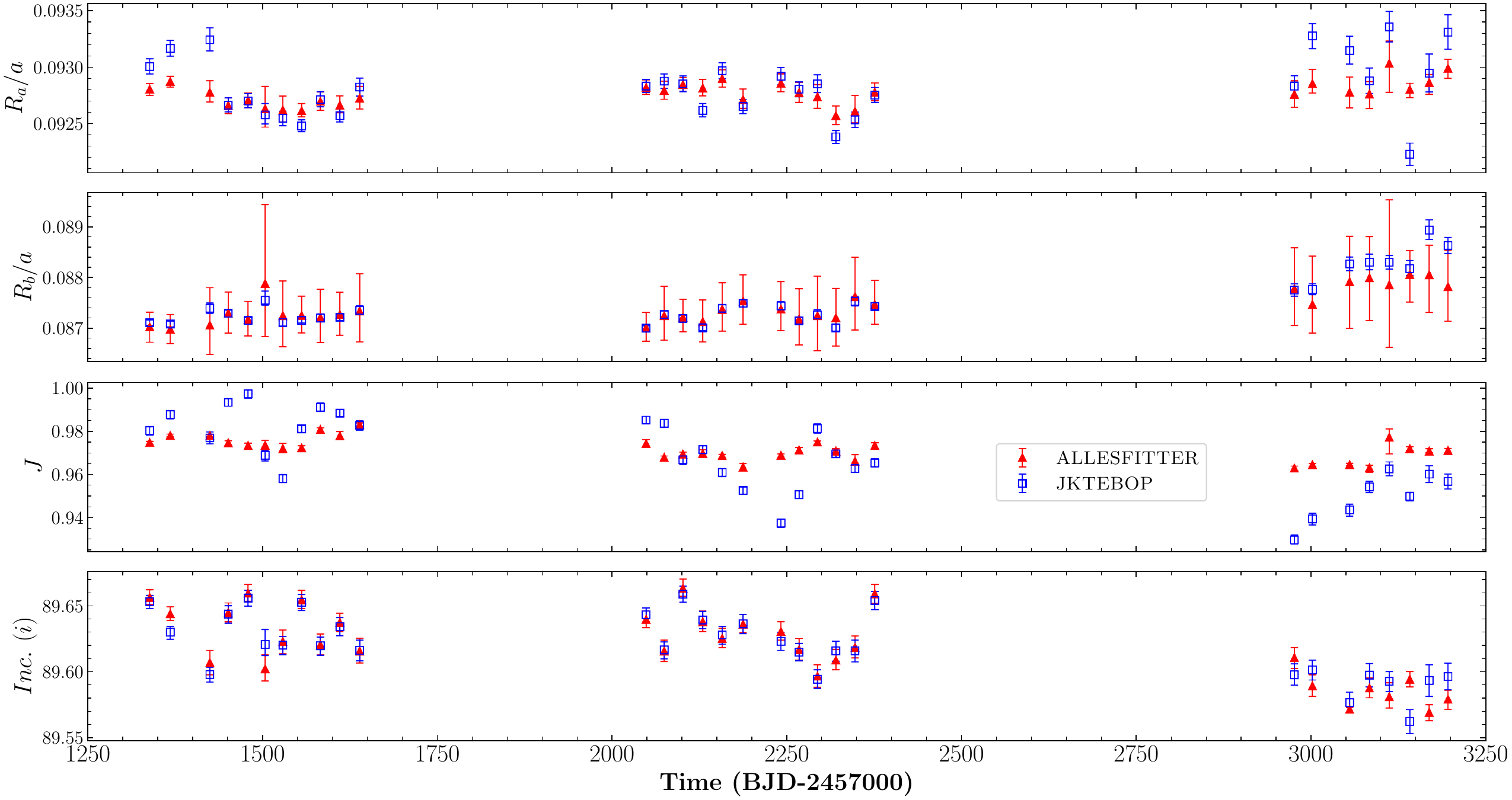}
       
    \caption{Parameters over each {\it TESS} sector from LC modelling. The top panels show the fractional radii of the primary and secondary, the surface‑brightness ratio (J) and the orbital inclination ($i$) obtained independently with \JKTEBOP{} (blue squares) and \ALLESFITTER{} (red triangles) for each sector.}
    \label{fig:parmssector}
\end{figure*}

\begin{figure}[htb]
    \centering
    \includegraphics[width=0.98\columnwidth]{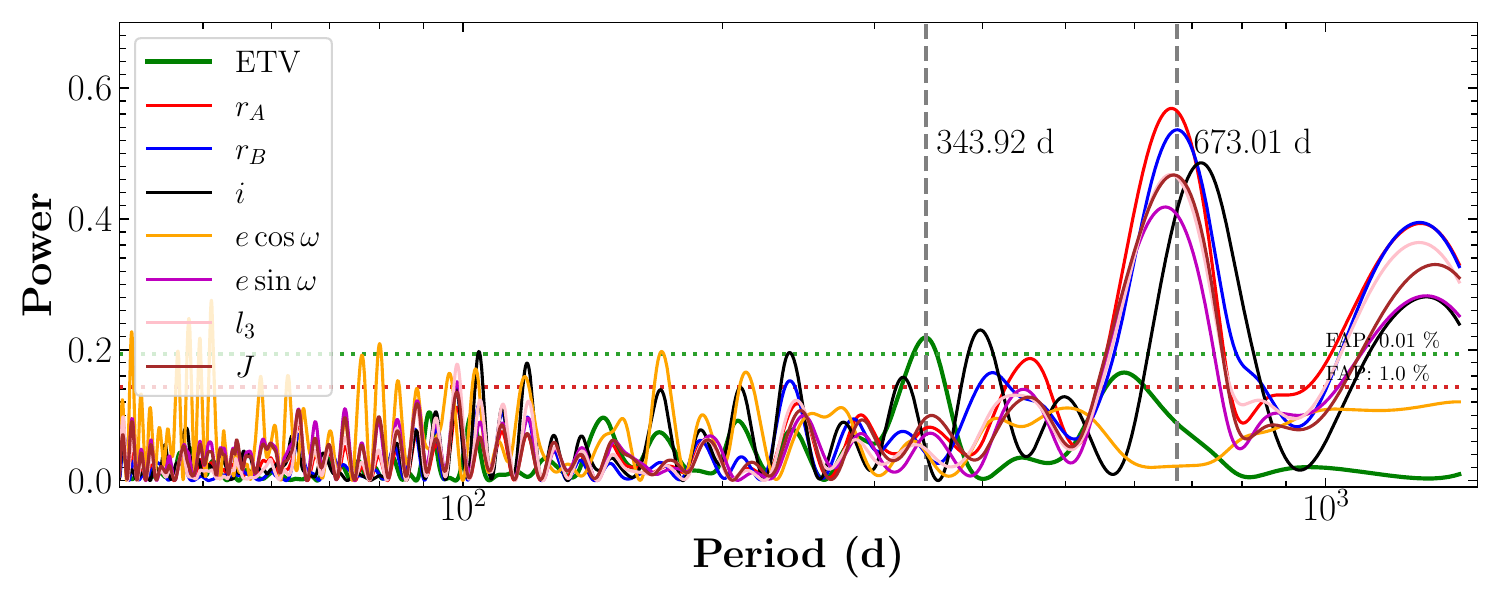}
    \caption{Periodogram of the ETV and stellar parameters of UX Men from each sector showing the dominant power peak near the $\sim$673‑day signal associated with stellar activity (additional peaks at $\simeq$344~d); the horizontal dashed lines indicate the false‑alarm probabilities of 0.01\% and 1.0\% respectively.
}
    \label{fig:etvperiodogram}
\end{figure}

\subsection{Effective temperature estimates}

\begin{figure}[htb]
    \centering
    \includegraphics[width=0.98\linewidth]{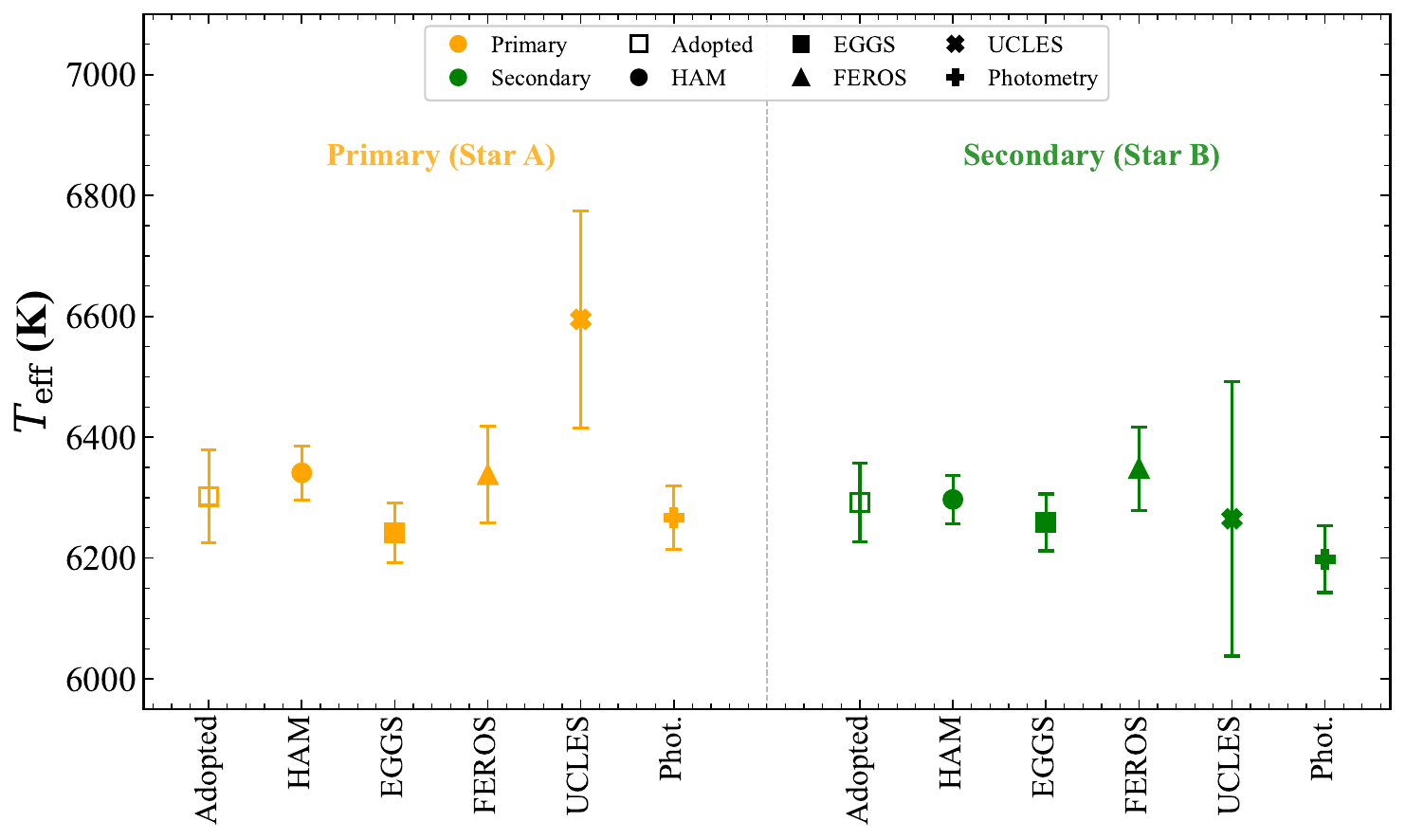}
    \caption{Measurement of \teff{} for primary (A) and secondary (B) stars using multiple spectrographs like HARPS with HAM and EGGS mode, FEROS, UCLES and the photometric method.}
    \label{fig:teff_inst}
\end{figure}

\begin{table*}[htb]
    \centering
    \caption{Weighted average of the atmospheric parameters measured using \ISPEC{} for all the spectrographs and modes. \label{tab:ipsec_allatmparms} } 
    
    \begin{tabular}{lccccccccc}
    \hline
    &&Primary &&&&Secondary \\
    \hline
    \hline
    Parameters &$wt_{avg_{A}}$&$wt_{err_{A}}$&$rms_{err_{A}}$ & $f_{err_{A}}$ &$wt_{avg_{B}}$&$wt_{err_{B}}$&$rms_{err_{B}}$ & $f_{err_{B}}$\\
    \hline
    $T_{\rm eff}$ (K) &6302.35&52.10&56.48& 76.84 &6292.68&47.63&44.60 & 65.25 \\
    $[M/H]$ (dex) &0.02&0.01&0.13& 0.13&0.07&0.01&0.10 & 0.10 \\
    $[\alpha/Fe]$ (dex) &0.00&0.04&0.08&0.09&-0.03&0.03&0.07 & 0.07\\
    $v_\mathrm{mic}$ (km/s) &1.58&0.18&0.07&0.19&1.64&0.16&0.13 & 0.21 \\
    $v_\mathrm{mac}$ (km/s)&5.57&1.62&0.90&1.85 &7.05&1.14&2.67 & 2.90\\
    $v \sin i$ (km/s) & $16.85\tablefootmark{b}$ &-&-&3.61& $15.19\tablefootmark{b}$&-&-&3.89& \\
    \hline       
    
    \end{tabular}
     \tablefoot{
            \tablefoottext{b}
            {Estimated from BF analysis.}
            }
    \label{tab:iSpec_wtavgparams}
\end{table*}

The effective surface temperatures (\teff{}) of the two components of UX~Men were derived by combining high-resolution spectroscopy from four instruments (HARPS with EGGS and HAM mode, FEROS, and UCLES) and multi-band photometric flux ratios. Each spectrum was analysed with the \ISPEC{} pipeline, fitting individual lines against synthetic spectra generated with MARCS atmospheres, and the resulting \teff{} values from the different spectrographs were found to be mutually consistent can be seen in Figure~\ref{fig:teff_inst} (see the spectroscopic analysis section in the text).
The results of all the \ISPEC{} fits are shown in Table~\ref{tab:ispeccomptable}. In Table~\ref{tab:iSpec_wtavgparams} we present the adopted values of \ISPEC-derived parameters, and the $v\sin{i}$ from the broadening function. We compute the adopted results similarly to those from \JKTEBOP{}, with weighted averages, $wt_{err}$ and $rms_{err}$. We use all spectrographs and modes except UCLES, which is excluded because its low SNR yields by far the largest uncertainties (its \teff{} errors are up to \textasciitilde5$\times$ those of the other instruments).
The final adopted spectroscopic effective temperatures and uncertainties, after combining the $wt_{err}$ and $rms_{err}$ terms, are therefore $T_{\rm eff,A}=6302\pm77$ K and $T_{\rm eff,B}=6292\pm65$ K. 
The photometric method, which employs the measured light‑ratio in the {\it TESS} band together with the spectroscopic \teff{} as priors, yields temperatures of $T_{\rm eff,A}=6267\pm 53$\,K and $T_{\rm eff,B}=6198\pm 55$\,K. They are consistent with \ISPEC{} within the errors, confirming that the photometric flux‑ratio does not introduce a bias for this nearly equal‑temperature system.

\subsection{Atmospheric parameters}\label{sec:atm_parms}
Apart from $T_{\rm eff}$, we have measured the atmospheric parameters of UX~Men, as shown in Table \ref{tab:iSpec_wtavgparams}. The final metallicities for primary and secondary stars are $\mathrm{[M/H]_{A}} $ = $0.02\pm{0.13}$ dex and $\mathrm{[M/H]_{B}} = $ $0.07\pm{0.10}$ dex, confirming that both components are essentially solar‑type with near zero $\alpha$ enhancement, $[\alpha/Fe]_{A}$ = $0.0\pm{0.09}$ dex and $[\alpha/Fe]_{B}$= $-0.03\pm{0.08}$ dex. The projected rotational velocities derived from the broadening‑function method \citep{bfsvd} are $v\sin{i}_{A} = 16.85\pm{3.61} $~km/s and $v\sin{i}_{B} = 15.19\pm{3.89}$~km/s.  Microturbulence velocities are $v_\mathrm{mic_{A}}$ = $1.58\pm{0.19}$ km/s and $v_\mathrm{mic_{B}}$ = $1.64\pm{0.21}$ km/s, while macroturbulence velocities are $v_\mathrm{mac_{A}}$ = $5.57\pm{1.85}$ km/s and $v_\mathrm{mac_{B}}$ = $7.05\pm{2.90}$ km/s. The agreement between results from each instrument and mode is very good, however the errors from UCLES measurements were by far the largest. The $rms_{err}$ terms usually, but not always, dominate over $wt_{err}$, but not that much as for LC-based parameters (only [$M/H$] stands out). 
These parameters, together with the precisely measured masses and radii, provide a robust atmospheric benchmark for solar‑type detached eclipsing binaries.

\subsection{Correlation of photometric fitted parameters}
\label{sec:corr_param}

The 31 {\it TESS} sectors were independently modelled with both \JKTEBOP{} and \ALLESFITTER{}. Figure \ref{fig:corel1} shows the correlation between the parameters derived from the two codes. For each parameter, the black marker represents the adopted weighted average of the two methods, with the error bars computed by adding the $wt_{err}$ and $rms_{err}$ terms in quadrature. The sector-by-sector light-curve solutions are broadly consistent between the two codes. The geometric parameters like fractional radii, third light and inclination are clustered tightly, whereas the surface-brightness ratio shows a noticeably larger spread, indicating
that this radiative parameter is the one most affected by spot-induced light-curve variations.

The eccentricity components $e\cos{\omega}$ and $e\sin{\omega}$ were additionally obtained from the ETVs of the 31 {\it TESS} sectors ($e\cos{\omega} = 0.000551\pm{0.000012}$ and $e\sin{\omega} = 0.00070^{+0.00044}_{-0.00011}$). 
The strong agreement between the methods and the consistent eccentric components derived from ETV demonstrates that, despite the activity-driven scatter, the combined analysis provides robust, accurate and high-precision orbital and stellar parameters.
\begin{figure}[ht!]
    \centering
    \includegraphics[width=0.5\textwidth]{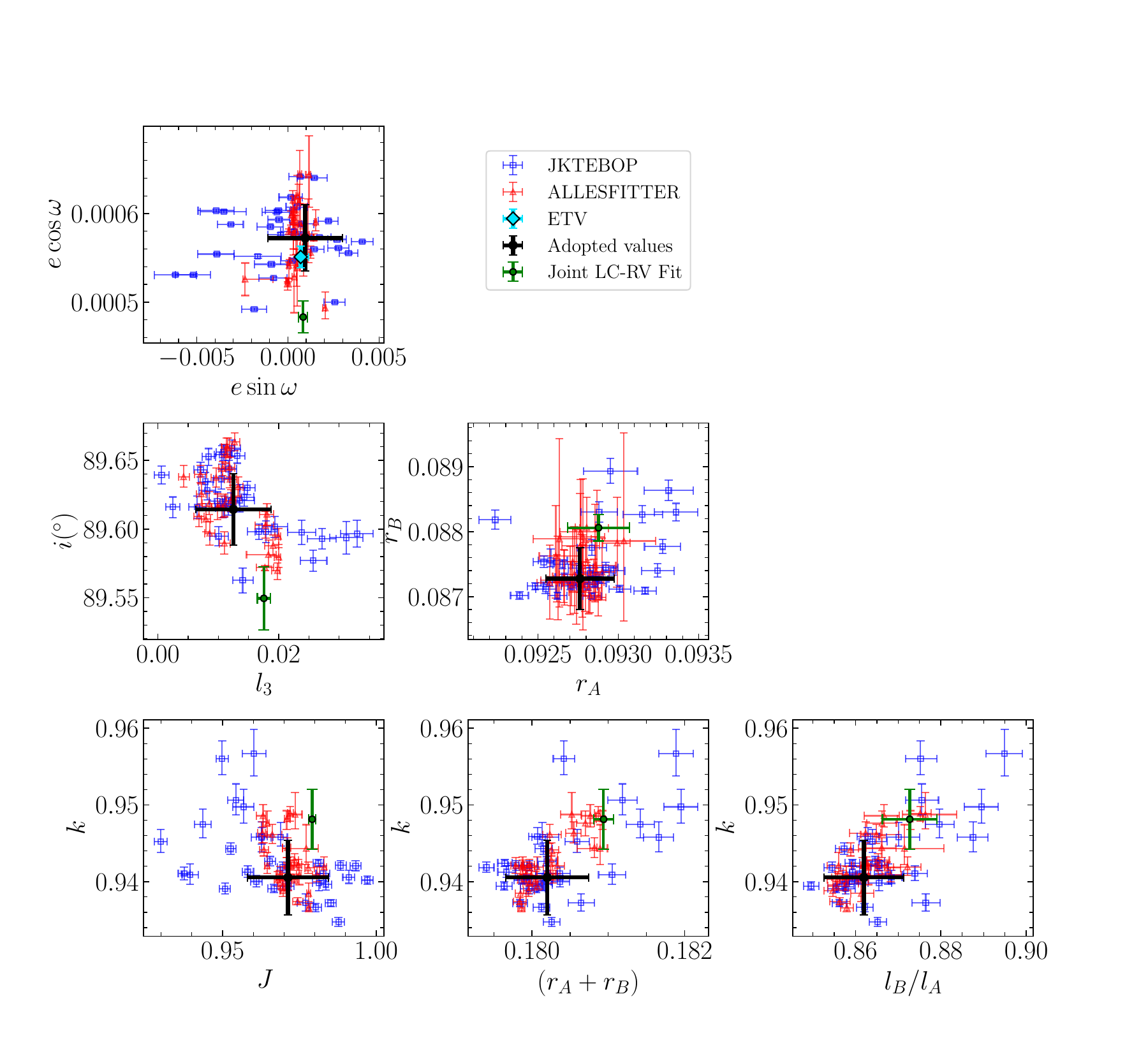}     
    \caption{Correlation of the parameters estimated from \ALLESFITTER{} (red) and \JKTEBOP{} (blue). Cyan -- estimated $e \cos{\omega}$ and $e \sin{\omega}$ from ETVs. Black -- the adopted weighted average of the two methods, accounting for quadrature errors. The single-sector joint LC-RV fit is shown in Green markers; it agrees with the separate-fit values within the errors.}
 
    \label{fig:corel1}
\end{figure}

\begin{table}[!ht]
    \centering
     \caption{Stellar and orbital parameters of the UX~Men system.}
    \label{tab:v2fit}
    \begin{tabular}{lcc}
    \hline
         Parameters &  Values \\
         \hline
         \hline
         \multicolumn{2}{c}{Binary orbit} \\
         \hline
         $P$  (d) & $4.181093974\pm{0.000000094}$  \\ 
         $T_{\rm 0}$ (BJD - 2450000) & $9189.8547388\pm{0.0000029}$ \\
         $e$ & $0.0011\pm{0.0018}$  \\ 
         $i\ (^\circ)$ & $89.6144\pm{0.0059}$ \\
         $\omega$ (deg) &$59 \pm{1}$ \\
         $K_{A}$ (km/s) & $87.473\pm{0.041}$     \\
         $K_{B}$ (km/s)&  $90.069\pm{0.036}$  \\
         $M_{\rm A} \sin^{3}{(i)}$ (M$_{\odot}$)&  $1.2300\pm{0.0012}$   \\
         $M_{\rm B} \sin^{3}{(i)}$ (M$_{\odot}$)&  $1.1945\pm{0.0012}$   \\
         $q$ &            $0.9712\pm{0.0006}$\\
         $a_{} \sin{i}$ (R$_{\odot}$) & $14.6667\pm{0.0045}$ \\
         $ \gamma_{} $ (km/s) & $48.965\pm{0.029}$ \\
         \hline
         \hline
         \multicolumn{2}{c}{Physical parameters} \\
         \hline
         $M_{\rm A}$ (M$_{\odot}$)&  $1.2300\pm{0.0012}$   \\
         $M_{\rm B}$ (M$_{\odot}$)&  $1.1946\pm{0.0012}$  \\
         $R_{\rm A}$ (R$_{\odot}$)&  $1.3605\pm{0.0031}$   \\
         $R_{\rm B}$ (R$_{\odot}$)&  $1.2801\pm{0.0069}$  \\
         $\log g_{\rm A}$ (cm/$s^2$)\tablefootmark{a} & $4.260\pm{0.002}$ \\
         $\log g_{\rm B}$ (cm/s$^2$)\tablefootmark{a} & $4.301\pm{0.005}$ \\
         $T_{\rm eff,A}$ {\rm (K)}\tablefootmark{b} & $6302\pm77$\\
         $T_{\rm eff,B}$ {\rm (K)}\tablefootmark{b} & $6292\pm65$\\
         $\mathrm{[M/H]_{A}}$ (dex) & \textbf{$0.02\pm{0.13}$} \\
         $\mathrm{[M/H]_{B}}$ (dex) & \textbf{$0.07\pm{0.10}$} \\
         $\log\,(L_{\rm A}$/L$_\odot)$ & $0.421\pm{0.021}$ \\
         $\log\,(L_{\rm B}$/L$_\odot)$ & $0.365\pm{0.019}$ \\
         $\tau$ (Gyr) & $2.75\pm{0.13}$ \\
         \hline
    \end{tabular}
        \vspace{0.2cm}
        \tablefoot{
            \tablefoottext{a}{Calculated from radii and masses.} 
            \tablefoottext{b}{From \ISPEC.}}\\
\end{table}

\subsection{Stellar Age}\label{sec:stellarage}

\begin{figure}
    \centering
    \includegraphics[width=0.9\columnwidth]{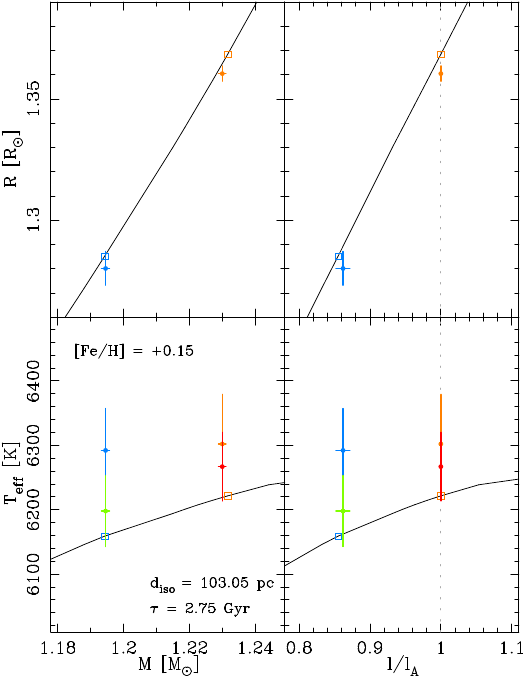}
    \caption{Results of our analysis of UX~Men compared with the best-fitting \MESA{} isochrone (2.75~Gyr, +0.15~dex; black line) on $M-R$, $l/l_A-R$, $M-T_{\rm eff}$, and $l/l_A-T_{\rm eff}$ planes. Orange and red symbols represent the primary, blue and green are for the secondary. Open squares mark the values corresponding to the best-fitting EEPs on the isochrone. Temperatures marked with orange and blue are from \ISPEC, while those in red and green are from the photometric method. For the primary the value of $l/l_A = 1$ by definition.}
    \label{fig:isochrone}
\end{figure}
To determine the age of the UX~Men system, we used \MESA{} Isochrone and Stellar Tracks (MIST; \citealt{mesadotter, mesachoi}) isochrones. We used the procedure {\sc isofitter}, described in \cite{kris2024isofitter}. It searches for a pair of equivalent evolutionary points (EEPs) on a single isochrone, that simultaneously reproduce various parameters of a binary system. We used the stellar properties from LC-RV modelling and spectroscopic analysis, namely masses, radii, metallicities and effective temperatures, as well as the flux ratio in the {\it TESS} band. We also used the GDR3 distance as an additional constraint. 

With {\sc isofitter} and \MESA{} isochrones we found UX~Men to be $\tau=2.75\pm0.13$~Gyr old ($\log(\tau)=9.44\pm0.02$), with initial metallicity of $+0.15\pm0.05$~dex. The best-fitting isochrone predicts the distance $d=103.05\pm0.53$~pc and $E(B-V)=0.036\pm0.011$~mag. Our value of $d$ is thus in a very good agreement with $103.24\pm0.16$~pc from {\it Gaia}. We found that both photometric and \ISPEC-derived temperatures give the same results, however the photometric values match the isochrone better. The fit is shown on Figure~\ref{fig:isochrone}. 
Our age estimate, $\tau$ = $2.75\pm{0.13} $ Gyr, corresponds to a relative precision of \textasciitilde4.7\%, i.e. below 10\%. UX Men therefore meets the PLATO benchmark requirements of \textasciitilde10\% in age and \textasciitilde1\% (here better) in mass and radius, reinforcing its value as a calibration target.


\section{Conclusions}\label{sec:conclusion}

In this work, we measured extremely precise and accurate fundamental parameters for the UX~Men system by combining {\it TESS} LC modelling, RV fitting, spectral disentangling and detailed spectroscopic analysis. Comparison of results obtained by \JKTEBOP{} and \ALLESFITTER{} shows that both codes produce very similar results. Uncertainty of LC-based parameters is inflated beyond the random, single-measurement errors, which we attribute to spot‑modulated stellar activity, even though its level seems to be very low. Consequently, the error budget must explicitly incorporate an activity‑driven systematic term. Otherwise, the claimed sub‑percent precision would be underestimated. Thus, the multi-sector {\it TESS} observation needs to be modelled to avoid underestimated errors. After accounting for this systematic component, the masses remain accurate to $<0.1\%$ and the radii to $0.5\%$. 

Spectroscopic analysis with \ISPEC{} also showed good agreement of results between instruments and modes. Their individual uncertainties varies with $SNR$ of the composite spectra used for disentangling. The resulting $T_{\rm eff}$  seem to be somewhat overestimated with respect to values obtained with the photometric method, and the best-fitting isochrone. 

In general, the fundamental properties of the UX~Men system have a precision of $<$1\%, despite the stellar activity that affects the light curve, confirming UX Men as a platypus-class high‑precision benchmark binary suitable for calibrating photometric pipelines of missions like {\it PLATO} and testing stellar‑evolution models. 

All extended Tables are available online in machine-readable format \footnotemark[7] \footnotetext[7]{\url{https://doi.org/10.5281/zenodo.18468054}}.

\begin{acknowledgements}

This work is funded by the Polish National Science Centre (NCN) through a grant \mbox{2024/53/N/ST9/03885}.
We acknowledge support provided by the Polish National Science Center through grants no. \mbox{2017/27/B/ST9/02727}, \mbox{2021/41/N/ST9/02746}, \mbox{2021/43/B/ST9/02972} and \mbox{2023/49/B/ST9/01671}. AM acknowledges support from the UK Science and Technology Facilities Council (STFC) under grant number ST/Y002563/1.
This research made use of Lightkurve, a Python package for Kepler and {\it TESS} data analysis \citep{lightkurve2018ascl.soft12013L}. This research uses data collected by the {\it TESS} mission, which are publicly available from the Mikulski Archive for Space Telescopes (MAST) at the Space Telescope Science Institute (STScI). Funding for the {\it TESS} mission is provided by NASA’s Science Mission Directorate. Finally, we wish to thank the referee Thibault Merle for comments that helped us to improve this paper.
\end{acknowledgements}
%
%
\bibliographystyle{aa}
\bibliography{aa}

@ARTICLE{2015JATIS...1a4003R,
       author = {{Ricker}, George R. and {Winn}, Joshua N. and {Vanderspek}, Roland and {Latham}, David W. and {Bakos}, G{\'a}sp{\'a}r {\'A}. and {Bean}, Jacob L. and {Berta-Thompson}, Zachory K. and {Brown}, Timothy M. and {Buchhave}, Lars and {Butler}, Nathaniel R. and {Butler}, R. Paul and {Chaplin}, William J. and {Charbonneau}, David and {Christensen-Dalsgaard}, J{\o}rgen and {Clampin}, Mark and {Deming}, Drake and {Doty}, John and {De Lee}, Nathan and {Dressing}, Courtney and {Dunham}, Edward W. and {Endl}, Michael and {Fressin}, Francois and {Ge}, Jian and {Henning}, Thomas and {Holman}, Matthew J. and {Howard}, Andrew W. and {Ida}, Shigeru and {Jenkins}, Jon M. and {Jernigan}, Garrett and {Johnson}, John Asher and {Kaltenegger}, Lisa and {Kawai}, Nobuyuki and {Kjeldsen}, Hans and {Laughlin}, Gregory and {Levine}, Alan M. and {Lin}, Douglas and {Lissauer}, Jack J. and {MacQueen}, Phillip and {Marcy}, Geoffrey and {McCullough}, Peter R. and {Morton}, Timothy D. and {Narita}, Norio and {Paegert}, Martin and {Palle}, Enric and {Pepe}, Francesco and {Pepper}, Joshua and {Quirrenbach}, Andreas and {Rinehart}, Stephen A. and {Sasselov}, Dimitar and {Sato}, Bun'ei and {Seager}, Sara and {Sozzetti}, Alessandro and {Stassun}, Keivan G. and {Sullivan}, Peter and {Szentgyorgyi}, Andrew and {Torres}, Guillermo and {Udry}, Stephane and {Villasenor}, Joel},
        title = "{Transiting Exoplanet Survey Satellite (TESS)}",
      journal = {Journal of Astronomical Telescopes, Instruments, and Systems},
         year = 2015,
        month = jan,
       volume = {1},
          eid = {014003},
        pages = {014003},
          doi = {10.1117/1.JATIS.1.1.014003},
       adsurl = {https://ui.adsabs.harvard.edu/abs/2015JATIS...1a4003R}
}

@MISC{lightkurve2018ascl.soft12013L,
    author = {{Lightkurve Collaboration} and {Cardoso}, J.~V.~d.~M. and
                {Hedges}, C. and {Gully-Santiago}, M. and {Saunders}, N. and
                {Cody}, A.~M. and {Barclay}, T. and {Hall}, O. and
                {Sagear}, S. and {Turtelboom}, E. and {Zhang}, J. and
                {Tzanidakis}, A. and {Mighell}, K. and {Coughlin}, J. and
                {Bell}, K. and {Berta-Thompson}, Z. and {Williams}, P. and
                {Dotson}, J. and {Barentsen}, G.},
    title = "{Lightkurve: Kepler and TESS time series analysis in Python}",
howpublished = {Astrophysics Source Code Library},
        year = 2018,
    month = dec,
archivePrefix = "ascl",
    eprint = {1812.013},
    adsurl = {http://adsabs.harvard.edu/abs/2018ascl.soft12013L},
}

@ARTICLE{wotan2019AJ....158..143H,
       author = {{Hippke}, Michael and {David}, Trevor J. and {Mulders}, Gijs D. and
         {Heller}, Ren{\'e}},
        title = "{W{\={o}}tan: Comprehensive Time-series Detrending in Python}",
      journal = {\aj},
         year = "2019",
        month = "Oct",
       volume = {158},
       number = {4},
          eid = {143},
        pages = {143},
          doi = {10.3847/1538-3881/ab3984},
archivePrefix = {arXiv},
       eprint = {1906.00966},
 primaryClass = {astro-ph.EP},
       adsurl = {https://ui.adsabs.harvard.edu/abs/2019AJ....158..143H}
}

@ARTICLE{todcor2005ApJ...626..431K,
       author = {{Konacki}, Maciej},
        title = "{Precision Radial Velocities of Double-lined Spectroscopic Binaries with an Iodine Absorption Cell}",
      journal = {\apj},
         year = 2005,
        month = jun,
       volume = {626},
       number = {1},
        pages = {431-438},
          doi = {10.1086/429880},
archivePrefix = {arXiv},
       eprint = {astro-ph/0410389},
 primaryClass = {astro-ph},
       adsurl = {https://ui.adsabs.harvard.edu/abs/2005ApJ...626..431K}
}

@dataset{2022yCat.1355....0G,
       author = {{Gaia Collaboration}},
        title = "{VizieR Online Data Catalog: Gaia DR3 Part 1. Main source (Gaia Collaboration, 2022)}",
 howpublished = {VizieR On-line Data Catalog: I/355.  Originally published in: doi:10.1051/0004-63},
         year = 2022,
        month = may,
          eid = {I/355},
          doi = {10.26093/cds/vizier.1355},
       adsurl = {https://ui.adsabs.harvard.edu/abs/2022yCat.1355....0G}
}

@ARTICLE{GBS2015A&A...582A..49H,
       author = {{Heiter}, U. and {Jofr{\'e}}, P. and {Gustafsson}, B. and {Korn}, A.~J. and {Soubiran}, C. and {Th{\'e}venin}, F.},
        title = "{Gaia FGK benchmark stars: Effective temperatures and surface gravities}",
      journal = {\aap},
         year = 2015,
        month = oct,
       volume = {582},
          eid = {A49},
        pages = {A49},
          doi = {10.1051/0004-6361/201526319},
archivePrefix = {arXiv},
       eprint = {1506.06095},
 primaryClass = {astro-ph.SR},
       adsurl = {https://ui.adsabs.harvard.edu/abs/2015A&A...582A..49H}
}

@ARTICLE{2019ARA&A..57..571J,
       author = {{Jofr{\'e}}, Paula and {Heiter}, Ulrike and {Soubiran}, Caroline},
        title = "{Accuracy and Precision of Industrial Stellar Abundances}",
      journal = {\araa},
         year = 2019,
        month = aug,
       volume = {57},
        pages = {571-616},
          doi = {10.1146/annurev-astro-091918-104509},
archivePrefix = {arXiv},
       eprint = {1811.08041},
 primaryClass = {astro-ph.SR},
       adsurl = {https://ui.adsabs.harvard.edu/abs/2019ARA&A..57..571J}
}

@ARTICLE{2017A&A...597A.137K,
       author = {{Kervella}, P. and {Bigot}, L. and {Gallenne}, A. and {Th{\'e}venin}, F.},
        title = "{The radii and limb darkenings of {\ensuremath{\alpha}} Centauri A and B . Interferometric measurements with VLTI/PIONIER}",
      journal = {\aap},
         year = 2017,
        month = jan,
       volume = {597},
          eid = {A137},
        pages = {A137},
          doi = {10.1051/0004-6361/201629505},
archivePrefix = {arXiv},
       eprint = {1610.06185},
 primaryClass = {astro-ph.SR},
       adsurl = {https://ui.adsabs.harvard.edu/abs/2017A&A...597A.137K}
}

@ARTICLE{2021AJ....162...14A,
       author = {{Akeson}, Rachel and {Beichman}, Charles and {Kervella}, Pierre and {Fomalont}, Edward and {Benedict}, G. Fritz},
        title = "{Precision Millimeter Astrometry of the {\ensuremath{\alpha}} Centauri AB System}",
      journal = {\aj},
         year = 2021,
        month = jul,
       volume = {162},
       number = {1},
          eid = {14},
        pages = {14},
          doi = {10.3847/1538-3881/abfaff},
archivePrefix = {arXiv},
       eprint = {2104.10086},
 primaryClass = {astro-ph.SR},
       adsurl = {https://ui.adsabs.harvard.edu/abs/2021AJ....162...14A}
}

@ARTICLE{2014ApJ...780...59G,
       author = {{Graczyk}, Dariusz and {Pietrzy{\'n}ski}, Grzegorz and {Thompson}, Ian B. and {Gieren}, Wolfgang and {Pilecki}, Bogumi{\l} and {Konorski}, Piotr and {Udalski}, Andrzej and {Soszy{\'n}ski}, Igor and {Villanova}, Sandro and {G{\'o}rski}, Marek and {Suchomska}, Ksenia and {Karczmarek}, Paulina and {Kudritzki}, Rolf-Peter and {Bresolin}, Fabio and {Gallenne}, Alexandre},
        title = "{The Araucaria Project. The Distance to the Small Magellanic Cloud from Late-type Eclipsing Binaries}",
      journal = {\apj},
         year = 2014,
        month = jan,
       volume = {780},
       number = {1},
          eid = {59},
        pages = {59},
          doi = {10.1088/0004-637X/780/1/59},
archivePrefix = {arXiv},
       eprint = {1311.2340},
 primaryClass = {astro-ph.CO},
       adsurl = {https://ui.adsabs.harvard.edu/abs/2014ApJ...780...59G}
}

@ARTICLE{2023MNRAS.521.1908M,
       author = {{Moharana}, Ayush and {He{\l}miniak}, K.~G. and {Marcadon}, F. and {Pawar}, T. and {Konacki}, M. and {Ukita}, N. and {Kambe}, E. and {Maehara}, H.},
        title = "{Detached eclipsing binaries in compact hierarchical triples: triple-lined systems BD+442258 and KIC 06525196}",
      journal = {\mnras},
         year = 2023,
        month = may,
       volume = {521},
       number = {2},
        pages = {1908-1923},
          doi = {10.1093/mnras/stad622},
archivePrefix = {arXiv},
       eprint = {2303.05272},
 primaryClass = {astro-ph.SR},
       adsurl = {https://ui.adsabs.harvard.edu/abs/2023MNRAS.521.1908M}
}

@ARTICLE{v2fit2010ApJ...719.1293K,
       author = {{Konacki}, Maciej and {Muterspaugh}, Matthew W. and {Kulkarni}, Shrinivas R. and {He{\l}miniak}, Krzysztof G.},
        title = "{High-precision Orbital and Physical Parameters of Double-lined Spectroscopic Binary Stars{\textemdash}HD78418, HD123999, HD160922, HD200077, and HD210027}",
      journal = {\apj},
         year = 2010,
        month = aug,
       volume = {719},
       number = {2},
        pages = {1293-1314},
          doi = {10.1088/0004-637X/719/2/1293},
archivePrefix = {arXiv},
       eprint = {0910.4482},
 primaryClass = {astro-ph.SR},
       adsurl = {https://ui.adsabs.harvard.edu/abs/2010ApJ...719.1293K}
}

@ARTICLE{shiftandadd2020A&A...639L...6S,
       author = {{Shenar}, T. and {Bodensteiner}, J. and {Abdul-Masih}, M. and {Fabry}, M. and {Mahy}, L. and {Marchant}, P. and {Banyard}, G. and {Bowman}, D.~M. and {Dsilva}, K. and {Hawcroft}, C. and {Reggiani}, M. and {Sana}, H.},
        title = "{The ``hidden'' companion in LB-1 unveiled by spectral disentangling}",
      journal = {\aap},
         year = 2020,
        month = jul,
       volume = {639},
          eid = {L6},
        pages = {L6},
          doi = {10.1051/0004-6361/202038275},
archivePrefix = {arXiv},
       eprint = {2004.12882},
 primaryClass = {astro-ph.SR},
       adsurl = {https://ui.adsabs.harvard.edu/abs/2020A&A...639L...6S}
}

@ARTICLE{shiftandadd22022A&A...665A.148S,
       author = {{Shenar}, T. and {Sana}, H. and {Mahy}, L. and {Ma{\'\i}z Apell{\'a}niz}, J. and {Crowther}, Paul A. and {Gromadzki}, M. and {Herrero}, A. and {Langer}, N. and {Marchant}, P. and {Schneider}, F.~R.~N. and {Sen}, K. and {Soszy{\'n}ski}, I. and {Toonen}, S.},
        title = "{The Tarantula Massive Binary Monitoring. VI. Characterisation of hidden companions in 51 single-lined O-type binaries: A flat mass-ratio distribution and black-hole binary candidates}",
      journal = {\aap},
         year = 2022,
        month = sep,
       volume = {665},
          eid = {A148},
        pages = {A148},
          doi = {10.1051/0004-6361/202244245},
archivePrefix = {arXiv},
       eprint = {2207.07674},
 primaryClass = {astro-ph.SR},
       adsurl = {https://ui.adsabs.harvard.edu/abs/2022A&A...665A.148S}
}

@ARTICLE{shiftandaddalgo2006A&A...448..283G,
       author = {{Gonz{\'a}lez}, J.~F. and {Levato}, H.},
        title = "{Separation of composite spectra: the spectroscopic detection of an eclipsing binary star}",
      journal = {\aap},
         year = 2006,
        month = mar,
       volume = {448},
       number = {1},
        pages = {283-292},
          doi = {10.1051/0004-6361:20053177},
       adsurl = {https://ui.adsabs.harvard.edu/abs/2006A&A...448..283G}
}

@ARTICLE{ispec12014A&A...569A.111B,
       author = {{Blanco-Cuaresma}, S. and {Soubiran}, C. and {Heiter}, U. and {Jofr{\'e}}, P.},
        title = "{Determining stellar atmospheric parameters and chemical abundances of FGK stars with iSpec}",
      journal = {\aap},
         year = 2014,
        month = sep,
       volume = {569},
          eid = {A111},
        pages = {A111},
          doi = {10.1051/0004-6361/201423945},
archivePrefix = {arXiv},
       eprint = {1407.2608},
 primaryClass = {astro-ph.IM},
       adsurl = {https://ui.adsabs.harvard.edu/abs/2014A&A...569A.111B}
}

@ARTICLE{ispec22019MNRAS.486.2075B,
       author = {{Blanco-Cuaresma}, Sergi},
        title = "{Modern stellar spectroscopy caveats}",
      journal = {\mnras},
         year = 2019,
        month = jun,
       volume = {486},
       number = {2},
        pages = {2075-2101},
          doi = {10.1093/mnras/stz549},
archivePrefix = {arXiv},
       eprint = {1902.09558},
 primaryClass = {astro-ph.SR},
       adsurl = {https://ui.adsabs.harvard.edu/abs/2019MNRAS.486.2075B}
}

@ARTICLE{jktebop2013A&A...557A.119S,
       author = {{Southworth}, J.},
        title = "{The solar-type eclipsing binary system LL Aquarii}",
      journal = {\aap},
         year = 2013,
        month = sep,
       volume = {557},
          eid = {A119},
        pages = {A119},
          doi = {10.1051/0004-6361/201322195},
archivePrefix = {arXiv},
       eprint = {1308.1320},
 primaryClass = {astro-ph.SR},
       adsurl = {https://ui.adsabs.harvard.edu/abs/2013A&A...557A.119S}
}

@ARTICLE{jktebop2004MNRAS.351.1277S,
       author = {{Southworth}, J. and {Maxted}, P.~F.~L. and {Smalley}, B.},
        title = "{Eclipsing binaries in open clusters - II. V453 Cyg in NGC 6871}",
      journal = {\mnras},
         year = 2004,
        month = jul,
       volume = {351},
       number = {4},
        pages = {1277-1289},
          doi = {10.1111/j.1365-2966.2004.07871.x},
archivePrefix = {arXiv},
       eprint = {astro-ph/0403572},
 primaryClass = {astro-ph},
       adsurl = {https://ui.adsabs.harvard.edu/abs/2004MNRAS.351.1277S}
}

@INPROCEEDINGS{ebop1981ASIC...69..111E,
       author = {{Etzel}, P.~B.},
        title = "{A Simple Synthesis Method for Solving the Elements of Well-Detached Eclipsing Systems}",
    booktitle = {Photometric and Spectroscopic Binary Systems},
         year = 1981,
       editor = {{Carling}, Ellen B. and {Kopal}, Zdenek},
       series = {NATO Advanced Study Institute (ASI) Series C},
       volume = {69},
        month = jan,
        pages = {111},
          doi = {10.1007/978-94-009-8486-8_3},
       adsurl = {https://ui.adsabs.harvard.edu/abs/1981ASIC...69..111E}
}

@ARTICLE{ebop21981AJ.....86..102P,
       author = {{Popper}, D.~M. and {Etzel}, P.~B.},
        title = "{Photometric orbits of seven detached eclipsing binaries.}",
      journal = {\aj},
         year = 1981,
        month = jan,
       volume = {86},
        pages = {102-120},
          doi = {10.1086/112862},
       adsurl = {https://ui.adsabs.harvard.edu/abs/1981AJ.....86..102P}
}

@ARTICLE{allesfitter-paper,
 author = {{G{\"u}nther}, Maximilian N. and {Daylan}, Tansu},
 title = "{Allesfitter: Flexible Star and Exoplanet Inference from Photometry and Radial Velocity}",
 journal = {\apjs},
 year = 2021,
 month = may,
 volume = {254},
 number = {1},
 eid = {13},
 pages = {13},
 doi = {10.3847/1538-4365/abe70e},
 archivePrefix = {arXiv},
 eprint = {2003.14371},
 primaryClass = {astro-ph.EP},
 adsurl = {https://ui.adsabs.harvard.edu/abs/2021ApJS..254...13G}
}

@MISC{allesfitter-code,
 author = {{G{\"u}nther}, Maximilian~N. and {Daylan}, Tansu},
 title = "{Allesfitter: Flexible Star and Exoplanet Inference From Photometry and Radial Velocity}",
 howpublished = {Astrophysics Source Code Library},
 year = 2019,
 month = mar,
 archivePrefix = "ascl",
 eprint = {1903.003},
 adsurl = {http://adsabs.harvard.edu/abs/2019ascl.soft03003G}
}

@ARTICLE{ellc2016A&A...591A.111M,
       author = {{Maxted}, P.~F.~L.},
        title = "{ellc: A fast, flexible light curve model for detached eclipsing binary stars and transiting exoplanets}",
      journal = {\aap},
         year = 2016,
        month = jun,
       volume = {591},
          eid = {A111},
        pages = {A111},
          doi = {10.1051/0004-6361/201628579},
archivePrefix = {arXiv},
       eprint = {1603.08484},
 primaryClass = {astro-ph.IM},
       adsurl = {https://ui.adsabs.harvard.edu/abs/2016A&A...591A.111M}
}

@ARTICLE{celerite2017AJ....154..220F,
       author = {{Foreman-Mackey}, Daniel and {Agol}, Eric and {Ambikasaran}, Sivaram and {Angus}, Ruth},
        title = "{Fast and Scalable Gaussian Process Modeling with Applications to Astronomical Time Series}",
      journal = {\aj},
         year = 2017,
        month = dec,
       volume = {154},
       number = {6},
          eid = {220},
        pages = {220},
          doi = {10.3847/1538-3881/aa9332},
archivePrefix = {arXiv},
       eprint = {1703.09710},
 primaryClass = {astro-ph.IM},
       adsurl = {https://ui.adsabs.harvard.edu/abs/2017AJ....154..220F}
}

@ARTICLE{dynesty2020MNRAS.493.3132S,
       author = {{Speagle}, Joshua S.},
        title = "{DYNESTY: a dynamic nested sampling package for estimating Bayesian posteriors and evidences}",
      journal = {\mnras},
         year = 2020,
        month = apr,
       volume = {493},
       number = {3},
        pages = {3132-3158},
          doi = {10.1093/mnras/staa278},
archivePrefix = {arXiv},
       eprint = {1904.02180},
 primaryClass = {astro-ph.IM},
       adsurl = {https://ui.adsabs.harvard.edu/abs/2020MNRAS.493.3132S}
}

@ARTICLE{emcee2013PASP..125..306F,
       author = {{Foreman-Mackey}, Daniel and {Hogg}, David W. and {Lang}, Dustin and {Goodman}, Jonathan},
        title = "{emcee: The MCMC Hammer}",
      journal = {\pasp},
         year = 2013,
        month = mar,
       volume = {125},
       number = {925},
        pages = {306},
          doi = {10.1086/670067},
archivePrefix = {arXiv},
       eprint = {1202.3665},
 primaryClass = {astro-ph.IM},
       adsurl = {https://ui.adsabs.harvard.edu/abs/2013PASP..125..306F}
}

@ARTICLE{spectrum1994AJ....107..742G,
       author = {{Gray}, R.~O. and {Corbally}, C.~J.},
        title = "{The Calibration of MK Spectral Classes Using Spectral Synthesis. I. The Effective Temperature Calibration of Dwarf Stars}",
      journal = {\aj},
         year = 1994,
        month = feb,
       volume = {107},
        pages = {742},
          doi = {10.1086/116893},
       adsurl = {https://ui.adsabs.harvard.edu/abs/1994AJ....107..742G}
}

@ARTICLE{marcs2008A&A...486..951G,
       author = {{Gustafsson}, B. and {Edvardsson}, B. and {Eriksson}, K. and {J{\o}rgensen}, U.~G. and {Nordlund}, {\r{A}}. and {Plez}, B.},
        title = "{A grid of MARCS model atmospheres for late-type stars. I. Methods and general properties}",
      journal = {\aap},
         year = 2008,
        month = aug,
       volume = {486},
       number = {3},
        pages = {951-970},
          doi = {10.1051/0004-6361:200809724},
archivePrefix = {arXiv},
       eprint = {0805.0554},
 primaryClass = {astro-ph},
       adsurl = {https://ui.adsabs.harvard.edu/abs/2008A&A...486..951G}
}

@ARTICLE{asplund2009ARA&A..47..481A,
       author = {{Asplund}, Martin and {Grevesse}, Nicolas and {Sauval}, A. Jacques and {Scott}, Pat},
        title = "{The Chemical Composition of the Sun}",
      journal = {\araa},
         year = 2009,
        month = sep,
       volume = {47},
       number = {1},
        pages = {481-522},
          doi = {10.1146/annurev.astro.46.060407.145222},
archivePrefix = {arXiv},
       eprint = {0909.0948},
 primaryClass = {astro-ph.SR},
       adsurl = {https://ui.adsabs.harvard.edu/abs/2009ARA&A..47..481A}
}

@ARTICLE{krisuxmen2009MNRAS.400..969H,
       author = {{He{\l}miniak}, K.~G. and {Konacki}, M. and {Ratajczak}, M. and {Muterspaugh}, M.~W.},
        title = "{Orbital and physical parameters of eclipsing binaries from the All-Sky Automated Survey catalogue - I. A sample of systems with components' masses between 1 and 2 M$_{solar}$}",
      journal = {\mnras},
         year = 2009,
        month = dec,
       volume = {400},
       number = {2},
        pages = {969-983},
          doi = {10.1111/j.1365-2966.2009.15513.x},
archivePrefix = {arXiv},
       eprint = {0908.3471},
 primaryClass = {astro-ph.SR},
       adsurl = {https://ui.adsabs.harvard.edu/abs/2009MNRAS.400..969H}
}

@ARTICLE{1966IBVS..140....1S,
       author = {{Strohmeier}, W. and {Fischer}, H. and {Ott}, H.},
        title = "{Bright Southern BV-Stars}",
      journal = {Information Bulletin on Variable Stars},
         year = 1966,
        month = jun,
       volume = {140},
        pages = {1},
       adsurl = {https://ui.adsabs.harvard.edu/abs/1966IBVS..140....1S}
}

@ARTICLE{1974A&A....32..429I,
       author = {{Imbert}, M.},
        title = "{Orbital elements and dimensions of two doublelines eclipsing binaries: W Gru and UX Men.}",
      journal = {\aap},
         year = 1974,
        month = jun,
       volume = {32},
        pages = {429},
       adsurl = {https://ui.adsabs.harvard.edu/abs/1974A&A....32..429I}
}

@ARTICLE{1976A&A....48...49C,
       author = {{Clausen}, J.~V. and {Gr{\o}nbech}, B.},
        title = "{Four-colour photometry of eclipsing binaries. IV: UX Mensae, light curves, photometric elements and absolute dimensions.}",
      journal = {\aap},
         year = 1976,
        month = apr,
       volume = {48},
        pages = {49-53},
       adsurl = {https://ui.adsabs.harvard.edu/abs/1976A&A....48...49C}
}

@ARTICLE{2015A&A...584A...8M,
        author = {{Mikul{\'a}{\v{s}}ek}, Zden{\v{e}}k},
        title = "{Phenomenological modelling of eclipsing system light curves}",
        journal = {\aap},
        year = 2015,
        month = dec,
        volume = {584},
        eid = {A8},
        pages = {A8},
        doi = {10.1051/0004-6361/201425244},
        archivePrefix = {arXiv},
        eprint = {1508.04827},
        primaryClass = {astro-ph.IM},
        adsurl = {https://ui.adsabs.harvard.edu/abs/2015A&A...584A...8M}
}

@ARTICLE{2026A&A...705A.251P,
       author = {{Pawar}, T. and {Miszuda}, A. and {He{\l}miniak}, K.~G. and {Marcadon}, F. and {Moharana}, A. and {Pawar}, G. and {Konacki}, M.},
        title = "{A comprehensive study of {\ensuremath{\delta}} Scuti-type pulsators in eclipsing binaries: Oscillating eclipsing Algols}",
      journal = {\aap},
         year = 2026,
        month = jan,
       volume = {705},
          eid = {A251},
        pages = {A251},
          doi = {10.1051/0004-6361/202554310},
archivePrefix = {arXiv},
       eprint = {2502.20626},
 primaryClass = {astro-ph.SR},
       adsurl = {https://ui.adsabs.harvard.edu/abs/2026A&A...705A.251P}
}

@ARTICLE{2024MNRAS.527...53M,
        author = {{Moharana}, A. and {He{\l}miniak}, K.~G. and {Marcadon}, F. and {Pawar}, T. and {Pawar}, G. and {Garczy{\'n}ski}, P. and {Per{\l}a}, J. and {Koz{\l}owski}, S.~K. and {Sybilski}, P. and {Ratajczak}, M. and {Konacki}, M.},
        title = "{Solaris photometric survey: Search for circumbinary companions using eclipse timing variations}",
        journal = {\mnras},
        year = 2024,
        month = jan,
        volume = {527},
        number = {1},
        pages = {53-65},
        doi = {10.1093/mnras/stad3117},
        archivePrefix = {arXiv},
        eprint = {2310.05890},
        primaryClass = {astro-ph.SR},
        adsurl = {https://ui.adsabs.harvard.edu/abs/2024MNRAS.527...53M}
}

@ARTICLE{2024ApJ...976..242M,
        author = {{Marcadon}, Fr{\'e}d{\'e}ric and {Pr{\v{s}}a}, Andrej},
        title = "{Precision Timing of Eclipsing Binaries from TESS Full Frame Images: Method and Performance}",
        journal = {\apj},
        year = 2024,
        month = dec,
        volume = {976},
        number = {2},
        eid = {242},
        pages = {242},
        doi = {10.3847/1538-4357/ad8571},
        archivePrefix = {arXiv},
        eprint = {2403.07694},
        primaryClass = {astro-ph.SR},
        adsurl = {https://ui.adsabs.harvard.edu/abs/2024ApJ...976..242M}
}

@ARTICLE{quel00,
       author = {{Queloz}, D. and {Mayor}, M. and {Weber}, L. and {Bl{\'e}cha}, A. and {Burnet}, M. and {Confino}, B. and {Naef}, D. and {Pepe}, F. and {Santos}, N. and {Udry}, S.},
        title = "{The CORALIE survey for southern extra-solar planets. I. A planet orbiting the star Gliese 86}",
      journal = {\aap},
         year = 2000,
        month = feb,
       volume = {354},
        pages = {99-102},
       adsurl = {https://ui.adsabs.harvard.edu/abs/2000A&A...354...99Q}
}

@ARTICLE{GaiaESOsurvery,
       author = {{Heiter}, U. and {Lind}, K. and {Bergemann}, M. and {Asplund}, M. and {Mikolaitis}, {\v{S}}. and {Barklem}, P.~S. and {Masseron}, T. and {de Laverny}, P. and {Magrini}, L. and {Edvardsson}, B. and {J{\"o}nsson}, H. and {Pickering}, J.~C. and {Ryde}, N. and {Bayo Ar{\'a}n}, A. and {Bensby}, T. and {Casey}, A.~R. and {Feltzing}, S. and {Jofr{\'e}}, P. and {Korn}, A.~J. and {Pancino}, E. and {Damiani}, F. and {Lanzafame}, A. and {Lardo}, C. and {Monaco}, L. and {Morbidelli}, L. and {Smiljanic}, R. and {Worley}, C. and {Zaggia}, S. and {Randich}, S. and {Gilmore}, G.~F.},
        title = "{Atomic data for the Gaia-ESO Survey}",
      journal = {\aap},
         year = 2021,
        month = jan,
       volume = {645},
          eid = {A106},
        pages = {A106},
          doi = {10.1051/0004-6361/201936291},
archivePrefix = {arXiv},
       eprint = {2011.02049},
 primaryClass = {astro-ph.IM},
       adsurl = {https://ui.adsabs.harvard.edu/abs/2021A&A...645A.106H}
}

@ARTICLE{2003A&A...408..681K,
       author = {{Kervella}, P. and {Th{\'e}venin}, F. and {Morel}, P. and {Bord{\'e}}, P. and {Di Folco}, E.},
        title = "{The interferometric diameter and internal structure of Sirius A}",
      journal = {\aap},
         year = 2003,
        month = sep,
       volume = {408},
        pages = {681-688},
          doi = {10.1051/0004-6361:20030994},
archivePrefix = {arXiv},
       eprint = {astro-ph/0306604},
 primaryClass = {astro-ph},
       adsurl = {https://ui.adsabs.harvard.edu/abs/2003A&A...408..681K}
}

@ARTICLE{maxted2018A&A...616A..39M,
       author = {{Maxted}, P.~F.~L.},
        title = "{Comparison of the power-2 limb-darkening law from the STAGGER-grid to Kepler light curves of transiting exoplanets}",
      journal = {\aap},
         year = 2018,
        month = aug,
       volume = {616},
          eid = {A39},
        pages = {A39},
          doi = {10.1051/0004-6361/201832944},
archivePrefix = {arXiv},
       eprint = {1804.07943},
 primaryClass = {astro-ph.SR},
       adsurl = {https://ui.adsabs.harvard.edu/abs/2018A&A...616A..39M}
}

@ARTICLE{Claret2010A&A...519A..57C,
       author = {{Claret}, A. and {Gim{\'e}nez}, A.},
        title = "{The apsidal-motion test of stellar structure and evolution: an update}",
      journal = {\aap},
         year = 2010,
        month = sep,
       volume = {519},
          eid = {A57},
        pages = {A57},
          doi = {10.1051/0004-6361/201014008},
       adsurl = {https://ui.adsabs.harvard.edu/abs/2010A&A...519A..57C}
}

@ARTICLE{Claret2021A&A...654A..17C,
       author = {{Claret}, A. and {Gim{\'e}nez}, A. and {Baroch}, D. and {Ribas}, I. and {Morales}, J.~C. and {Anglada-Escud{\'e}}, G.},
        title = "{Analysis of apsidal motion in eclipsing binaries using TESS data. II. A test of internal stellar structure}",
      journal = {\aap},
         year = 2021,
        month = oct,
       volume = {654},
          eid = {A17},
        pages = {A17},
          doi = {10.1051/0004-6361/202141484},
archivePrefix = {arXiv},
       eprint = {2107.10765},
 primaryClass = {astro-ph.SR},
       adsurl = {https://ui.adsabs.harvard.edu/abs/2021A&A...654A..17C}
}

@ARTICLE{2017ApJ...840...70B,
       author = {{Bond}, Howard E. and {Schaefer}, Gail H. and {Gilliland}, Ronald L. and {Holberg}, Jay B. and {Mason}, Brian D. and {Lindenblad}, Irving W. and {Seitz-McLeese}, Miranda and {Arnett}, W. David and {Demarque}, Pierre and {Spada}, Federico and {Young}, Patrick A. and {Barstow}, Martin A. and {Burleigh}, Matthew R. and {Gudehus}, Donald},
        title = "{The Sirius System and Its Astrophysical Puzzles: Hubble Space Telescope and Ground-based Astrometry}",
      journal = {\apj},
         year = 2017,
        month = may,
       volume = {840},
       number = {2},
          eid = {70},
        pages = {70},
          doi = {10.3847/1538-4357/aa6af8},
archivePrefix = {arXiv},
       eprint = {1703.10625},
 primaryClass = {astro-ph.SR},
       adsurl = {https://ui.adsabs.harvard.edu/abs/2017ApJ...840...70B}
}

@ARTICLE{2026A&A...710A.394M,
       author = {{Marcadon}, F. and {Moharana}, A. and {Pawar}, T.~B. and {Pawar}, G. and {He{\l}miniak}, K.~G. and {Marques}, J.~P. and {Konacki}, M.},
        title = "{RX Gru: A short-period pre-main-sequence eclipsing binary with a distant circumbinary companion}",
      journal = {\aap},
         year = 2026,
        month = jul,
       volume = {710},
          eid = {A394},
        pages = {A394},
          doi = {10.1051/0004-6361/202557336},
archivePrefix = {arXiv},
       eprint = {2509.17011},
 primaryClass = {astro-ph.SR},
       adsurl = {https://ui.adsabs.harvard.edu/abs/2026A&A...710A.394M}
}

@ARTICLE{2013AN....334....4T,
       author = {{Torres}, G.},
        title = "{Fundamental properties of lower main-sequence stars}",
      journal = {Astronomische Nachrichten},
         year = 2013,
        month = feb,
       volume = {334},
       number = {1-2},
        pages = {4},
          doi = {10.1002/asna.201211743},
archivePrefix = {arXiv},
       eprint = {1209.1279},
 primaryClass = {astro-ph.SR},
       adsurl = {https://ui.adsabs.harvard.edu/abs/2013AN....334....4T}
}

@ARTICLE{2018AJ....155..225K,
       author = {{Kesseli}, Aurora Y. and {Muirhead}, Philip S. and {Mann}, Andrew W. and {Mace}, Greg},
        title = "{Magnetic Inflation and Stellar Mass. II. On the Radii of Single, Rapidly Rotating, Fully Convective M-Dwarf Stars}",
      journal = {\aj},
         year = 2018,
        month = jun,
       volume = {155},
       number = {6},
          eid = {225},
        pages = {225},
          doi = {10.3847/1538-3881/aabccb},
archivePrefix = {arXiv},
       eprint = {1804.04133},
 primaryClass = {astro-ph.SR},
       adsurl = {https://ui.adsabs.harvard.edu/abs/2018AJ....155..225K}
}

@ARTICLE{2021A&A...649A.109G,
       author = {{Graczyk}, D. and {Pietrzy{\'n}ski}, G. and {Galan}, C. and {Gieren}, W. and {Tkachenko}, A. and {Anderson}, R.~I. and {Gallenne}, A. and {G{\'o}rski}, M. and {Hajdu}, G. and {Ka{\l}uszy{\'n}ski}, M. and {Karczmarek}, P. and {Kervella}, P. and {Maxted}, P.~F.~L. and {Nardetto}, N. and {Narloch}, W. and {Pavlovski}, K. and {Pilecki}, B. and {Pych}, W. and {Southworth}, J. and {Storm}, J. and {Suchomska}, K. and {Taormina}, M. and {Villanova}, S. and {Wielg{\'o}rski}, P. and {Zgirski}, B. and {Konorski}, P.},
        title = "{The surface brightness-colour relations based on eclipsing binary stars and calibrated with Gaia EDR3}",
      journal = {\aap},
         year = 2021,
        month = may,
       volume = {649},
          eid = {A109},
        pages = {A109},
          doi = {10.1051/0004-6361/202140571},
archivePrefix = {arXiv},
       eprint = {2103.02077},
 primaryClass = {astro-ph.SR},
       adsurl = {https://ui.adsabs.harvard.edu/abs/2021A&A...649A.109G}
}

@ARTICLE{2014ApJ...797...31T,
       author = {{Torres}, Guillermo and {Sandberg Lacy}, Claud H. and {Pavlovski}, Kre{\v{s}}imir and {Feiden}, Gregory A. and {Sabby}, Jeffrey A. and {Bruntt}, Hans and {Viggo Clausen}, Jens},
        title = "{The G+M Eclipsing Binary V530 Orionis: A Stringent Test of Magnetic Stellar Evolution Models for Low-mass Stars}",
      journal = {\apj},
         year = 2014,
        month = dec,
       volume = {797},
       number = {1},
          eid = {31},
        pages = {31},
          doi = {10.1088/0004-637X/797/1/31},
archivePrefix = {arXiv},
       eprint = {1410.6170},
 primaryClass = {astro-ph.SR},
       adsurl = {https://ui.adsabs.harvard.edu/abs/2014ApJ...797...31T}
}

@ARTICLE{2017ApJ...837....7G,
       author = {{Graczyk}, Dariusz and {Konorski}, Piotr and {Pietrzy{\'n}ski}, Grzegorz and {Gieren}, Wolfgang and {Storm}, Jesper and {Nardetto}, Nicolas and {Gallenne}, Alexandre and {Maxted}, Pierre F.~L. and {Kervella}, Pierre and {Ko{\l}aczkowski}, Zbigniew},
        title = "{The Surface Brightness-color Relations Based on Eclipsing Binary Stars: Toward Precision Better than 1\% in Angular Diameter Predictions}",
      journal = {\apj},
         year = 2017,
        month = mar,
       volume = {837},
       number = {1},
          eid = {7},
        pages = {7},
          doi = {10.3847/1538-4357/aa5d56},
archivePrefix = {arXiv},
       eprint = {1611.09976},
 primaryClass = {astro-ph.SR},
       adsurl = {https://ui.adsabs.harvard.edu/abs/2017ApJ...837....7G}
}

@ARTICLE{1989A&A...211..346A,
       author = {{Andersen}, J. and {Clausen}, J.~V. and {Magain}, P.},
        title = "{Absolute dimensions of eclipsing binaries. XIV. UX Mensae.}",
      journal = {\aap},
         year = 1989,
        month = mar,
       volume = {211},
        pages = {346-352},
       adsurl = {https://ui.adsabs.harvard.edu/abs/1989A&A...211..346A}
}

@ARTICLE{demove2006S&C....16..239T,
       author = {{Ter Braak}, Cajo J.~F.},
        title = "{A Markov Chain Monte Carlo version of the genetic algorithm Differential Evolution: easy Bayesian computing for real parameter spaces}",
      journal = {Statistics and Computing},
         year = 2006,
        month = sep,
       volume = {16},
       number = {3},
        pages = {239-249},
          doi = {10.1007/s11222-006-8769-1},
       adsurl = {https://ui.adsabs.harvard.edu/abs/2006S&C....16..239T}
}

@ARTICLE{mesadotter,
       author = {{Dotter}, Aaron},
        title = "{MESA Isochrones and Stellar Tracks (MIST) 0: Methods for the Construction of Stellar Isochrones}",
      journal = {\apjs},
         year = 2016,
        month = jan,
       volume = {222},
       number = {1},
          eid = {8},
        pages = {8},
          doi = {10.3847/0067-0049/222/1/8},
archivePrefix = {arXiv},
       eprint = {1601.05144},
 primaryClass = {astro-ph.SR},
       adsurl = {https://ui.adsabs.harvard.edu/abs/2016ApJS..222....8D}
}

@ARTICLE{mesachoi,
       author = {{Choi}, Jieun and {Dotter}, Aaron and {Conroy}, Charlie and {Cantiello}, Matteo and {Paxton}, Bill and {Johnson}, Benjamin D.},
        title = "{Mesa Isochrones and Stellar Tracks (MIST). I. Solar-scaled Models}",
      journal = {\apj},
         year = 2016,
        month = jun,
       volume = {823},
       number = {2},
          eid = {102},
        pages = {102},
          doi = {10.3847/0004-637X/823/2/102},
archivePrefix = {arXiv},
       eprint = {1604.08592},
 primaryClass = {astro-ph.SR},
       adsurl = {https://ui.adsabs.harvard.edu/abs/2016ApJ...823..102C}
}

@INPROCEEDINGS{bfsvd,
        author = {{Rucinski}, S.},
        title = "{Determination of Broadening Functions Using the Singular-Value Decomposition (SVD) Technique}",
        booktitle = {IAU Colloq. 170: Precise Stellar Radial Velocities},
        year = 1999,
        editor = {{Hearnshaw}, J.~B. and {Scarfe}, Colin David},
        series = {Astronomical Society of the Pacific Conference Series},
        volume = {185},
        month = jan,
        pages = {82},
        archivePrefix = {arXiv},
        eprint = {astro-ph/9807327},
        primaryClass = {astro-ph},
        adsurl = {https://ui.adsabs.harvard.edu/abs/1999ASPC..185...82R}
}

@ARTICLE{2010A&ARv..18...67T,
       author = {{Torres}, G. and {Andersen}, J. and {Gim{\'e}nez}, A.},
        title = "{Accurate masses and radii of normal stars: modern results and applications}",
      journal = {\aapr},
         year = 2010,
        month = feb,
       volume = {18},
       number = {1-2},
        pages = {67-126},
          doi = {10.1007/s00159-009-0025-1},
archivePrefix = {arXiv},
       eprint = {0908.2624},
 primaryClass = {astro-ph.SR},
       adsurl = {https://ui.adsabs.harvard.edu/abs/2010A&ARv..18...67T}
}

@ARTICLE{phoebe,
       author = {{Pr{\v{s}}a}, A. and {Conroy}, K.~E. and {Horvat}, M. and {Pablo}, H. and {Kochoska}, A. and {Bloemen}, S. and {Giammarco}, J. and {Hambleton}, K.~M. and {Degroote}, P.},
        title = "{Physics Of Eclipsing Binaries. II. Toward the Increased Model Fidelity}",
      journal = {\apjs},
         year = 2016,
        month = dec,
       volume = {227},
       number = {2},
          eid = {29},
        pages = {29},
          doi = {10.3847/1538-4365/227/2/29},
archivePrefix = {arXiv},
       eprint = {1609.08135},
 primaryClass = {astro-ph.SR},
       adsurl = {https://ui.adsabs.harvard.edu/abs/2016ApJS..227...29P}
}

@ARTICLE{krishides2017,
       author = {{He{\l}miniak}, K.~G. and {Ukita}, N. and {Kambe}, E. and {Koz{\l}owski}, S.~K. and {Sybilski}, P. and {Maehara}, H. and {Ratajczak}, M. and {Konacki}, M. and {Paw{\l}aszek}, R.~K.},
        title = "{HIDES spectroscopy of bright detached eclipsing binaries from the Kepler field - II. Double- and triple-lined objects}",
      journal = {\mnras},
         year = 2017,
        month = jun,
       volume = {468},
       number = {2},
        pages = {1726-1746},
          doi = {10.1093/mnras/stx385},
archivePrefix = {arXiv},
       eprint = {1702.03311},
 primaryClass = {astro-ph.SR},
       adsurl = {https://ui.adsabs.harvard.edu/abs/2017MNRAS.468.1726H}
}

@ARTICLE{zuckermazeh,
       author = {{Zucker}, S. and {Mazeh}, T.},
        title = "{Study of Spectroscopic Binaries with TODCOR. I. A New Two-dimensional Correlation Algorithm to Derive the Radial Velocities of the Two Components}",
      journal = {\apj},
         year = 1994,
        month = jan,
       volume = {420},
        pages = {806},
          doi = {10.1086/173605},
       adsurl = {https://ui.adsabs.harvard.edu/abs/1994ApJ...420..806Z}
}

@ARTICLE{claret2004A&A...428.1001C,
       author = {{Claret}, A.},
        title = "{A new non-linear limb-darkening law for LTE stellar atmosphere models III. Sloan filters: Calculations for -5.0 {\ensuremath{\leq}} log [M/H] {\ensuremath{\leq}} +1, 2000 K {\ensuremath{\leq}} T$_{eff}$ {\ensuremath{\leq}} 50 000 K at several surface gravities}",
      journal = {\aap},
         year = 2004,
        month = dec,
       volume = {428},
        pages = {1001-1005},
          doi = {10.1051/0004-6361:20041673},
       adsurl = {https://ui.adsabs.harvard.edu/abs/2004A&A...428.1001C}
}

@ARTICLE{2000A&A...363.1081C,
       author = {{Claret}, A.},
        title = "{A new non-linear limb-darkening law for LTE stellar atmosphere models. Calculations for -5.0 <= log[M/H] <= +1, 2000 K <= T$_{eff}$ <= 50000 K at several surface gravities}",
      journal = {\aap},
         year = 2000,
        month = nov,
       volume = {363},
        pages = {1081-1190},
       adsurl = {https://ui.adsabs.harvard.edu/abs/2000A&A...363.1081C}
}

@ARTICLE{feros,
       author = {{Kaufer}, A. and {Stahl}, O. and {Tubbesing}, S. and {N{\o}rregaard}, P. and {Avila}, G. and {Francois}, P. and {Pasquini}, L. and {Pizzella}, A.},
        title = "{Commissioning FEROS, the new high-resolution spectrograph at La-Silla.}",
      journal = {The Messenger},
         year = 1999,
        month = mar,
       volume = {95},
        pages = {8-12},
       adsurl = {https://ui.adsabs.harvard.edu/abs/1999Msngr..95....8K}
}

@ARTICLE{harps,
       author = {{Mayor}, M. and {Pepe}, F. and {Queloz}, D. and {Bouchy}, F. and {Rupprecht}, G. and {Lo Curto}, G. and {Avila}, G. and {Benz}, W. and {Bertaux}, J. -L. and {Bonfils}, X. and {Dall}, Th. and {Dekker}, H. and {Delabre}, B. and {Eckert}, W. and {Fleury}, M. and {Gilliotte}, A. and {Gojak}, D. and {Guzman}, J.~C. and {Kohler}, D. and {Lizon}, J. -L. and {Longinotti}, A. and {Lovis}, C. and {Megevand}, D. and {Pasquini}, L. and {Reyes}, J. and {Sivan}, J. -P. and {Sosnowska}, D. and {Soto}, R. and {Udry}, S. and {van Kesteren}, A. and {Weber}, L. and {Weilenmann}, U.},
        title = "{Setting New Standards with HARPS}",
      journal = {The Messenger},
         year = 2003,
        month = dec,
       volume = {114},
        pages = {20-24},
       adsurl = {https://ui.adsabs.harvard.edu/abs/2003Msngr.114...20M}
}

@ARTICLE{2006A&A...450..681T,
       author = {{Tokovinin}, A. and {Thomas}, S. and {Sterzik}, M. and {Udry}, S.},
        title = "{Tertiary companions to close spectroscopic binaries}",
      journal = {\aap},
         year = 2006,
        month = may,
       volume = {450},
       number = {2},
        pages = {681-693},
          doi = {10.1051/0004-6361:20054427},
archivePrefix = {arXiv},
       eprint = {astro-ph/0601518},
 primaryClass = {astro-ph},
       adsurl = {https://ui.adsabs.harvard.edu/abs/2006A&A...450..681T}
}

@ARTICLE{2023A&A...674A...1G,
       author = {{Gaia Collaboration} and {Vallenari}, A. and {Brown}, A.~G.~A. and {Prusti}, T. and {de Bruijne}, J.~H.~J. and {Arenou}, F. and {Babusiaux}, C. and {Biermann}, M. and {Creevey}, O.~L. and {Ducourant}, C. and {Evans}, D.~W. and {Eyer}, L. and {Guerra}, R. and {Hutton}, A. and {Jordi}, C. and {Klioner}, S.~A. and {Lammers}, U.~L. and {Lindegren}, L. and {Luri}, X. and {Mignard}, F. and {Panem}, C. and {Pourbaix}, D. and {Randich}, S. and {Sartoretti}, P. and {Soubiran}, C. and {Tanga}, P. and {Walton}, N.~A. and {Bailer-Jones}, C.~A.~L. and {Bastian}, U. and {Drimmel}, R. and {Jansen}, F. and {Katz}, D. and {Lattanzi}, M.~G. and {van Leeuwen}, F. and {Bakker}, J. and {Cacciari}, C. and {Casta{\~n}eda}, J. and {De Angeli}, F. and {Fabricius}, C. and {Fouesneau}, M. and {Fr{\'e}mat}, Y. and {Galluccio}, L. and {Guerrier}, A. and {Heiter}, U. and {Masana}, E. and {Messineo}, R. and {Mowlavi}, N. and {Nicolas}, C. and {Nienartowicz}, K. and {Pailler}, F. and {Panuzzo}, P. and {Riclet}, F. and {Roux}, W. and {Seabroke}, G.~M. and {Sordo}, R. and {Th{\'e}venin}, F. and {Gracia-Abril}, G. and {Portell}, J. and {Teyssier}, D. and {Altmann}, M. and {Andrae}, R. and {Audard}, M. and {Bellas-Velidis}, I. and {Benson}, K. and {Berthier}, J. and {Blomme}, R. and {Burgess}, P.~W. and {Busonero}, D. and {Busso}, G. and {C{\'a}novas}, H. and {Carry}, B. and {Cellino}, A. and {Cheek}, N. and {Clementini}, G. and {Damerdji}, Y. and {Davidson}, M. and {de Teodoro}, P. and {Nu{\~n}ez Campos}, M. and {Delchambre}, L. and {Dell'Oro}, A. and {Esquej}, P. and {Fern{\'a}ndez-Hern{\'a}ndez}, J. and {Fraile}, E. and {Garabato}, D. and {Garc{\'\i}a-Lario}, P. and {Gosset}, E. and {Haigron}, R. and {Halbwachs}, J.-L. and {Hambly}, N.~C. and {Harrison}, D.~L. and {Hern{\'a}ndez}, J. and {Hestroffer}, D. and {Hodgkin}, S.~T. and {Holl}, B. and {Jan{\ss}en}, K. and {Jevardat de Fombelle}, G. and {Jordan}, S. and {Krone-Martins}, A. and {Lanzafame}, A.~C. and {L{\"o}ffler}, W. and {Marchal}, O. and {Marrese}, P.~M. and {Moitinho}, A. and {Muinonen}, K. and {Osborne}, P. and {Pancino}, E. and {Pauwels}, T. and {Recio-Blanco}, A. and {Reyl{\'e}}, C. and {Riello}, M. and {Rimoldini}, L. and {Roegiers}, T. and {Rybizki}, J. and {Sarro}, L.~M. and {Siopis}, C. and {Smith}, M. and {Sozzetti}, A. and {Utrilla}, E. and {van Leeuwen}, M. and {Abbas}, U. and {{\'A}brah{\'a}m}, P. and {Abreu Aramburu}, A. and {Aerts}, C. and {Aguado}, J.~J. and {Ajaj}, M. and {Aldea-Montero}, F. and {Altavilla}, G. and {{\'A}lvarez}, M.~A. and {Alves}, J. and {Anders}, F. and {Anderson}, R.~I. and {Anglada Varela}, E. and {Antoja}, T. and {Baines}, D. and {Baker}, S.~G. and {Balaguer-N{\'u}{\~n}ez}, L. and {Balbinot}, E. and {Balog}, Z. and {Barache}, C. and {Barbato}, D. and {Barros}, M. and {Barstow}, M.~A. and {Bartolom{\'e}}, S. and {Bassilana}, J.-L. and {Bauchet}, N. and {Becciani}, U. and {Bellazzini}, M. and {Berihuete}, A. and {Bernet}, M. and {Bertone}, S. and {Bianchi}, L. and {Binnenfeld}, A. and {Blanco-Cuaresma}, S. and {Blazere}, A. and {Boch}, T. and {Bombrun}, A. and {Bossini}, D. and {Bouquillon}, S. and {Bragaglia}, A. and {Bramante}, L. and {Breedt}, E. and {Bressan}, A. and {Brouillet}, N. and {Brugaletta}, E. and {Bucciarelli}, B. and {Burlacu}, A. and {Butkevich}, A.~G. and {Buzzi}, R. and {Caffau}, E. and {Cancelliere}, R. and {Cantat-Gaudin}, T. and {Carballo}, R. and {Carlucci}, T. and {Carnerero}, M.~I. and {Carrasco}, J.~M. and {Casamiquela}, L. and {Castellani}, M. and {Castro-Ginard}, A. and {Chaoul}, L. and {Charlot}, P. and {Chemin}, L. and {Chiaramida}, V. and {Chiavassa}, A. and {Chornay}, N. and {Comoretto}, G. and {Contursi}, G. and {Cooper}, W.~J. and {Cornez}, T. and {Cowell}, S. and {Crifo}, F. and {Cropper}, M. and {Crosta}, M. and {Crowley}, C. and {Dafonte}, C. and {Dapergolas}, A. and {David}, M. and {David}, P. and {de Laverny}, P. and {De Luise}, F. and {De March}, R.},
        title = "{Gaia Data Release 3. Summary of the content and survey properties}",
      journal = {\aap},
         year = 2023,
        month = jun,
       volume = {674},
          eid = {A1},
        pages = {A1},
          doi = {10.1051/0004-6361/202243940},
archivePrefix = {arXiv},
       eprint = {2208.00211},
 primaryClass = {astro-ph.GA},
       adsurl = {https://ui.adsabs.harvard.edu/abs/2023A&A...674A...1G}
}

@ARTICLE{2020MNRAS.497.2899M,
       author = {{Miller}, N.~J. and {Maxted}, P.~F.~L. and {Smalley}, B.},
        title = "{Fundamental effective temperature measurements for eclipsing binary stars - I. Development of the method and application to AI Phoenicis}",
      journal = {\mnras},
         year = 2020,
        month = sep,
       volume = {497},
       number = {3},
        pages = {2899-2909},
          doi = {10.1093/mnras/staa2167},
archivePrefix = {arXiv},
       eprint = {2004.04568},
 primaryClass = {astro-ph.SR},
       adsurl = {https://ui.adsabs.harvard.edu/abs/2020MNRAS.497.2899M}
}

@ARTICLE{2008A&A...482.1031H,
       author = {{Hensberge}, H. and {Iliji{\'c}}, S. and {Torres}, K.~B.~V.},
        title = "{On the separation of component spectra in binary and higher-multiplicity stellar systems: bias progression and spurious patterns}",
      journal = {\aap},
         year = 2008,
        month = may,
       volume = {482},
       number = {3},
        pages = {1031-1051},
          doi = {10.1051/0004-6361:20079038},
       adsurl = {https://ui.adsabs.harvard.edu/abs/2008A&A...482.1031H}
}

@BOOK{1959cbs..book.....K,
        author = {{Kopal}, Zdenek},
        title = "{Close binary systems}",
        year = 1959,
        adsurl = {https://ui.adsabs.harvard.edu/abs/1959cbs..book.....K}
}

@BOOK{2001icbs.book.....H,
        author = {{Hilditch}, Ronald W.},
        title = "{An Introduction to Close Binary Stars}",
        year = 2001,
        adsurl = {https://ui.adsabs.harvard.edu/abs/2001icbs.book.....H}
}

@ARTICLE{2021A&A...649A...2L,
        author = {{Lindegren}, L. and {Klioner}, S.~A. and {Hern{\'a}ndez}, J. and {Bombrun}, A. and {Ramos-Lerate}, M. and {Steidelm{\"u}ller}, H. and {Bastian}, U. and {Biermann}, M. and {de Torres}, A. and {Gerlach}, E. and {Geyer}, R. and {Hilger}, T. and {Hobbs}, D. and {Lammers}, U. and {McMillan}, P.~J. and {Stephenson}, C.~A. and {Casta{\~n}eda}, J. and {Davidson}, M. and {Fabricius}, C. and {Gracia-Abril}, G. and {Portell}, J. and {Rowell}, N. and {Teyssier}, D. and {Torra}, F. and {Bartolom{\'e}}, S. and {Clotet}, M. and {Garralda}, N. and {Gonz{\'a}lez-Vidal}, J.~J. and {Torra}, J. and {Abbas}, U. and {Altmann}, M. and {Anglada Varela}, E. and {Balaguer-N{\'u}{\~n}ez}, L. and {Balog}, Z. and {Barache}, C. and {Becciani}, U. and {Bernet}, M. and {Bertone}, S. and {Bianchi}, L. and {Bouquillon}, S. and {Brown}, A.~G.~A. and {Bucciarelli}, B. and {Busonero}, D. and {Butkevich}, A.~G. and {Buzzi}, R. and {Cancelliere}, R. and {Carlucci}, T. and {Charlot}, P. and {Cioni}, M.-R.~L. and {Crosta}, M. and {Crowley}, C. and {del Peloso}, E.~F. and {del Pozo}, E. and {Drimmel}, R. and {Esquej}, P. and {Fienga}, A. and {Fraile}, E. and {Gai}, M. and {Garcia-Reinaldos}, M. and {Guerra}, R. and {Hambly}, N.~C. and {Hauser}, M. and {Jan{\ss}en}, K. and {Jordan}, S. and {Kostrzewa-Rutkowska}, Z. and {Lattanzi}, M.~G. and {Liao}, S. and {Licata}, E. and {Lister}, T.~A. and {L{\"o}ffler}, W. and {Marchant}, J.~M. and {Masip}, A. and {Mignard}, F. and {Mints}, A. and {Molina}, D. and {Mora}, A. and {Morbidelli}, R. and {Murphy}, C.~P. and {Pagani}, C. and {Panuzzo}, P. and {Pe{\~n}alosa Esteller}, X. and {Poggio}, E. and {Re Fiorentin}, P. and {Riva}, A. and {Sagrist{\`a} Sell{\'e}s}, A. and {Sanchez Gimenez}, V. and {Sarasso}, M. and {Sciacca}, E. and {Siddiqui}, H.~I. and {Smart}, R.~L. and {Souami}, D. and {Spagna}, A. and {Steele}, I.~A. and {Taris}, F. and {Utrilla}, E. and {van Reeven}, W. and {Vecchiato}, A.},
        title = "{Gaia Early Data Release 3. The astrometric solution}",
        journal = {\aap},
        year = 2021,
        month = may,
        volume = {649},
        eid = {A2},
        pages = {A2},
        doi = {10.1051/0004-6361/202039709},
        archivePrefix = {arXiv},
        eprint = {2012.03380},
        primaryClass = {astro-ph.IM},
        adsurl = {https://ui.adsabs.harvard.edu/abs/2021A&A...649A...2L}
}

@ARTICLE{maxted2020,
       author = {{Maxted}, P.~F.~L. and {Gaulme}, Patrick and {Graczyk}, D. and {He{\l}miniak}, K.~G. and {Johnston}, C. and {Orosz}, Jerome A. and {Pr{\v{s}}a}, Andrej and {Southworth}, John and {Torres}, Guillermo and {Davies}, Guy R. and {Ball}, Warrick and {Chaplin}, William J.},
        title = "{The TESS light curve of AI Phoenicis}",
      journal = {\mnras},
         year = 2020,
        month = oct,
       volume = {498},
       number = {1},
        pages = {332-343},
          doi = {10.1093/mnras/staa1662},
archivePrefix = {arXiv},
       eprint = {2003.09295},
 primaryClass = {astro-ph.SR},
       adsurl = {https://ui.adsabs.harvard.edu/abs/2020MNRAS.498..332M}
}

@ARTICLE{kris2024isofitter,
       author = {{He{\l}miniak}, K.~G. and {Olszewska}, J.~M. and {Puciata-Mroczynska}, M. and {Pawar}, T.},
        title = "{High-resolution spectroscopy of detached eclipsing binaries during total eclipses}",
      journal = {\aap},
         year = 2024,
        month = nov,
       volume = {691},
          eid = {A170},
        pages = {A170},
          doi = {10.1051/0004-6361/202450607},
archivePrefix = {arXiv},
       eprint = {2409.20144},
 primaryClass = {astro-ph.SR},
       adsurl = {https://ui.adsabs.harvard.edu/abs/2024A&A...691A.170H}
}

@ARTICLE{tycHog,
       author = {{H{\o}g}, E. and {Fabricius}, C. and {Makarov}, V.~V. and {Urban}, S. and {Corbin}, T. and {Wycoff}, G. and {Bastian}, U. and {Schwekendiek}, P. and {Wicenec}, A.},
        title = "{The Tycho-2 catalogue of the 2.5 million brightest stars}",
      journal = {\aap},
         year = 2000,
        month = mar,
       volume = {355},
        pages = {L27-L30},
       adsurl = {https://ui.adsabs.harvard.edu/abs/2000A&A...355L..27H}
}

@ARTICLE{KiragASAS,
       author = {{Kiraga}, M.},
        title = "{ASAS Photometry of ROSAT Sources. I. Periodic Variable Stars Coincident with Bright Sources from the ROSAT All Sky Survey}",
      journal = {\actaa},
         year = 2012,
        month = mar,
       volume = {62},
       number = {1},
        pages = {67-95},
          doi = {10.48550/arXiv.1204.3825},
archivePrefix = {arXiv},
       eprint = {1204.3825},
 primaryClass = {astro-ph.SR},
       adsurl = {https://ui.adsabs.harvard.edu/abs/2012AcA....62...67K}
}

@dataset{2mass,
       author = {{Cutri}, R.~M. and {Skrutskie}, M.~F. and {van Dyk}, S. and {Beichman}, C.~A. and {Carpenter}, J.~M. and {Chester}, T. and {Cambresy}, L. and {Evans}, T. and {Fowler}, J. and {Gizis}, J. and {Howard}, E. and {Huchra}, J. and {Jarrett}, T. and {Kopan}, E.~L. and {Kirkpatrick}, J.~D. and {Light}, R.~M. and {Marsh}, K.~A. and {McCallon}, H. and {Schneider}, S. and {Stiening}, R. and {Sykes}, M. and {Weinberg}, M. and {Wheaton}, W.~A. and {Wheelock}, S. and {Zacarias}, N.},
        title = "{VizieR Online Data Catalog: 2MASS All-Sky Catalog of Point Sources (Cutri+ 2003)}",
 howpublished = {VizieR On-line Data Catalog: II/246.  Originally published in: University of Massachusetts and Infrared Processing and Analysis Center, (IPAC/California Institute of Technology) (2003)},
         year = 2003,
        month = jun,
          eid = {II/246},
       adsurl = {https://ui.adsabs.harvard.edu/abs/2003yCat.2246....0C}
}

@ARTICLE{2025MNRAS.544.4611M,
       author = {{Maxted}, P.~F.~L. and {Miller}, N.~J. and {Baycroft}, T.~A. and {Sebastian}, D. and {Triaud}, A.~H.~M.~J. and {Martin}, D.~V.},
        title = "{Fundamental effective temperature measurements for eclipsing binary stars ─ VI. Improved methodology and application to the circumbinary planet host star BEBOP-3}",
      journal = {\mnras},
         year = 2025,
        month = dec,
       volume = {544},
       number = {4},
        pages = {4611-4620},
          doi = {10.1093/mnras/staf1856},
archivePrefix = {arXiv},
       eprint = {2510.23218},
 primaryClass = {astro-ph.SR},
       adsurl = {https://ui.adsabs.harvard.edu/abs/2025MNRAS.544.4611M}
}

@ARTICLE{1997A&A...318..269M,
       author = {{Munari}, U. and {Zwitter}, T.},
        title = "{Equivalent width of NA I and K I lines and reddening.}",
      journal = {\aap},
         year = 1997,
        month = feb,
       volume = {318},
        pages = {269-274},
       adsurl = {https://ui.adsabs.harvard.edu/abs/1997A&A...318..269M}
}

@ARTICLE{2025RNAAS...9..146M,
       author = {{Maxted}, Pierre F.~L.},
        title = "{Equivalent Width of the Interstellar Na I D$_{1}$ and D$_{2}$ Absorption Lines for Stars with E(B ‑ V) < 0.15}",
      journal = {Research Notes of the American Astronomical Society},
         year = 2025,
        month = jun,
       volume = {9},
       number = {6},
          eid = {146},
        pages = {146},
          doi = {10.3847/2515-5172/ade39b},
       adsurl = {https://ui.adsabs.harvard.edu/abs/2025RNAAS...9..146M}
}

@ARTICLE{2014AJ....147..127B,
       author = {{Bohlin}, R.~C.},
        title = "{Hubble Space Telescope CALSPEC Flux Standards: Sirius (and Vega)}",
      journal = {\aj},
         year = 2014,
        month = jun,
       volume = {147},
       number = {6},
          eid = {127},
        pages = {127},
          doi = {10.1088/0004-6256/147/6/127},
       adsurl = {https://ui.adsabs.harvard.edu/abs/2014AJ....147..127B}
}

@ARTICLE{2013MSAIS..24..128A,
       author = {{Allard}, F. and {Homeier}, D. and {Freytag}, B. and {Schaffenberger}, W. and {Rajpurohit}, A.~S.},
        title = "{Progress in modeling very low mass stars, brown dwarfs, and planetary mass objects.}",
      journal = {Memorie della Societa Astronomica Italiana Supplementi},
         year = 2013,
        month = jan,
       volume = {24},
        pages = {128},
          doi = {10.48550/arXiv.1302.6559},
archivePrefix = {arXiv},
       eprint = {1302.6559},
 primaryClass = {astro-ph.SR},
       adsurl = {https://ui.adsabs.harvard.edu/abs/2013MSAIS..24..128A}
}

@ARTICLE{2021Univ....7..369S,
       author = {{Southworth}, John},
        title = "{Space-Based Photometry of Binary Stars: From Voyager to TESS}",
      journal = {Universe},
         year = 2021,
        month = sep,
       volume = {7},
       number = {10},
          eid = {369},
        pages = {369},
          doi = {10.3390/universe7100369},
archivePrefix = {arXiv},
       eprint = {2110.03543},
 primaryClass = {astro-ph.SR},
       adsurl = {https://ui.adsabs.harvard.edu/abs/2021Univ....7..369S}
}

%
%

\begin{appendix} 

\section{Measured parameters -- \JKTEBOP{}}
We provide the full set of light‑curve parameters obtained with \JKTEBOP{} for each of the 31 sectors.
The weighted averages and combined error budget are available in the Table \ref{tab:params_lc} of the main text. 
\begin{table*}[h!]
\centering
    \caption{The set of parameters obtained from modelling with \JKTEBOP{} for each of the {\it TESS} 31 sectors}
    \label{tab:params_jktebop}

\begin{sideways}

\begin{tabular}{ccccccc}
\hline
Sector & $r_{A}$ & $r_{B}$  & $r_{A}+ r_{B}$  & $i$ & $k$ & $J$\\
\hline
1 & $0.093008^{+0.000068}_{-0.000066}$ & $0.087114^{+0.000055}_{-0.000052}$ & $0.180122^{+0.000111}_{-0.000107}$ & $89.653^{+0.005}_{-0.005}$ & $0.936635^{+0.000540}_{-0.000555}$ & $0.980288^{+0.001950}_{-0.001979}$ \\
2 & $0.093167^{+0.000071}_{-0.000069}$ & $0.087090^{+0.000055}_{-0.000053}$ & $0.180256^{+0.000111}_{-0.000109}$ & $89.630^{+0.005}_{-0.005}$ & $0.934746^{+0.000611}_{-0.000588}$ & $0.987734^{+0.001975}_{-0.001983}$ \\
4 & $0.093244^{+0.000103}_{-0.000100}$ & $0.087397^{+0.000107}_{-0.000097}$ & $0.180642^{+0.000173}_{-0.000167}$ & $89.598^{+0.006}_{-0.006}$ & $0.937247^{+0.001168}_{-0.001083}$ & $0.977006^{+0.002623}_{-0.002754}$ \\
5 & $0.092666^{+0.000063}_{-0.000059}$ & $0.087299^{+0.000066}_{-0.000061}$ & $0.179966^{+0.000112}_{-0.000103}$ & $89.644^{+0.006}_{-0.007}$ & $0.942049^{+0.000665}_{-0.000611}$ & $0.993305^{+0.001825}_{-0.001799}$ \\
6 & $0.092703^{+0.000064}_{-0.000058}$ & $0.087161^{+0.000057}_{-0.000057}$ & $0.179865^{+0.000108}_{-0.000103}$ & $89.656^{+0.006}_{-0.006}$ & $0.940188^{+0.000542}_{-0.000530}$ & $0.997163^{+0.001822}_{-0.001883}$ \\
7 & $0.092580^{+0.000103}_{-0.000079}$ & $0.087558^{+0.000172}_{-0.000099}$ & $0.180136^{+0.000211}_{-0.000140}$ & $89.621^{+0.011}_{-0.008}$ & $0.945795^{+0.001945}_{-0.001156}$ & $0.968947^{+0.002119}_{-0.002627}$ \\
8 & $0.092548^{+0.000067}_{-0.000062}$ & $0.087117^{+0.000073}_{-0.000069}$ & $0.179666^{+0.000117}_{-0.000114}$ & $89.620^{+0.007}_{-0.007}$ & $0.941260^{+0.000808}_{-0.000695}$ & $0.958192^{+0.001875}_{-0.001826}$ \\
9 & $0.092483^{+0.000053}_{-0.000050}$ & $0.087159^{+0.000053}_{-0.000051}$ & $0.179643^{+0.000095}_{-0.000092}$ & $89.653^{+0.006}_{-0.006}$ & $0.942420^{+0.000494}_{-0.000471}$ & $0.981179^{+0.001612}_{-0.001613}$ \\
10 & $0.092716^{+0.000069}_{-0.000067}$ & $0.087208^{+0.000079}_{-0.000071}$ & $0.179927^{+0.000121}_{-0.000120}$ & $89.620^{+0.007}_{-0.007}$ & $0.940557^{+0.000887}_{-0.000780}$ & $0.991096^{+0.001961}_{-0.001954}$ \\
11 & $0.092574^{+0.000059}_{-0.000056}$ & $0.087218^{+0.000061}_{-0.000060}$ & $0.179793^{+0.000105}_{-0.000100}$ & $89.635^{+0.007}_{-0.007}$ & $0.942111^{+0.000648}_{-0.000598}$ & $0.988443^{+0.001755}_{-0.001746}$ \\
12 & $0.092825^{+0.000081}_{-0.000079}$ & $0.087356^{+0.000091}_{-0.000083}$ & $0.180183^{+0.000143}_{-0.000135}$ & $89.617^{+0.007}_{-0.008}$ & $0.941055^{+0.001033}_{-0.000921}$ & $0.982720^{+0.002172}_{-0.002300}$ \\
27 & $0.092836^{+0.000059}_{-0.000056}$ & $0.087008^{+0.000049}_{-0.000048}$ & $0.179844^{+0.000099}_{-0.000093}$ & $89.643^{+0.005}_{-0.005}$ & $0.937207^{+0.000483}_{-0.000476}$ & $0.985214^{+0.001781}_{-0.001775}$ \\
28 & $0.092878^{+0.000066}_{-0.000063}$ & $0.087273^{+0.000068}_{-0.000065}$ & $0.180152^{+0.000113}_{-0.000107}$ & $89.617^{+0.006}_{-0.006}$ & $0.939624^{+0.000769}_{-0.000716}$ & $0.983650^{+0.001844}_{-0.001865}$ \\
29 & $0.092855^{+0.000070}_{-0.000069}$ & $0.087201^{+0.000063}_{-0.000059}$ & $0.180057^{+0.000120}_{-0.000117}$ & $89.659^{+0.006}_{-0.006}$ & $0.939104^{+0.000559}_{-0.000548}$ & $0.966753^{+0.001918}_{-0.002025}$ \\
30 & $0.092621^{+0.000059}_{-0.000059}$ & $0.087013^{+0.000054}_{-0.000054}$ & $0.179635^{+0.000103}_{-0.000101}$ & $89.639^{+0.007}_{-0.007}$ & $0.939436^{+0.000536}_{-0.000516}$ & $0.971419^{+0.001806}_{-0.001796}$ \\
31 & $0.092969^{+0.000071}_{-0.000068}$ & $0.087397^{+0.000069}_{-0.000065}$ & $0.180367^{+0.000121}_{-0.000117}$ & $89.628^{+0.007}_{-0.007}$ & $0.940028^{+0.000762}_{-0.000677}$ & $0.960987^{+0.001875}_{-0.001941}$ \\
32 & $0.092653^{+0.000063}_{-0.000062}$ & $0.087494^{+0.000070}_{-0.000069}$ & $0.180147^{+0.000115}_{-0.000108}$ & $89.636^{+0.007}_{-0.007}$ & $0.944278^{+0.000779}_{-0.000707}$ & $0.952614^{+0.001674}_{-0.001664}$ \\
34 & $0.092921^{+0.000076}_{-0.000076}$ & $0.087447^{+0.000082}_{-0.000077}$ & $0.180367^{+0.000135}_{-0.000130}$ & $89.623^{+0.007}_{-0.007}$ & $0.941052^{+0.000881}_{-0.000800}$ & $0.937474^{+0.001896}_{-0.001912}$ \\
35 & $0.092806^{+0.000067}_{-0.000066}$ & $0.087153^{+0.000069}_{-0.000062}$ & $0.179960^{+0.000114}_{-0.000110}$ & $89.615^{+0.007}_{-0.007}$ & $0.939062^{+0.000747}_{-0.000670}$ & $0.950785^{+0.001770}_{-0.001848}$ \\
36 & $0.092855^{+0.000080}_{-0.000077}$ & $0.087270^{+0.000098}_{-0.000085}$ & $0.180129^{+0.000137}_{-0.000128}$ & $89.595^{+0.007}_{-0.007}$ & $0.939829^{+0.001201}_{-0.001040}$ & $0.981345^{+0.002062}_{-0.002113}$ \\
37 & $0.092386^{+0.000055}_{-0.000056}$ & $0.087016^{+0.000063}_{-0.000061}$ & $0.179403^{+0.000099}_{-0.000097}$ & $89.616^{+0.007}_{-0.008}$ & $0.941848^{+0.000722}_{-0.000631}$ & $0.969583^{+0.001668}_{-0.001649}$ \\
38 & $0.092536^{+0.000067}_{-0.000066}$ & $0.087533^{+0.000095}_{-0.000086}$ & $0.180070^{+0.000125}_{-0.000116}$ & $89.616^{+0.008}_{-0.008}$ & $0.945887^{+0.001196}_{-0.001031}$ & $0.962771^{+0.001768}_{-0.001754}$ \\
39 & $0.092755^{+0.000068}_{-0.000065}$ & $0.087435^{+0.000070}_{-0.000063}$ & $0.180191^{+0.000122}_{-0.000115}$ & $89.654^{+0.007}_{-0.007}$ & $0.942634^{+0.000635}_{-0.000580}$ & $0.965422^{+0.001862}_{-0.001814}$ \\
61 & $0.092835^{+0.000093}_{-0.000089}$ & $0.087753^{+0.000125}_{-0.000116}$ & $0.180590^{+0.000159}_{-0.000154}$ & $89.598^{+0.008}_{-0.008}$ & $0.945223^{+0.001600}_{-0.001438}$ & $0.929879^{+0.002113}_{-0.002077}$ \\
62 & $0.093277^{+0.000109}_{-0.000114}$ & $0.087770^{+0.000110}_{-0.000106}$ & $0.181049^{+0.000177}_{-0.000177}$ & $89.602^{+0.008}_{-0.008}$ & $0.940904^{+0.001434}_{-0.001279}$ & $0.939406^{+0.002627}_{-0.002606}$ \\
64 & $0.093149^{+0.000127}_{-0.000118}$ & $0.088266^{+0.000138}_{-0.000129}$ & $0.181417^{+0.000188}_{-0.000178}$ & $89.577^{+0.008}_{-0.008}$ & $0.947485^{+0.001964}_{-0.001786}$ & $0.943576^{+0.002720}_{-0.002758}$ \\
65 & $0.092880^{+0.000115}_{-0.000111}$ & $0.088300^{+0.000160}_{-0.000151}$ & $0.181185^{+0.000193}_{-0.000192}$ & $89.598^{+0.009}_{-0.009}$ & $0.950627^{+0.002118}_{-0.001945}$ & $0.954321^{+0.002462}_{-0.002574}$ \\
66 & $0.093359^{+0.000135}_{-0.000134}$ & $0.088300^{+0.000135}_{-0.000133}$ & $0.181662^{+0.000190}_{-0.000199}$ & $89.593^{+0.007}_{-0.008}$ & $0.945767^{+0.001996}_{-0.001900}$ & $0.962681^{+0.003108}_{-0.003271}$ \\
67 & $0.092234^{+0.000096}_{-0.000100}$ & $0.088182^{+0.000152}_{-0.000141}$ & $0.180415^{+0.000144}_{-0.000136}$ & $89.563^{+0.009}_{-0.009}$ & $0.956021^{+0.002325}_{-0.002089}$ & $0.949843^{+0.002012}_{-0.001965}$ \\
68 & $0.092950^{+0.000169}_{-0.000167}$ & $0.088933^{+0.000195}_{-0.000188}$ & $0.181889^{+0.000225}_{-0.000227}$ & $89.594^{+0.012}_{-0.012}$ & $0.956700^{+0.003138}_{-0.002878}$ & $0.960223^{+0.003842}_{-0.003810}$ \\
69 & $0.093313^{+0.000153}_{-0.000152}$ & $0.088633^{+0.000156}_{-0.000158}$ & $0.181951^{+0.000219}_{-0.000226}$ & $89.597^{+0.010}_{-0.010}$ & $0.949752^{+0.002315}_{-0.002143}$ & $0.956832^{+0.003403}_{-0.003547}$ \\

\hline

\end{tabular}
\end{sideways}

\end{table*}

\begin{table*}[h!]
\centering
    \caption{Continued parameters from \JKTEBOP{}.}
    \label{tab:params_jktebop2}

\begin{sideways}

\begin{tabular}{ccccccc}
\hline
Sector & $e\cos{\omega}$ & $e\sin{\omega}$ & $l_{3}$ &  $l_{B}/l_{A}$ & {\bf $P$} \\
\hline

1 & $0.000492^{+0.000002}_{-0.000002}$ & $-0.001849^{+0.000667}_{-0.000698}$ & $0.013134^{+0.001280}_{-0.001234}$ & $0.862108^{+0.001983}_{-0.001878}$ & $4.181076^{+0.000001}_{-0.000001}$ \\
2 & $0.000588^{+0.000001}_{-0.000001}$ & $-0.003138^{+0.000689}_{-0.000722}$ & $0.014782^{+0.001279}_{-0.001221}$ & $0.865178^{+0.002026}_{-0.001895}$ & $4.181105^{+0.000001}_{-0.000001}$ \\
4 & $0.000531^{+0.000002}_{-0.000002}$ & $-0.005185^{+0.000950}_{-0.001005}$ & $0.016728^{+0.002015}_{-0.001937}$ & $0.876443^{+0.003429}_{-0.003218}$ & $4.181085^{+0.000001}_{-0.000001}$ \\
5 & $0.000607^{+0.000002}_{-0.000002}$ & $0.000512^{+0.000571}_{-0.000614}$ & $0.011707^{+0.001317}_{-0.001172}$ & $0.865377^{+0.002117}_{-0.001907}$ & $4.181073^{+0.000001}_{-0.000001}$ \\
6 & $0.000642^{+0.000002}_{-0.000002}$ & $0.000679^{+0.000598}_{-0.000628}$ & $0.010882^{+0.001250}_{-0.001198}$ & $0.860483^{+0.001929}_{-0.001875}$ & $4.181098^{+0.000001}_{-0.000001}$ \\
7 & $0.000640^{+0.000002}_{-0.000002}$ & $0.001463^{+0.000690}_{-0.000946}$ & $0.013544^{+0.002474}_{-0.001617}$ & $0.870082^{+0.004886}_{-0.002887}$ & $4.181095^{+0.000001}_{-0.000001}$ \\
8 & $0.000573^{+0.000002}_{-0.000002}$ & $0.001740^{+0.000619}_{-0.000640}$ & $0.009826^{+0.001338}_{-0.001321}$ & $0.859275^{+0.002303}_{-0.002183}$ & $4.181102^{+0.000001}_{-0.000001}$ \\
9 & $0.000571^{+0.000002}_{-0.000002}$ & $0.002693^{+0.000496}_{-0.000519}$ & $0.008405^{+0.001119}_{-0.001058}$ & $0.859289^{+0.001719}_{-0.001673}$ & $4.181097^{+0.000001}_{-0.000001}$ \\
10 & $0.000547^{+0.000002}_{-0.000002}$ & $0.000246^{+0.000648}_{-0.000658}$ & $0.012077^{+0.001386}_{-0.001370}$ & $0.868458^{+0.002460}_{-0.002288}$ & $4.181102^{+0.000001}_{-0.000001}$ \\
11 & $0.000592^{+0.000002}_{-0.000002}$ & $0.002225^{+0.000532}_{-0.000567}$ & $0.007878^{+0.001211}_{-0.001157}$ & $0.865653^{+0.002018}_{-0.001913}$ & $4.181088^{+0.000001}_{-0.000001}$ \\
12 & $0.000602^{+0.000002}_{-0.000002}$ & $-0.000620^{+0.000726}_{-0.000789}$ & $0.010979^{+0.001640}_{-0.001565}$ & $0.873914^{+0.002883}_{-0.002647}$ & $4.181099^{+0.000001}_{-0.000001}$ \\
27 & $0.000575^{+0.000002}_{-0.000002}$ & $0.000744^{+0.000579}_{-0.000592}$ & $0.007067^{+0.001137}_{-0.001077}$ & $0.856199^{+0.001720}_{-0.001626}$ & $4.181102^{+0.000001}_{-0.000001}$ \\
28 & $0.000593^{+0.000002}_{-0.000002}$ & $-0.000488^{+0.000591}_{-0.000611}$ & $0.011651^{+0.001277}_{-0.001235}$ & $0.857419^{+0.002172}_{-0.002065}$ & $4.181080^{+0.000001}_{-0.000001}$ \\
29 & $0.000580^{+0.000002}_{-0.000002}$ & $0.000325^{+0.000688}_{-0.000711}$ & $0.012277^{+0.001402}_{-0.001350}$ & $0.855474^{+0.002118}_{-0.002028}$ & $4.181091^{+0.000001}_{-0.000001}$ \\
30 & $0.000561^{+0.000002}_{-0.000002}$ & $0.002771^{+0.000575}_{-0.000592}$ & $0.000617^{+0.001210}_{-0.001191}$ & $0.849577^{+0.001830}_{-0.001723}$ & $4.181101^{+0.000001}_{-0.000001}$ \\
31 & $0.000576^{+0.000002}_{-0.000002}$ & $-0.000406^{+0.000651}_{-0.000681}$ & $0.008154^{+0.001434}_{-0.001353}$ & $0.858515^{+0.002223}_{-0.002110}$ & $4.181089^{+0.000001}_{-0.000001}$ \\
32 & $0.000559^{+0.000002}_{-0.000002}$ & $0.001432^{+0.000553}_{-0.000591}$ & $0.010499^{+0.001343}_{-0.001270}$ & $0.857335^{+0.002161}_{-0.002021}$ & $4.181092^{+0.000001}_{-0.000001}$ \\
34 & $0.000603^{+0.000002}_{-0.000002}$ & $-0.000519^{+0.000715}_{-0.000733}$ & $0.014276^{+0.001570}_{-0.001524}$ & $0.860957^{+0.002532}_{-0.002430}$ & $4.181082^{+0.000001}_{-0.000001}$ \\
35 & $0.000618^{+0.000002}_{-0.000002}$ & $0.000161^{+0.000623}_{-0.000650}$ & $0.010797^{+0.001334}_{-0.001274}$ & $0.857705^{+0.002189}_{-0.002028}$ & $4.181085^{+0.000001}_{-0.000001}$ \\
36 & $0.000585^{+0.000002}_{-0.000002}$ & $-0.000959^{+0.000687}_{-0.000723}$ & $0.010035^{+0.001584}_{-0.001488}$ & $0.865513^{+0.002936}_{-0.002591}$ & $4.181095^{+0.000001}_{-0.000001}$ \\
37 & $0.000555^{+0.000002}_{-0.000002}$ & $0.003324^{+0.000525}_{-0.000519}$ & $0.002456^{+0.001155}_{-0.001146}$ & $0.854512^{+0.002014}_{-0.001869}$ & $4.181098^{+0.000001}_{-0.000001}$ \\
38 & $0.000500^{+0.000002}_{-0.000002}$ & $0.002567^{+0.000576}_{-0.000598}$ & $0.006480^{+0.001478}_{-0.001342}$ & $0.862880^{+0.002662}_{-0.002496}$ & $4.181080^{+0.000001}_{-0.000001}$ \\
39 & $0.000588^{+0.000002}_{-0.000002}$ & $0.000770^{+0.000647}_{-0.000657}$ & $0.010629^{+0.001444}_{-0.001344}$ & $0.864611^{+0.002247}_{-0.002064}$ & $4.181097^{+0.000001}_{-0.000001}$ \\
61 & $0.000527^{+0.000002}_{-0.000002}$ & $-0.000787^{+0.000743}_{-0.000804}$ & $0.017867^{+0.001868}_{-0.001818}$ & $0.863990^{+0.003367}_{-0.003137}$ & $4.181109^{+0.000001}_{-0.000001}$ \\
62 & $0.000603^{+0.000002}_{-0.000002}$ & $-0.003935^{+0.000979}_{-0.000998}$ & $0.019329^{+0.002128}_{-0.002091}$ & $0.867659^{+0.003275}_{-0.003257}$ & $4.181105^{+0.000001}_{-0.000001}$ \\
64 & $0.000554^{+0.000001}_{-0.000001}$ & $-0.003905^{+0.000952}_{-0.001048}$ & $0.025682^{+0.002280}_{-0.002126}$ & $0.879650^{+0.003477}_{-0.003337}$ & $4.181100^{+0.000001}_{-0.000001}$ \\
65 & $0.000543^{+0.000002}_{-0.000002}$ & $-0.000914^{+0.000893}_{-0.000934}$ & $0.023786^{+0.002333}_{-0.002270}$ & $0.875534^{+0.003904}_{-0.003832}$ & $4.181067^{+0.000001}_{-0.000001}$ \\
66 & $0.000531^{+0.000001}_{-0.000002}$ & $-0.006166^{+0.001132}_{-0.001159}$ & $0.027167^{+0.002288}_{-0.002375}$ & $0.887527^{+0.003450}_{-0.003636}$ & $4.181092^{+0.000001}_{-0.000001}$ \\
67 & $0.000568^{+0.000002}_{-0.000002}$ & $0.004079^{+0.000590}_{-0.000594}$ & $0.014027^{+0.001721}_{-0.001631}$ & $0.875237^{+0.003762}_{-0.003559}$ & $4.181092^{+0.000001}_{-0.000001}$ \\
68 & $0.000552^{+0.000002}_{-0.000002}$ & $-0.001651^{+0.001277}_{-0.001319}$ & $0.031204^{+0.002739}_{-0.002786}$ & $0.894937^{+0.004124}_{-0.004307}$ & $4.181089^{+0.000001}_{-0.000001}$ \\
69 & $0.000602^{+0.000002}_{-0.000002}$ & $-0.003520^{+0.001239}_{-0.001285}$ & $0.032946^{+0.002657}_{-0.002713}$ & $0.889541^{+0.003925}_{-0.004082}$ & $4.181091^{+0.000001}_{-0.000001}$ \\

\hline

\end{tabular}
\end{sideways}

\end{table*}

\section{Measured parameters -- \ALLESFITTER{}}
The provided table for the \ALLESFITTER{} analysis, including the posterior distributions from the MCMC runs (100 walkers, 10,000 steps, 2,000 burn‑in). The correlations between parameters derived from the two codes are illustrated in Figure \ref{fig:corel1}.
\begin{table*}[h!]
\centering
    \caption{The set of parameters obtained from modelling with \ALLESFITTER{} for each of the {\it TESS} 31 sectors}
    \label{tab:params_lc_alles}

\begin{sideways}

\begin{tabular}{ccccccc}
\hline
Sector & $r_{A}$ & $r_{B}$  & $r_{A}+ r_{B}$  & $i$ & $k$ & $J$\\
\hline

1 & $0.092807^{+0.000050}_{-0.000054}$ & $0.087035^{+0.000284}_{-0.000299}$ & $0.179843^{+0.000050}_{-0.000056}$ & $89.657^{+0.006}_{-0.006}$ & $0.937806^{+0.000503}_{-0.000526}$ & $0.974808^{+0.000521}_{-0.000941}$ \\
2 & $0.092875^{+0.000048}_{-0.000049}$ & $0.086986^{+0.000290}_{-0.000284}$ & $0.179861^{+0.000045}_{-0.000048}$ & $89.644^{+0.005}_{-0.005}$ & $0.936601^{+0.000519}_{-0.000504}$ & $0.978042^{+0.000612}_{-0.000673}$ \\
4 & $0.092779^{+0.000102}_{-0.000085}$ & $0.087069^{+0.000732}_{-0.000579}$ & $0.179848^{+0.000073}_{-0.000067}$ & $89.607^{+0.009}_{-0.009}$ & $0.938450^{+0.001348}_{-0.001056}$ & $0.978032^{+0.000572}_{-0.000631}$ \\
5 & $0.092662^{+0.000070}_{-0.000070}$ & $0.087304^{+0.000418}_{-0.000394}$ & $0.179966^{+0.000066}_{-0.000072}$ & $89.645^{+0.007}_{-0.007}$ & $0.942181^{+0.000743}_{-0.000692}$ & $0.974665^{+0.001021}_{-0.001459}$ \\
6 & $0.092710^{+0.000064}_{-0.000066}$ & $0.087180^{+0.000355}_{-0.000322}$ & $0.179890^{+0.000067}_{-0.000076}$ & $89.660^{+0.007}_{-0.006}$ & $0.940356^{+0.000623}_{-0.000550}$ & $0.973354^{+0.001204}_{-0.001687}$ \\
7 & $0.092634^{+0.000197}_{-0.000163}$ & $0.087887^{+0.001543}_{-0.001047}$ & $0.180521^{+0.000122}_{-0.000144}$ & $89.603^{+0.011}_{-0.009}$ & $0.948754^{+0.002837}_{-0.001868}$ & $0.973600^{+0.002259}_{-0.002108}$ \\
8 & $0.092624^{+0.000121}_{-0.000104}$ & $0.087254^{+0.000677}_{-0.000612}$ & $0.179877^{+0.000125}_{-0.000102}$ & $89.623^{+0.009}_{-0.009}$ & $0.942022^{+0.001188}_{-0.001085}$ & $0.971958^{+0.002368}_{-0.001686}$ \\
9 & $0.092617^{+0.000064}_{-0.000055}$ & $0.087257^{+0.000383}_{-0.000339}$ & $0.179874^{+0.000061}_{-0.000050}$ & $89.655^{+0.007}_{-0.007}$ & $0.942132^{+0.000681}_{-0.000607}$ & $0.972228^{+0.000983}_{-0.000717}$ \\
10 & $0.092701^{+0.000082}_{-0.000080}$ & $0.087220^{+0.000550}_{-0.000495}$ & $0.179921^{+0.000067}_{-0.000073}$ & $89.621^{+0.008}_{-0.008}$ & $0.940875^{+0.000997}_{-0.000887}$ & $0.980742^{+0.000811}_{-0.001372}$ \\
11 & $0.092666^{+0.000084}_{-0.000068}$ & $0.087271^{+0.000442}_{-0.000403}$ & $0.179937^{+0.000093}_{-0.000066}$ & $89.637^{+0.007}_{-0.007}$ & $0.941787^{+0.000765}_{-0.000716}$ & $0.977896^{+0.001997}_{-0.001115}$ \\
12 & $0.092727^{+0.000106}_{-0.000096}$ & $0.087341^{+0.000732}_{-0.000598}$ & $0.180068^{+0.000083}_{-0.000086}$ & $89.616^{+0.010}_{-0.009}$ & $0.941910^{+0.001333}_{-0.001071}$ & $0.983049^{+0.000864}_{-0.001393}$ \\
27 & $0.092824^{+0.000068}_{-0.000062}$ & $0.087015^{+0.000303}_{-0.000263}$ & $0.179839^{+0.000085}_{-0.000079}$ & $89.640^{+0.006}_{-0.006}$ & $0.937413^{+0.000506}_{-0.000434}$ & $0.974434^{+0.001717}_{-0.001679}$ \\
28 & $0.092795^{+0.000083}_{-0.000075}$ & $0.087253^{+0.000575}_{-0.000479}$ & $0.180048^{+0.000064}_{-0.000065}$ & $89.616^{+0.008}_{-0.008}$ & $0.940285^{+0.001049}_{-0.000863}$ & $0.967948^{+0.000664}_{-0.001040}$ \\
29 & $0.092847^{+0.000063}_{-0.000057}$ & $0.087223^{+0.000351}_{-0.000286}$ & $0.180070^{+0.000065}_{-0.000066}$ & $89.663^{+0.007}_{-0.006}$ & $0.939423^{+0.000616}_{-0.000489}$ & $0.969445^{+0.000858}_{-0.001342}$ \\
30 & $0.092816^{+0.000078}_{-0.000067}$ & $0.087137^{+0.000428}_{-0.000400}$ & $0.179952^{+0.000081}_{-0.000063}$ & $89.638^{+0.008}_{-0.007}$ & $0.938814^{+0.000750}_{-0.000713}$ & $0.969673^{+0.001737}_{-0.000881}$ \\
31 & $0.092899^{+0.000081}_{-0.000072}$ & $0.087376^{+0.000522}_{-0.000427}$ & $0.180275^{+0.000070}_{-0.000069}$ & $89.626^{+0.008}_{-0.007}$ & $0.940547^{+0.000940}_{-0.000757}$ & $0.968714^{+0.000689}_{-0.001270}$ \\
32 & $0.092716^{+0.000091}_{-0.000078}$ & $0.087539^{+0.000520}_{-0.000456}$ & $0.180255^{+0.000092}_{-0.000077}$ & $89.636^{+0.007}_{-0.006}$ & $0.944154^{+0.000912}_{-0.000805}$ & $0.963438^{+0.001563}_{-0.001343}$ \\
34 & $0.092861^{+0.000082}_{-0.000076}$ & $0.087382^{+0.000539}_{-0.000421}$ & $0.180243^{+0.000069}_{-0.000080}$ & $89.631^{+0.007}_{-0.007}$ & $0.940998^{+0.000974}_{-0.000737}$ & $0.968840^{+0.000677}_{-0.001456}$ \\
35 & $0.092776^{+0.000094}_{-0.000085}$ & $0.087187^{+0.000596}_{-0.000507}$ & $0.179963^{+0.000084}_{-0.000080}$ & $89.617^{+0.008}_{-0.008}$ & $0.939765^{+0.001071}_{-0.000904}$ & $0.971262^{+0.001143}_{-0.001554}$ \\
36 & $0.092739^{+0.000108}_{-0.000099}$ & $0.087253^{+0.000778}_{-0.000681}$ & $0.179992^{+0.000077}_{-0.000076}$ & $89.597^{+0.009}_{-0.008}$ & $0.940849^{+0.001428}_{-0.001242}$ & $0.974974^{+0.000573}_{-0.000890}$ \\
37 & $0.092574^{+0.000085}_{-0.000079}$ & $0.087212^{+0.000577}_{-0.000557}$ & $0.179786^{+0.000068}_{-0.000060}$ & $89.609^{+0.008}_{-0.008}$ & $0.942084^{+0.001050}_{-0.001019}$ & $0.970893^{+0.001121}_{-0.000873}$ \\
38 & $0.092611^{+0.000140}_{-0.000103}$ & $0.087623^{+0.000774}_{-0.000648}$ & $0.180234^{+0.000147}_{-0.000094}$ & $89.619^{+0.009}_{-0.008}$ & $0.946135^{+0.001347}_{-0.001156}$ & $0.966162^{+0.003070}_{-0.001739}$ \\
39 & $0.092779^{+0.000083}_{-0.000074}$ & $0.087474^{+0.000473}_{-0.000389}$ & $0.180253^{+0.000084}_{-0.000082}$ & $89.659^{+0.007}_{-0.007}$ & $0.942820^{+0.000831}_{-0.000671}$ & $0.973411^{+0.001281}_{-0.001663}$ \\
61 & $0.092763^{+0.000119}_{-0.000113}$ & $0.087784^{+0.000802}_{-0.000720}$ & $0.180546^{+0.000098}_{-0.000102}$ & $89.611^{+0.008}_{-0.008}$ & $0.946325^{+0.001444}_{-0.001283}$ & $0.962972^{+0.000954}_{-0.001568}$ \\
62 & $0.092855^{+0.000129}_{-0.000083}$ & $0.087476^{+0.000943}_{-0.000561}$ & $0.180331^{+0.000090}_{-0.000066}$ & $89.590^{+0.008}_{-0.008}$ & $0.942065^{+0.001729}_{-0.001018}$ & $0.964535^{+0.000543}_{-0.000914}$ \\
64 & $0.092779^{+0.000137}_{-0.000138}$ & $0.087920^{+0.000892}_{-0.000913}$ & $0.180699^{+0.000119}_{-0.000117}$ & $89.572^{+0.001}_{-0.002}$ & $0.947620^{+0.001595}_{-0.001636}$ & $0.964612^{+0.000508}_{-0.000585}$ \\
65 & $0.092766^{+0.000108}_{-0.000130}$ & $0.087997^{+0.000804}_{-0.000839}$ & $0.180763^{+0.000074}_{-0.000115}$ & $89.588^{+0.008}_{-0.007}$ & $0.948593^{+0.001466}_{-0.001495}$ & $0.963175^{+0.001021}_{-0.002092}$ \\
66 & $0.093035^{+0.000198}_{-0.000255}$ & $0.087856^{+0.001666}_{-0.001229}$ & $0.180891^{+0.000098}_{-0.000303}$ & $89.581^{+0.011}_{-0.008}$ & $0.944330^{+0.003097}_{-0.002070}$ & $0.977239^{+0.003929}_{-0.007725}$ \\
67 & $0.092804^{+0.000057}_{-0.000070}$ & $0.088062^{+0.000468}_{-0.000538}$ & $0.180865^{+0.000030}_{-0.000046}$ & $89.595^{+0.006}_{-0.006}$ & $0.948903^{+0.000865}_{-0.000984}$ & $0.971959^{+0.000998}_{-0.000787}$ \\
68 & $0.092867^{+0.000071}_{-0.000105}$ & $0.088053^{+0.000586}_{-0.000737}$ & $0.180920^{+0.000037}_{-0.000081}$ & $89.570^{+0.006}_{-0.007}$ & $0.948172^{+0.001083}_{-0.001332}$ & $0.970874^{+0.001043}_{-0.000695}$ \\
69 & $0.092993^{+0.000079}_{-0.000089}$ & $0.087825^{+0.000708}_{-0.000680}$ & $0.180818^{+0.000031}_{-0.000056}$ & $89.580^{+0.007}_{-0.008}$ & $0.944426^{+0.001326}_{-0.001249}$ & $0.971074^{+0.001034}_{-0.001576}$ \\

\hline

\end{tabular}
\end{sideways}

\end{table*}

\begin{table*}[h!]
\centering
    \caption{Continued parameters from \ALLESFITTER{}}
    \label{tab:alles_continue}

\begin{sideways}

\begin{tabular}{ccccccc}
\hline
Sector & $e\cos{\omega}$ & $e\sin{\omega}$ & $l_{3}$ & $l_{B}/l_{A}$ & {\bf $P$} \\
\hline

1 & $0.000486^{+0.000008}_{-0.000014}$ & $0.000240^{+0.000073}_{-0.000032}$ & $0.010167^{+0.000638}_{-0.000689}$ & $0.857323^{+0.001377}_{-0.001790}$ & $4.181076^{+0.000001}_{-0.000001}$ \\
2 & $0.000581^{+0.000001}_{-0.000004}$ & $0.000057^{+0.000047}_{-0.000046}$ & $0.010719^{+0.000558}_{-0.000600}$ & $0.857959^{+0.001488}_{-0.001515}$ & $4.181106^{+0.000001}_{-0.000001}$ \\
4 & $0.000522^{+0.000001}_{-0.000005}$ & $-0.000005^{+0.000051}_{-0.000041}$ & $0.007950^{+0.000785}_{-0.000757}$ & $0.861341^{+0.002977}_{-0.002494}$ & $4.181082^{+0.000001}_{-0.000001}$ \\
5 & $0.000613^{+0.000021}_{-0.000026}$ & $0.000599^{+0.000076}_{-0.000069}$ & $0.011903^{+0.000797}_{-0.000856}$ & $0.865215^{+0.002270}_{-0.002566}$ & $4.181071^{+0.000001}_{-0.000001}$ \\
6 & $0.000644^{+0.000027}_{-0.000033}$ & $0.000647^{+0.000095}_{-0.000094}$ & $0.011385^{+0.000800}_{-0.000911}$ & $0.860708^{+0.002206}_{-0.002499}$ & $4.181098^{+0.000001}_{-0.000002}$ \\
7 & $0.000644^{+0.000044}_{-0.000037}$ & $0.001156^{+0.000084}_{-0.000175}$ & $0.017805^{+0.001393}_{-0.001698}$ & $0.876370^{+0.007274}_{-0.005349}$ & $4.181098^{+0.000001}_{-0.000001}$ \\
8 & $0.000576^{+0.000040}_{-0.000031}$ & $0.001119^{+0.000060}_{-0.000197}$ & $0.012375^{+0.001422}_{-0.001153}$ & $0.862520^{+0.004277}_{-0.003482}$ & $4.181106^{+0.000001}_{-0.000001}$ \\
9 & $0.000573^{+0.000004}_{-0.000004}$ & $0.001413^{+0.000009}_{-0.000018}$ & $0.011185^{+0.000774}_{-0.000669}$ & $0.862963^{+0.002119}_{-0.001748}$ & $4.181097^{+0.000001}_{-0.000001}$ \\
10 & $0.000542^{+0.000018}_{-0.000024}$ & $0.000471^{+0.000083}_{-0.000056}$ & $0.012131^{+0.000772}_{-0.000847}$ & $0.868197^{+0.002558}_{-0.002851}$ & $4.181103^{+0.000001}_{-0.000001}$ \\
11 & $0.000590^{+0.000014}_{-0.000010}$ & $0.001526^{+0.000020}_{-0.000078}$ & $0.009631^{+0.001087}_{-0.000821}$ & $0.867357^{+0.003180}_{-0.002308}$ & $4.181087^{+0.000001}_{-0.000001}$ \\
12 & $0.000604^{+0.000016}_{-0.000028}$ & $0.000382^{+0.000104}_{-0.000071}$ & $0.009848^{+0.000992}_{-0.000990}$ & $0.872156^{+0.003234}_{-0.003218}$ & $4.181102^{+0.000001}_{-0.000001}$ \\
27 & $0.000576^{+0.000035}_{-0.000033}$ & $0.000856^{+0.000070}_{-0.000148}$ & $0.007054^{+0.000978}_{-0.000945}$ & $0.856277^{+0.002434}_{-0.002269}$ & $4.181102^{+0.000001}_{-0.000001}$ \\
28 & $0.000596^{+0.000010}_{-0.000017}$ & $0.000329^{+0.000071}_{-0.000045}$ & $0.010608^{+0.000747}_{-0.000752}$ & $0.855798^{+0.002496}_{-0.002490}$ & $4.181079^{+0.000001}_{-0.000001}$ \\
29 & $0.000580^{+0.000018}_{-0.000026}$ & $0.000447^{+0.000087}_{-0.000068}$ & $0.012689^{+0.000828}_{-0.000820}$ & $0.855551^{+0.001880}_{-0.002075}$ & $4.181091^{+0.000001}_{-0.000002}$ \\
30 & $0.000563^{+0.000024}_{-0.000022}$ & $0.000824^{+0.000030}_{-0.000168}$ & $0.004252^{+0.000994}_{-0.000824}$ & $0.854642^{+0.002896}_{-0.002074}$ & $4.181102^{+0.000001}_{-0.000001}$ \\
31 & $0.000576^{+0.000012}_{-0.000023}$ & $0.000296^{+0.000105}_{-0.000053}$ & $0.007181^{+0.000815}_{-0.000813}$ & $0.856952^{+0.002323}_{-0.002503}$ & $4.181090^{+0.000001}_{-0.000001}$ \\
32 & $0.000555^{+0.000028}_{-0.000025}$ & $0.000845^{+0.000050}_{-0.000125}$ & $0.011935^{+0.001092}_{-0.000948}$ & $0.858835^{+0.003053}_{-0.002661}$ & $4.181092^{+0.000001}_{-0.000001}$ \\
34 & $0.000615^{+0.000011}_{-0.000025}$ & $0.000277^{+0.000132}_{-0.000053}$ & $0.013291^{+0.000824}_{-0.000966}$ & $0.857885^{+0.002375}_{-0.002632}$ & $4.181084^{+0.000001}_{-0.000001}$ \\
35 & $0.000620^{+0.000024}_{-0.000033}$ & $0.000523^{+0.000096}_{-0.000097}$ & $0.011038^{+0.000961}_{-0.000969}$ & $0.857778^{+0.002964}_{-0.003023}$ & $4.181087^{+0.000001}_{-0.000001}$ \\
36 & $0.000591^{+0.000007}_{-0.000012}$ & $0.000279^{+0.000062}_{-0.000029}$ & $0.008566^{+0.000872}_{-0.000861}$ & $0.863044^{+0.003128}_{-0.003066}$ & $4.181095^{+0.000001}_{-0.000001}$ \\
37 & $0.000556^{+0.000007}_{-0.000006}$ & $0.001245^{+0.000015}_{-0.000030}$ & $0.006782^{+0.000889}_{-0.000779}$ & $0.861688^{+0.002916}_{-0.002638}$ & $4.181099^{+0.000001}_{-0.000001}$ \\
38 & $0.000493^{+0.000018}_{-0.000012}$ & $0.002043^{+0.000031}_{-0.000132}$ & $0.008498^{+0.001751}_{-0.001154}$ & $0.864881^{+0.005211}_{-0.003671}$ & $4.181080^{+0.000001}_{-0.000001}$ \\
39 & $0.000588^{+0.000029}_{-0.000034}$ & $0.000669^{+0.000086}_{-0.000108}$ & $0.011587^{+0.001010}_{-0.000985}$ & $0.865274^{+0.002663}_{-0.002710}$ & $4.181097^{+0.000001}_{-0.000001}$ \\
61 & $0.000528^{+0.000021}_{-0.000040}$ & $0.000345^{+0.000134}_{-0.000108}$ & $0.018040^{+0.001154}_{-0.001178}$ & $0.862372^{+0.003486}_{-0.003744}$ & $4.181118^{+0.000002}_{-0.000002}$ \\
62 & $0.000598^{+0.000004}_{-0.000011}$ & $0.000182^{+0.000075}_{-0.000034}$ & $0.010993^{+0.001021}_{-0.000803}$ & $0.856012^{+0.003625}_{-0.002660}$ & $4.181097^{+0.000001}_{-0.000001}$ \\
64 & $0.000547^{+0.000001}_{-0.000005}$ & $0.000101^{+0.000036}_{-0.000032}$ & $0.017706^{+0.001458}_{-0.001423}$ & $0.866206^{+0.003372}_{-0.003515}$ & $4.181100^{+0.000001}_{-0.000001}$ \\
65 & $0.000539^{+0.000031}_{-0.000043}$ & $0.000508^{+0.000158}_{-0.000091}$ & $0.019077^{+0.000905}_{-0.001395}$ & $0.866692^{+0.003598}_{-0.004613}$ & $4.181066^{+0.000001}_{-0.000001}$ \\
66 & $0.000526^{+0.000007}_{-0.000018}$ & $-0.002350^{+0.001545}_{-0.000107}$ & $0.018289^{+0.001258}_{-0.003628}$ & $0.871462^{+0.009220}_{-0.010709}$ & $4.181092^{+0.000001}_{-0.000001}$ \\
67 & $0.000541^{+0.000004}_{-0.000015}$ & $0.000059^{+0.000069}_{-0.000118}$ & $0.019966^{+0.000328}_{-0.000574}$ & $0.875168^{+0.002494}_{-0.002523}$ & $4.181093^{+0.000001}_{-0.000001}$ \\
68 & $0.000524^{+0.000003}_{-0.000011}$ & $-0.000022^{+0.000067}_{-0.000106}$ & $0.019788^{+0.000450}_{-0.000963}$ & $0.872845^{+0.002933}_{-0.003078}$ & $4.181089^{+0.000001}_{-0.000001}$ \\
69 & $0.000604^{+0.000015}_{-0.000044}$ & $0.000306^{+0.000147}_{-0.000137}$ & $0.020006^{+0.000287}_{-0.000726}$ & $0.866140^{+0.003355}_{-0.003696}$ & $4.181090^{+0.000001}_{-0.000001}$ \\

\hline

\end{tabular}
\end{sideways}

\end{table*}

\section{Times of minima}\label{app:ecl_times}

In Table \ref{tab:ecl_times}, we provide the times of primary and secondary minima of UX~Men derived from the {\it TESS} sector 1--69 data using the procedure described in Section~\ref{sec:etv}. 

\begin{table*}[]
    \centering
    \caption{Times of primary and secondary minima of UX~Men. Half-integer cycle numbers refer to secondary eclipses. $O-C$ refers to the fitting residuals.}
    \begin{tabular}{ccccc}
    \hline
    Time & Cycle & 1$\sigma$ error  & $\Delta_{\rm obs}$ & $O-C$ \\
    BJD--2457000 & no. & (d) & (s) & (s) \\
    \hline
1326.46011 & $-$392.5 & 0.00028 & 111.7 & $-$16.2 \\
1328.54948 & $-$392.0 & 0.00023 & 10.0  & 10.0    \\
1330.64137 & $-$391.5 & 0.00026 & 125.9 & $-$1.9  \\
1332.73067 & $-$391.0 & 0.00026 & 18.4  & 18.4    \\
1334.82268 & $-$390.5 & 0.00026 & 144.1 & 16.3    \\
.&.&.&.&.\\
.&.&.&.&.\\
.&.&.&.&.\\
.&.&.&.&.\\
.&.&.&.&.\\
.&.&.&.&.\\
3197.49838 & 55.0 & 0.00021 & 0.7     & 0.7     \\
3199.59025 & 55.5 & 0.00023 & 114.7   & $-$13.1 \\
3201.67900 & 56.0 & 0.00040 & $-$40.4 & $-$40.4 \\
3203.77146 & 56.5 & 0.00022 & 124.8   & $-$3.0  \\
3205.86055 & 57.0 & 0.00023 & $-$1.3  & $-$1.3  \\

\hline
\end{tabular}
    \label{tab:ecl_times}
    \tablefoot{In case of a circular ($e=0$) orbit the secondary eclipse would be at exactly half-integer cycle, and $\Delta_{\rm obs}$ would be equal to $O-C$.}
\end{table*}

\section{Radial velocity measurements}
We present RV measurements used in this analysis in the Table~\ref{tab:RV_data}. The UCLES measurements and errors come from \citet{krisuxmen2009MNRAS.400..969H}, but were re-fitted in this work. The CORALIE spectra were used for RVs only, and not for the disentangling.

\begin{table*}[]
    \centering
    \caption{RV measurements of UX~Men, their errors ($\sigma$) and residuals of the fit ($O-C$), all in km/s.}
    \begin{tabular}{cccccccc}
    \hline \hline
      JD-2450000 & $v_{A}$ & $\sigma_{A}$ & $(O-C)_A$ & $v_{B}$ & $\sigma_{B}$ & $(O-C)_B$ & Instrument \\
    \hline
4726.231200 &  10.474 &  0.225 &  0.513 &  88.834 &  0.160 &  0.015 & UCLES \\
4727.271400 & 126.622 &  0.220 & -0.208 & -31.545 &  0.166 & -0.059 & UCLES \\
4747.263900 &  25.567 &  0.225 &  0.036 &  72.924 &  0.171 &  0.143 & UCLES \\
4748.277400 & 131.771 &  0.243 & -0.133 & -36.657 &  0.157 &  0.048 & UCLES \\
4837.127000 &  75.675 &  0.218 & -0.166 &  21.225 &  0.151 &  0.246 & UCLES \\
4838.110600 & -31.653 &  0.218 & -0.097 & 131.483 &  0.151 & -0.120 & UCLES \\
4838.133400 & -32.599 &  0.219 &  0.090 & 132.385 &  0.151 & -0.386 & UCLES \\
4840.095200 & 122.517 &  0.221 & -0.032 & -26.918 &  0.165 &  0.164 & UCLES \\
5060.858812 &  27.338 &  0.156 & -0.024 &  71.098 &  0.191 & -0.061 & HARPS-EGGS \\
5120.685024 & 135.900 &  0.209 &  0.033 & -40.726 &  0.215 & -0.209 & HARPS-EGGS \\
5469.757188 & -38.377 &  0.211 & -0.011 & 139.082 &  0.194 &  0.194 & HARPS-EGGS \\
5469.764144 & -38.461 &  0.224 & -0.144 & 138.847 &  0.191 &  0.009 & HARPS-EGGS \\
5469.800127 & -37.916 &  0.223 & -0.002 & 138.515 &  0.178 &  0.093 & HARPS-EGGS \\
5469.877696 & -36.309 &  0.134 & -0.123 & 136.713 &  0.121 &  0.072 & HARPS-EGGS \\
5479.683361 & 113.196 &  0.156 &  0.006 & -17.296 &  0.159 & -0.105 & HARPS-EGGS \\
5811.829170 &  11.111 &  0.138 &  0.203 &  88.153 &  0.182 &  0.044 & HARPS-HAM \\
5811.877977 &   5.468 &  0.186 &  0.233 &  93.866 &  0.139 & -0.087 & HARPS-HAM \\
5812.824328 & -32.313 &  0.138 &  0.088 & 132.809 &  0.167 &  0.070 & HARPS-HAM \\
5812.897221 & -28.166 &  0.208 &  0.256 & 128.645 &  0.157 &  0.008 & HARPS-HAM \\
5813.843502 &  77.579 &  0.163 &  0.015 &  19.429 &  0.161 & -0.041 & HARPS-HAM \\
6137.939175 &  12.927 &  0.173 &  0.222 &  86.453 &  0.208 &  0.196 & HARPS-HAM \\
6138.912883 & -34.379 &  0.146 & -0.347 & 134.316 &  0.152 & -0.104 & HARPS-HAM \\
6178.864549 & 117.916 &  0.148 &  0.110 & -21.945 &  0.191 & -0.005 & HARPS-HAM \\
6178.915822 & 113.646 &  0.162 &  0.211 & -17.701 &  0.156 & -0.258 & HARPS-HAM \\
6179.914996 &  -5.848 &  0.188 & -0.128 & 105.343 &  0.142 &  0.102 & HARPS-HAM \\
6192.872382 & -35.496 &  0.117 & -0.006 & 135.418 &  0.134 & -0.123 & FEROS \\
6193.892757 &  22.048 &  0.128 &  0.197 &  76.387 &  0.149 & -0.067 & FEROS \\
6194.888912 & 129.904 &  0.238 &  0.172 & -34.624 &  0.104 & -0.036 & FEROS \\
6213.879657 & -38.104 &  0.215 & -0.264 & 138.241 &  0.152 & -0.105 & HARPS-EGGS \\
6237.772605 &  78.278 &  0.265 &  0.116 &  18.562 &  0.203 &  0.038 & CORALIE \\
6238.829160 & -34.114 &  0.301 &  0.089 & 133.773 &  0.357 & -0.494 & CORALIE \\
6239.713035 &   1.443 &  0.334 & -0.019 &  97.631 &  0.212 &  0.121 & CORALIE \\
6241.754266 & 100.954 &  0.376 & -0.250 &  -5.143 &  0.319 &  0.045 & CORALIE \\
6242.761856 & -18.385 &  0.350 & -0.084 & 117.864 &  0.263 & -0.012 & CORALIE \\
6290.613094 &  92.177 &  0.213 & -0.081 &   4.180 &  0.311 &  0.215 & FEROS \\
6291.621462 & 126.682 &  0.142 & -0.215 & -31.820 &  0.155 & -0.148 & FEROS \\
6292.733562 &   1.542 &  0.166 & -0.056 &  97.569 &  0.121 &  0.250 & FEROS \\
6606.663048 & -28.589 &  0.188 &  0.038 & 128.736 &  0.144 & -0.112 & HARPS-EGGS \\
    \hline
    \end{tabular}
    \label{tab:RV_data}
\end{table*}

\section{\ISPEC{} results}
Table~\ref{tab:ispeccomptable} presents individual results of \ISPEC{} spectral analysis of each disentangled spectrum, i.e. for the primary (A) and secondary (B) components deconvolved from HARPS-HAM, HARPS-EGGS, FEROS and UCLES.

\begin{table*}
\caption{Measurement of \teff{},$[M/H]$, $[\alpha/Fe]$, $v_\mathrm{mic}$ and $v_\mathrm{mac}$ for primary (A) and secondary (B) for each spectrograph after distentangling the compsite spectra.}
\label{tab:ispeccomptable}
\centering
\begin{tabular}{lccccc}
\hline

Spectrograph   &$T_{\rm eff}$  & $[M/H]$  & $[\alpha/Fe]$ & $v_\mathrm{mic}$ & $v_\mathrm{mac}$ \\
  &(K) & (dex)  & (dex) & (km/s) & (km/s) \\
\hline

${\rm HARPS_{HAM,A}}$  & $6341.47 \pm 45.41$   & $0.03  \pm  0.01$ & $-0.02 \pm  0.03$   & $1.63 \pm  0.15$  & $5.44  \pm  1.39$  \\
${\rm HARPS_{HAM,B}}$  & $6297.72 \pm  40.24$  & $0.07  \pm  0.01$ & $-0.03 \pm  0.03$   & $1.71 \pm  0.14 $ & $6.97  \pm  0.98$  \\
${\rm HARPS_{EGGS,A}}$& $6242.62 \pm  49.35$  & $0.01 \pm  0.02$ & $0.01  \pm  0.04$   & $1.54 \pm  0.17$  & $6.03  \pm  1.53$  \\
${\rm HARPS_{EGGS,B}}$& $6259.50 \pm  47.89$  & $0.10 \pm  0.02$ & $-0.05 \pm  0.03$   & $1.60 \pm  0.16$  & $7.66  \pm  1.08$  \\
${\rm FEROS_{A}}$      & $6338.16 \pm  80.27$  & $-0.16 \pm  0.09$ & $0.11  \pm  0.09$   & $1.50 \pm  0.28$  & $4.38  \pm  2.98$ \\
${\rm FEROS_{B}}$      & $6346.09 \pm  68.57$  & $-0.07 \pm  0.07$ & $0.06  \pm  0.08$   & $1.47 \pm  0.26$  & $3.32  \pm  2.88$  \\
${\rm UCLES_{A}}$      & $6595.27 \pm  180.14$ & $0.45  \pm  0.08$ & $0.00  \pm  0.08$   & $1.68 \pm  0.33$  & $10.82 \pm  2.33$  \\
${\rm UCLES_{B}}$      & $6265.22 \pm  227.18$ & $0.12  \pm  0.13$ & $-0.01 \pm  0.15$   & $1.53 \pm  0.46$  & $6.79  \pm  3.39$  \\

\hline
\end{tabular}
\end{table*}

\section{Comparison with previous studies}
 The new analysis achieves sub-1\% precision on masses and radii, improving upon \cite{krisuxmen2009MNRAS.400..969H} (\textasciitilde$0.1\%$ for masses, \textasciitilde$1\%$ for radii) and markedly better than \cite{1989A&A...211..346A} (\textasciitilde$0.5\%$ for masses, $>2\%$ for radii) as seen in the Table \ref{tab:comparison}. Temperatures and metallicities are fully consistent across all three studies, with the present work providing the smallest uncertainties due to the extensive 31‑sector {\it TESS} photometry and high‑resolution spectroscopy. The systemic velocity agrees to better than 0.1 km/s, confirming the long‑term stability of the system. 

\begin{table*}
\caption{Comparison of orbital, physical, and atmospheric parameters from previous studies and this work.}
\label{tab:comparison}
\centering
\begin{tabular}{lccc}
\hline
Parameter 
& He{\l}miniak et al. (2009) 
& Andersen et al. (1989) 
& This work \\
\hline
$P$ (d) 
& $4.181096 \pm 0.000003$ 
& $4.181100 \pm 0.000001$ 
& $4.181093974\pm{0.000000094}$ \\

$K_{A}$ (km\,s$^{-1}$) 
& $87.31 \pm 0.12$ 
& $87.41 \pm 0.25$ 
& $87.473 \pm 0.041$ \\

$K_{B}$ (km\,s$^{-1}$) 
& $89.90 \pm 0.08$ 
& $90.28 \pm 0.17$ 
& $90.069 \pm 0.036$ \\

$\gamma$ (km\,s$^{-1}$) 
& $48.86 \pm 0.05$ 
& $48.47 \pm 0.17$ 
& $48.965 \pm 0.029$ \\

$q = M_B/M_A$ 
& $0.9712 \pm 0.0016$ 
& $0.968 \pm 0.003$ 
& $0.9712 \pm 0.0006$ \\

$M_{\rm A}$ (M$_{\odot}$)
& $1.2229\pm 0.0015$
& $1.238\pm 0.006$
&  $1.2300\pm{0.0012}$   \\

$M_{\rm B}$ (M$_{\odot}$)
& $1.1878 \pm 0.0015$
& $1.198 \pm 0.007$
& $1.1946\pm{0.0012}$  \\

$a$ ($R_\odot$) 
& $14.639 \pm 0.012$ 
& $14.678 \pm 0.025$ 
& $14.6668 \pm 0.0045$ \\

$e\sin\omega$ 
& $0$ (fixed) 
& $0.0025 \pm 0.0050$ 
& $0.00095 \pm 0.00205$ \\

$e\cos\omega$ 
& $0$ (fixed) 
& $0.00083 \pm 0.00007$ 
& $0.000572 \pm 0.000037$ \\

$i$ (deg) 
& $89.86 \pm 0.15$ 
& $89.6 \pm 0.1$ 
& $89.614 \pm 0.026$ \\

$R_A$ ($R_\odot$) 
& $1.321 \pm 0.036$ 
& $1.348 \pm 0.013$ 
& $1.3605 \pm 0.0031$ \\

$R_B$ ($R_\odot$) 
& $1.285 \pm 0.037$ 
& $1.274 \pm 0.013$ 
& $1.2801 \pm 0.0069$ \\

$T_{\rm eff,B}/T_{\rm eff,A}$ 
& $0.992 \pm 0.018$ 
& $0.993 \pm 0.032$ 
& $0.998 \pm 0.016$ / $0.988\pm0.009$ \\

$T_{\rm eff,A}$ (K) 
& -- 
& $6200 \pm 100$ 
& $6302 \pm 77$ / $6267 \pm 53$\\

$T_{\rm eff,B}$ (K) 
& -- 
& $6150 \pm 100$ 
& $6292 \pm 65$ / $6198 \pm55$\\

$[{\rm M/H}]_A$ (dex) 
& -- 
& $+0.04 \pm 0.10$ 
& $+0.02 \pm 0.13$ \\

$[{\rm M/H}]_B$ (dex) 
& -- 
& $+0.04 \pm 0.10$ 
& $+0.07 \pm 0.10$ \\

$v\sin i_A$ (km\,s$^{-1}$) 
& -- 
& $16.4 \pm 0.3$ 
& $16.85 \pm 3.61 $ \\

$v\sin i_B$ (km\,s$^{-1}$) 
& -- 
& $15.1 \pm {0.3}$
& $15.19 \pm 3.89$\\

$\tau$ (Gyr) 
& $2.75$ 
& $2.7 \pm 0.3$ 
& $2.75 \pm 0.13$ \\
\hline
\end{tabular}
\end{table*}

\end{appendix}

\end{document}